\documentclass[dvipsnames, twocolumn, table]{aastex631}

\usepackage{amsmath}
\usepackage{booktabs}
\usepackage{bm}
\usepackage{natbib}
\usepackage{mathtools}
\usepackage{multirow}
\usepackage{xcolor}

\definecolor{mgreen}{rgb}{0.1,0.7,0.1}
\newcommand{\nkjm}[1]{{\textcolor{mgreen}{\sf{[NKJ-M: #1]}}}}

\setcitestyle{notesep={ }}

\DeclareMathOperator\sign{sign}

\begin{document}

\title{Effect of Deviations from General Relativity on Searches for Gravitational Wave Microlensing and Type II Strong Lensing}

\author{Mick Wright}
\affiliation{SUPA, School of Physics and Astronomy, University of Glasgow, Glasgow, Scotland}

\author{Justin Janquart}
\affiliation{Department of Physics, Institute for Gravitational and Subatomic Physics (GRASP), Utrecht University, Princetonplein 1, NL-3584 CC Utrecht, The Netherlands}
\affiliation{Nikhef---National Institute for Subatomic Physics, Science Park, NL-1098 XG Amsterdam, The Netherlands}

\author{Nathan K. Johnson-McDaniel}
\affiliation{Department of Physics and Astronomy, The University of Mississippi, University, Mississippi 38677, USA}

\begin{abstract}
    As the gravitational wave detector network is upgraded and the sensitivity of the detectors improves, 
novel scientific avenues open for exploration. For example, tests of general relativity 
will become more accurate as smaller deviations can be probed. 
Additionally, the detection of lensed gravitational waves becomes more likely. 
However, these new avenues could also interact with each other, and a gravitational 
wave event presenting deviations from general relativity could be mistaken for a lensed one. 
Here, we explore how phenomenological deviations from general relativity or binaries of exotic 
compact objects could impact those lensing searches focusing on a single event. 
We consider strong lensing, millilensing, and microlensing and find that certain phenomenological deviations 
from general relativity may be mistaken for all of these types of lensing. 
Therefore, our study shows that future candidate lensing events would need 
to be carefully examined to avoid a false claim of lensing where instead a deviation 
from general relativity has been seen.

\end{abstract}

\section{Introduction}\label{sec:intro}
General Relativity (GR) is the most successful description of gravity to date, 
able to predict the behaviours of many astrophysical phenomena with great accuracy~\citep{astrophysical_gr_testing}
and passing many laboratory tests~\citep{lab_testing_gr}. One such prediction of GR is the existence of gravitational 
waves (GWs), which were first observed in 2015~\citep{gw150914_detection} by the Advanced LIGO 
detectors~\citep{ligo_detector_ref}, later joined by Advanced Virgo~\citep{virgo_detector_ref}, and have 
been observed with increasing regularity ever since---with $\sim$90 detections reported in the latest 
catalog published by the LIGO, Virgo, and KAGRA (LVK) collaborations, GWTC-3~\citep{gwtc3}.
Many more detections are anticipated during the fourth LVK observing run (O4), 
currently underway, and in the future, when existing detectors become more sensitive
and additional ground-based detectors come online \citep[see, e.g.,][]{et_detector,
evans2021horizonstudycosmicexplorer}.

As the detection of GWs moves from exceptional to commonplace, the focus is shifting to investigate a 
range of phenomena that they may allow us to probe. In particular, two active areas of study are the 
search for gravitational lensing of GW signals~\citep{o2_lensing, o3a_lensing, o3_lensing, o3_lensing_followup} 
and more detailed investigations as to whether there is evidence for any deviations from the predictions of 
GR~\citep{gw150914_tgr, gw170817_tgr, gwtc1_tgr, gwtc2_tgr, gwtc3_tgr} in the dynamic, 
highly non-linear regime to which GW signals from compact binary mergers grant access.

To briefly describe GW lensing: similarly to light, GW signals that pass by
massive objects have their propagation modified by that object. These
modifications take a variety of forms depending upon the mass scale of the lens
in question. This leads to a categorisation of lensing phenomena based upon
these scales. In this work, we focus on three of these scales which are of
interest for the lensing of GW signals and define them in terms of the
phenomenology expected to occur to these signals with current ground based
detectors. These scales are \textit{strong lensing}, which results in multiple
distinct signals---called images---which each experience a particular
magnification, phase shift, and time delay~\citep{wang_strong_lensing,
dai_strong_lensing, ezquiaga_strong_lensing}; \textit{millilensing}, in which
the time delay is sufficiently short that the signals overlap resulting in
beating patterns but may still be treated
geometrically~\citep{liu_millilensing}; and \textit{microlensing}, in which the
overlap is sufficient to result in significant frequency-dependent wave-optics
effects in the beating patterns of the signal~\citep{deguichi_microlensing,
takahashi_microlensing, cao_microlensing}.

As already stated, GR has passed all of the laboratory and astrophysical tests that 
have been conducted so far. GW signals offer the ability to test gravity in the dynamical, 
highly non-linear regime. In general, tests using GW signals have focused on placing constraints
 on potential deviations from GR rather than on affirmative testing of non-GR theories. 
 This has been due to a lack of accurate waveform models for inspiral, merger, and 
 ringdown (IMR) in these non-GR theories. 
 Modifications to gravity will broadly affect three areas of the GW\@: 
 generation, with different post-Newtonian coefficients \citep[e.g.,][]{Tahura:2018zuq,bernard_scalar_polarisation,Shiralilou:2021mfl}, as well as differences in the merger-ringdown phase \citep[e.g.,][]{Okounkova:2019zjf,Corman:2022xqg} and quasinormal mode spectrum of the final black hole \citep[e.g.,][]{Srivastava:2021imr,Wagle:2021tam,Cano:2023tmv};
propagation~\citep[e.g.,][]{mirshekari_dispersion, kosteleck_tgr_propagation, mewes_tgr_propagation, Zhu:2023rrx}; 
and the polarisation modes---with a general theory of gravity allowing for up to 
six modes: two tensorial, two vectorial, and two scalar, compared with only the two 
tensorial (termed \textit{plus} and \textit{cross}) modes predicted 
by GR~\citep{eardley_polarisation_modes, eardley_polarisation_modes_letter}.

Some specific examples of additional effects on GWs that modifications to 
gravity may yield would be dispersion~\citep{mirshekari_dispersion} due to 
a massive graviton or Lorentz violations, amplitude birefringence~\citep{okounkova_birefingence}---the explicit 
enhancement or suppression of the left or right circular polarization modes---, or amplitude
dampening~\citep{nishizawa_amplitude_damping, belgacem_amplitude_damping}. 

As of the end of the third observing run (O3) and the construction of GWTC-3, analysis of the detected GW signals 
has not yielded any confident detections of GW 
lensing~\citep{o2_lensing, dai_lensing_search, liu_lensing_search, o3a_lensing, o3_lensing, o3_lensing_followup} nor 
any significant support for physics beyond GR~\citep{gw150914_tgr,gw170817_tgr,gwtc1_tgr,gwtc2_tgr,gwtc3_tgr}. However, with O4 underway and the expected increase 
in the sensitivity of the detectors~\citep{Aasi:2013wya}, there is the potential
for the detection of lensed GW events~(see e.g.~\citep{Ng_2018}) as well as the 
prospect of even greater sensitivity to small deviations from
GR~\cite{gwtc3_tgr}.

Both lensing and deviations from GR have the capacity to cause alterations to GW waveforms. Moreover, 
in some cases, the effect of lensing can be leveraged to test deviations from GR. For instance,
one can probe extra polarisations~\citep{Goyal:2020bkm, MaganaHernandez:2022ayv}, 
birefringence~\citep{Ezquiaga:2020dao, Goyal:2023uvm}, or modified dispersion 
relations including the massive graviton hypothesis~\citep{Finke:2021znb, Narola:2023viz}. 
It is, therefore, vital in the individual searches for these phenomena to be aware of the effect 
that each phenomenon could 
have on the other, and in particular to understand whether false positives may arise in one search as a result 
of the other alteration. 
Here we investigate the effect of some phenomenological deviations from GR or proxies for signals from black hole mimickers
on the searches for GW microlensing and Type II strong lensing---searches which consider only a single GW event at a time.
Specifically, we consider the effect of dispersive propagation from a massive graviton, a modified energy and angular momentum
flux, a modified quasinormal mode spectrum in the ringdown portion of the signal, an additional scalar polarization,
and numerical relativity binary neutron star waveforms scaled to binary black hole-like masses as a proxy for the signal from a
binary of black hole mimickers. We also note that lensing 
could bias 
the result of analyses looking for deviations from GR\@. This complementary approach has been 
investigated in~\citet{Mishra:2023vzo}, where they show that in some cases, microlensed events can be mistaken 
for deviations from GR\@. We refer the reader
to~\cite{gupta2024possiblecausesfalsegeneral} for a review of other potential
phenomena that may be mistaken for deviations from GR.

The rest of the paper adopts the following structure. Section~\ref{sec:lensing}
introduces GW lensing including the three considered regimes.
Section~\ref{sec:tgr} details the potential deviations from GR explored in this
work. Section~\ref{sec:results} discusses the tests performed and their
results. Section~\ref{sec:conclusion} then gives the conclusions and future
prospects from our investigations. Appendix~\ref{app:non-GR_constr} gives
technical details about the construction of the non-GR signals and illustrates
the waveforms we consider. Appendix~\ref{app:prior} outlines the complete prior
choices for the analyses in this work. Appendices~\ref{app:microlensing},
\ref{app:millilensing}, and~\ref{app:SL_extra_res} give additional results from
the microlensing, millilensing, and Type II strong lensing investigations,
respectively.

\section{Gravitational Lensing}\label{sec:lensing}
As previously discussed, the lensing of a GW signal occurs when it passes by a massive object which
modifies its propagation. The exact nature of the modification depends upon the mass scale of the
lensing object, however, the effect in all cases may be described by a relatively simple
relationship. A standard, unlensed GW signal may be described by its frequency-domain strain,
$h^{\textrm{U}}(f)$. After undergoing the process of lensing, the resultant lensed signal strain,
$h^{\textrm{L}}(f)$, is described in terms of the unlensed strain and an \textit{amplification
factor}, $F(f)$, by 
\begin{equation}
	h^{\textrm{L}}(f) = F(f) \times h^{\textrm{U}}(f).
	\label{eq:lensed-strains}
\end{equation}

The amplification factor depends upon the mass density profile of the lensing object. However, 
for any profile it may be calculated from the following general 
expression~\citep{schneider_gravitational_lenses, takahashi_microlensing}:
\begin{equation}
	F(w, \bm{y}) = \frac{w}{2\pi{}i} \int d^{2}x \exp{\left[ iwT(\bm{x},\bm{y}) \right]},
	\label{eq:general-amplification}
\end{equation}
where $w$ is the dimensionless frequency, $\bm{y}$ is the dimensionless
displacement of the source from the optical axis, $\bm{x}$ is the dimensionless
displacement of the image from the optical axis, and the function $T$ yields
the dimensionless time delay of the resulting image. The dimensionless
displacements may be related to the physical positions of the \textit{impact
patameter}, $\bm{\xi}$, and the \textit{source position}, $\bm{\eta}$ by
defining two scales, one for the lens plane, $\xi_{0}$---which is defined on a
per-lens-model basis---and one for the source plane, $\eta_{0} =
\xi_{0}D_{OL}/D_{OS}$, where the quantities $D_{XY}$ represent the angular
distances between the observer $O$, lens $L$, and source $S$. The definitions for $\xi_{0}$
for the two specific lensing models considered in this work are $\xi_{0} =
(4M_{L}D_{OL}D_{LS}/D_{OS})^{1/2}$ for an isolated point mass
lens~\citep{schneider_gravitational_lenses}---where $M_{L}$ is the mass of the
lens---and $\xi_{0} = 4\pi \sigma_{v}^{2} D_{OL}D_{LS}/D_{OS}$ for a singular
isothermal sphere (SIS) lens~\citep{binney_tremaine_galactic_dynamics}, where
$\sigma_v$ is the lens velocity dispersion. With
these definitions, the dimensionless displacements are then given by
\begin{equation}
	\bm{x} = \frac{\bm{\xi}}{\xi_{0}}, \qquad \bm{y} = \frac{\bm{\eta}}{\eta_{0}}.
	\label{eq:dimensionless-scales}
\end{equation}
The conversion from frequency, $f$, to dimensionless frequency, $w$, is given by
\begin{equation}
  w = \frac{D_{OS}}{D_{OL}D_{LS}} \xi_{0}^{2} (1 + z_{L}) 2\pi{}f,
	\label{eq:dimensionless-frequency}
\end{equation}
for a lens at redshift $z_{L}$. In the case of strong lensing, $w \gg 1$ for
the frequencies at which a signal is within the detectable band of the current
ground-based detector network. This allows a simplification of
Equation~\eqref{eq:general-amplification} by taking the geometric optics
approximation. 

In the geometric optics approximation, only the stationary points of the dimensionless time delay function, 
$T(\bm{x},\bm{y})$, contribute, reducing Equation~\eqref{eq:general-amplification} to a summation over these 
points~\citep{nakamura_deguchi_wave_optics}. The stationary points correspond to the individual, 
resolvable strong lensing images, and the $j^{\textrm{th}}$ such image will have its individual 
amplification given by~\citep{nakamura_deguchi_wave_optics, takahashi_microlensing, ezquiaga_strong_lensing}:
\begin{equation}
	F_{j}(f) = |\mu_{j}|^{1/2} \exp
		\left[ 2i\pi t_{j}f - i\pi{}n_{j}\sign(f) \right].
	\label{eq:geometric-amplification}
\end{equation}
Equation~\eqref{eq:geometric-amplification} indicates that the effect of strong lensing is threefold. 
It is the combination of a magnification, $\mu_{j}$, scaling the signal's amplitude; a temporal delay, 
$t_{j}$, of the signal due to both the increased path length and Shapiro delay; and an overall phase 
shift termed the Morse phase. For the latter, the Morse factor, $n_{j}$, may take one of three specific 
values: $0$, $1/2$, or $1$ depending upon whether the stationary point corresponds to a minimum, saddle point, 
or maximum of the time delay function. These cases are referred to as type I, II, and III images, respectively. 

Evidence for an individual signal being type II may be searched for directly, as in this case the phase 
shift yields a frequency dependent time delay for different frequency components. This may be detectable 
in the case of a signal with sufficiently measurable higher order 
modes~\citep{ezquiaga_strong_lensing, Wang:2021kzt, Janquart:2021nus, 2023PhRvD.108d3036V, o3_lensing, o3_lensing_followup}. 
An implementation of such a waveform is implemented in the \textsc{GOLUM} package~\citep{golum, git_golum, 10.1093/mnras/stad2838}.  

The millilensing case is similar to the strong lensing case in that the lens mass is still sufficiently high that the geometric optics approximation may be applied. 
However, the lens is decreased in scale from that of galaxies and galaxy clusters to that of sub-galactic structure. Thus,
the induced time delays are small enough that the resulting signals overlap, resulting in beating patterns within the waveform with the total amplification being described by the summation of each of the images from Equation~\eqref{eq:geometric-amplification}. Implementation of this as a phenomenological modification to a waveform with which to perform parameter estimation based searches was done by~\cite{liu_millilensing}, using the~\textsc{Gravelamps} package~\citep{gravelamps, gravelamps_software}.  

In the case of microlensing, the mass scales of the lensing object are sufficiently low that some or all of the frequencies of the signal inside of the detectable band will not yield $w \gg 1$.  
This means that the geometric optics approximation will not be valid 
for all of the detectable frequencies with current setups and the full wave optics treatment must be 
considered. Equation~\eqref{eq:general-amplification} must then be calculated for whichever mass density 
profile the lens is being used to conduct the investigation.

The simplest class of profiles that may be considered for a lensing object are axially symmetric lenses. 
In this case, Equation~\eqref{eq:general-amplification} may be simplified to~\citep{takahashi_microlensing}: 
\begin{multline}
	F(w, y) = -iwe^{iwy^{2}/2} \\
	\times \int_{0}^{\infty} dx x J_{0} (wxy) 
	\exp{\left[{iw \left( \frac{1}{2}x^{2} - \psi(x) + \phi_{m}(y) \right)}\right]}, 
	\label{eq:axisymmetric-amplification}
\end{multline}
where $J_0$ is a Bessel function, $\psi(x)$ is the lensing potential, and $\phi_{m}(y)$ is the phase 
shift required to enforce that the minimal time delay induced by the lensing is zero. 

The simplest of the axisymmetric lenses is the isolated point mass lens. In this case, the 
lensing potential is given by $\psi(x) = \ln x$, and the phase term is given by 
$\phi_{m}(y) = {(x_{m} - y)}^{2}/2 - \ln x_{m}$, where the image position yielding the minimum
 time delay, $x_{m}$, is given by $y + \sqrt{y^{2} + 4}$. Inserting these into 
 Equation~\eqref{eq:axisymmetric-amplification} yields an expression that may be analytically 
 solved, leading to~\citep{takahashi_microlensing}:
\begin{multline}
	F(w, y) = \exp\left[ \frac{\pi{}w}{4} 
		  + i\frac{w}{2}\ln \left( \frac{w}{2} - 2\phi_{m}(y) \right) \right] \\
		  \times \Gamma \left(1 - \frac{i}{2}w \right) 
		  {}_{1}F_{1} \left( \frac{i}{2}w; 1; \frac{i}{2}wy^{2} \right),
	\label{eq:point-amplification}
\end{multline}
where ${}_1F_{1}$ is the confluent hypergeometric function of the first kind. 
Due to this analytic form of the amplification factor, the isolated point mass is the most 
computationally tractable profile with which to conduct microlensing searches and as such 
has been used in such searches previously~\citep{o2_lensing, o3a_lensing, o3_lensing,o3_lensing_followup}. 
Even where more realistic profiles may be deployed, the isolated point mass profile makes for a useful first 
pass to determine candidates worthy of follow-up~\citep{o3_lensing_followup} to limit the need for 
the use of these more computationally intensive profiles, such as the SIS~\citep{binney_tremaine_galactic_dynamics}
or Navarro, Frenk, White (NFW,~\citealt{nfw_profile}) lens profiles. An implementation of these various profiles 
may be found in the \textsc{Gravelamps} package~\citep{gravelamps, gravelamps_software} leveraging the C arbitrary-precision library \textsc{arb}~\citep{arb}.

\section{Deviations from General Relativity and Generation of Modified Waveforms}\label{sec:tgr}
In general, alternative theories of gravity will affect three aspects
of the GW signal itself: its generation, propagation, and
polarizations---such modifications may affect the detection of the
GW~\citep[see, e.g.]{Narola_2023, Sharma_2024}. We consider phenomenological modifications
to all of these, as well as proxies for waveforms from a binary of black hole mimickers (in GR). Specifically, we consider:
\begin{itemize}
	\item dispersion due to a massive graviton\,;
	\item self-consistent modifications to the energy flux and the quasi-normal mode (QNM) spectrum as proxies for general modifications arising from non-GR theories\,;
	\item addition of a scalar polarization\,;
	\item scaled waveforms from a binary neutron star (BNS) system as a proxy for a black hole mimicker.
\end{itemize}
The philosophy of the non-GR waveform construction is to include an interesting
non-GR effect in a controlled manner, not intending to emulate any specific
alternative theory, except for the massive graviton case (where this still is
just a phenomenological introduction of dispersive propagation). These waveform
constructions are also used in a mock data challenge being carried out by the
LVK testing GR group. These phenomenologies are some of those that are possible
in extended theories, but are by no means exhausitive---for instance, they do
not include birefringent propagation effects. These cases were, however, deemed a
sufficient selection for an initial investigation. We now describe exactly
how these waveforms are constructed.  These waveforms are all illustrated in
Appendix~\ref{subsec:app_wf_illustration}.

All of the non-GR waveforms are constructed based on
the \texttt{TEOBResumS-v3-GIOTTO} effective-one-body GR waveform model~\citep{Damour:2014sva,Nagar:2015xqa,nagar_teobresums,Nagar:2019wds,Akcay:2020qrj,Riemenschneider:2021ppj,Gamba:2021ydi}. 
We use this as a model for precessing binary black holes on quasicircular orbits. 
It contains the dominant $(2,\pm 2)$ modes as well as 
the $(2, \pm 1)$, $(3, \pm 3)$, $(3, \pm 2)$, $(4, \pm 4)$, and $(5, \pm 5)$ higher modes 
(in the coprecessing frame) with a robust merger-ringdown model~\citep{Nagar:2020pcj}. 
All these modes are included in our injections.

\subsection{Dispersion and Massive Graviton}\label{subsec:massive-graviton-theory}

In GR, the propagation of GWs is non-dispersive and consequently follows the relation $E = p^{2}c^{2}$ 
between the energy, $E$, and momentum, $p$, of the wave. In non-GR theories, the propagation may be
dispersive. The following parametric relation describes the leading effect of dispersion in a number of non-GR
scenarios~\citep{mirshekari_dispersion}:
\begin{equation}
	E^{2} = p^{2}c^{2} + A_{\alpha}p^{\alpha}c^{\alpha},
	\label{eq:dispersion-relation}
\end{equation}
where the phenomenological parameters $\alpha$ and $A_{\alpha}$ describe the dispersion.

In the specific case of a massive graviton with no other dispersive effects, special relativity 
predicts~\citep{will_massive_graviton}:
\begin{equation}
	E^{2} = p^{2}c^{2} + m_{g}^{2}c^{4},
	\label{eq:dispersion-massive-graviton}
\end{equation}
which is a special case of Equation~\eqref{eq:dispersion-relation} with $\alpha = 0$ and $\sqrt{A_{0}} = m_{g}c^{2} > 0$. 
The effect of this dispersion is to introduce a dephasing of the entire signal in the frequency domain compared 
with the GR prediction~\citep{will_massive_graviton}: 
\begin{equation}
	\delta\Psi(f) = \frac{\pi c D_{0}}{\lambda_{g}^{2}f}.
	\label{eq:massive-graviton-dephasing}
\end{equation}
Here, $\lambda_{g} = h/(m_{g}c)$ is the Compton wavelength of the graviton and $D_{0}$ is 
a distance measure given by~\citep{mirshekari_dispersion}:
\begin{equation}
	D_{0} = \frac{1 + z}{H_{0}} \int_{0}^{z} \frac{dz'}{{(1+z')}^{2}\sqrt{\Omega_{M}{(1+z')}^{3} + \Omega_{\Lambda}}},
	\label{eq:distance-measure}
\end{equation}
where $z$ is the binary's redshift, $H_0 = 67.90 \text{ km s}^{-1} \text{ Mpc}^{-1}$ is the Hubble constant, $\Omega_M = 0.3065$ is the matter density, and
$\Omega_\Lambda = 0.6935$ is the dark energy density. Here we use the TT+lowP+lensing+ext results from \cite{Ade:2015xua} as in the LVK analysis.
Using all the high significance binary black hole detections through the O3 observing run, the LVK Collaboration has set bounds on the possible mass of the 
graviton by constraining this dispersive effect, giving $m_g \leq 1.27 \times 10^{-23} \text{ eV}/c^2$ at 90\% credibility~\citep{gwtc3_tgr},
recently improved to $m_g \leq 9.6 \times 10^{-24} \text{ eV}/c^2$ by simultaneously modeling the astrophysical population of compact binaries~\citep{Payne:2023kwj}.

\subsection{Modifications to Energy Flux and QNM Spectrum}\label{subsec:alpha2-qnm-theory}

Non-GR theories often lead to differences in all of the GW energy and angular momentum fluxes (and also the linear momentum flux), the binary's binding energy,
and the QNM spectrum of the final black hole.\footnote{See the discussion in Section~\ref{sec:intro}, where the first two deviations lead to the deviations in post-Newtonian
coefficients discussed there.} However, for the purposes of these investigations, we isolate the effect of modifications of 
the energy flux and QNM spectrum specifically, recalling that we are not trying to emulate any specific theory. In both cases, we set the mass and spin of the 
final black hole used to obtain the QNM frequencies and damping times in the model iteratively, so that they satisfy energy and angular momentum balance (since the
energy and angular momentum carried away by the gravitational waves are modified in both cases).

For the modification to the energy and angular momentum flux, we generalize the construction used in~\citet{ghosh_modified_energy_flux, ghosh_modified_energy_flux_2, johnson_mcdaniel_modified_energy_flux}, which applied it to an older nonspinning effective-one-body model, to the modern \texttt{TEOBResumS-v3-GIOTTO} model.
Specifically, we modify the energy flux by multiplying those waveform modes which enter at the 
second post-Newtonian ($2$PN) order by a factor $a_{2}^{1/2}$, so their contribution to the energy and angular momentum fluxes is
multiplied by a factor $a_{2}$. One thus gets modifications to the GW phasing at $2$PN, $3$PN, and all higher PN orders. This is similar to what one expects for the modifications
to the GW phasing in non-GR theories. For instance, in dynamical Chern-Simons theory, the GW phasing also starts to differ from GR at $2$PN \citep{Yagi:2011xp}.
The modes that first enter at $2$PN are the $(3, \pm2), (4, \pm4)$, and $(4, \pm2)$ modes. We only modify the
$(3, \pm 2)$ and $(4, \pm 4)$ modes, since those are the ones that first enter the energy flux at $2$PN and have a robust merger-ringdown model in
\texttt{TEOBResumS-v3-GIOTTO}. The contribution from the omitted $(4, \pm 2)$ modes is much smaller than that from the $(3, \pm 2)$ and $(4, \pm 4)$ modes
(only $\sim 5\times 10^{-4}$ of the total Newtonian energy flux from this set of modes).

We consider both the case where the $(3, \pm2)$ and $(4, \pm4)$ modes in the GW signal are multiplied by $a_{2}^{1/2}$, as well as the case where they are not (so the
additional energy and angular momentum is radiated in a field that does not couple to GW detectors). We refer to these as the cases with and without waveform scaling,
respectively. We consider the latter case in order to have larger deviations in the PN coefficients, length of the waveform (from a given frequency), and final mass and spin
(since such larger deviations were found to be necessary to obtain a detectable GR deviation in the earlier studies that just analyzed the dominant modes) without having the
significantly non-GR waveform features caused by significantly increasing the amplitude of the higher modes. The waveforms are illustrated in Appendix~\ref{subsec:app_wf_illustration}, and Appendix~\ref{subsec:app_Mfaf} shows the final mass and spin obtained in this construction and with the modified QNM spectrum.

As a proxy for the modified QNM spectrum one would obtain in a non-GR theory, we modify the QNM spectrum from that of Kerr to that of Kerr-Newman, which includes a charge
parameter in addition to the mass and spin. We use the results for the $(2,2,0)$ and $(3,3,0)$ QNMs computed in \cite{dias_qnm} \citep[see also][]{Carullo:2021oxn}. Here the
$(\ell, m, n)$ notation includes the overtone index $n$ in addition to the multipolar indices $\ell$ and $m$. The additional QNMs computed in \cite{Dias:2022oqm} were not available
when these waveforms were first constructed, so we do not use them here, but instead construct the additional QNMs needed for the \texttt{TEOBResumS-v3-GIOTTO} ringdown model
(the $n = 0, 1$ QNMs for each $\ell, m$ mode included) of these using simple expressions based on the eikonal approximation or spin expansion, as discussed in
Appendix~\ref{subsec:app_KN_QNM_scaling}. 
As shown in \cite{johnson_mcdaniel_modified_energy_flux}, a number of standard tests of GR applied by the LVK \citep[in, e.g.,][]{gwtc3_tgr} can detect sufficiently large
modified energy fluxes. There are also standard tests sensitive to modified QNM spectra, including dedicated ringdown analyses \citep[see, e.g.,][]{gwtc3_tgr}. None of these tests has
detected a deviation from GR. However, it is not straightforward to translate either of these constraints into constraints on $a_2$ or the charge of the final black hole in these waveforms.

\subsection{Addition of Scalar Polarization}\label{subsec:scalar-polarization-theory}

Here we want to add a scalar polarization to a GR waveform. We would expect any alternative theory that radiates in additional polarizations to have a different tensor waveform
than in GR, due, e.g., to the change in phasing from the extra energy radiated in the non-tensor modes. However, here we leave the tensor polarization waveform unchanged,
in order to isolate the effects of the additional polarization. We use the leading tensor-scalar waveform results from \cite{bernard_scalar_polarisation} to scale the tensor polarizations,
specializing to the case where the sensitivities of the two objects are equal. In this case, there is no dipole radiation, which is consistent with our construction, since there
is no dipole radiation present in the tensor waveform. This choice
gives nonzero scalar modes for the spherical harmonic modes where $\ell + m$ is even, where the specific expressions we use are given in Appendix~\ref{subsec:app_ST_scaling}. The contribution from scalar modes is parameterized by a ratio $\mathcal{A}$ of dominant mode scalar to tensor contributions to the response of the detector with the largest tensor mode SNR.

Additionally, we still apply this scaling to the modes where $\ell + m$ is even in the inertial frame even for precessing systems as we consider here. This is a further simplification of the
behavior one would expect in a precessing system emitting scalar radiation, where all the scalar modes in the inertial frame will be nonzero, due to the binary's precession, even though
only the scalar modes where $\ell + m$ is even are nonzero in the coprecessing frame. However, it is in keeping with the general philosophy of having a simple somewhat physically
motivated addition of the scalar modes to the signal.

Again, the LVK Collaboration has looked for possible extra polarizations in observed BBH events, finding no
support for possible additional ones \citep[see][for the latest results]{gwtc3_tgr}. However, we do not try to
translate those constraints into any constraints on these scalar-tensor waveforms.

\subsection{Black Hole Mimickers}\label{subsec:scaled-bns-theory}

We now consider proxies for waveforms from black hole mimickers. Black hole mimickers can be created by
additional fields in GR, such as boson stars~\citep{Liebling:2012fv}, or more generally dark matter stars~\citep{giudice_dark_matter_star},
or by more exotic non-GR phenomena, such as firewalls~\citep{almheiri_firewalls} or gravastars~\citep{mazur_gravstars}.
See \cite{Cardoso:2019rvt} for a review. We would ideally consider waveforms from numerical simulations of binaries of black hole mimickers.
However, such simulations have not yet reached the same level of sophistication as simulations of binaries of black holes and/or neutron stars, even
for binary boson stars, the best-modeled black hole mimickers \citep[but see][for recent work on simulations of mergers of quasicircular binary boson stars]{Siemonsen:2023age}. We thus consider BNS waveforms scaled to BBH-like masses, as in \cite{Johnson-McDaniel:2018cdu}, since waveforms from binaries of black hole mimickers
are also expected to have the effects of a non-zero tidal deformability \citep[see, e.g.,][]{Cardoso:2017cfl} and non-Kerr spin-induced multipole moments \citep[see, e.g.,][]{Uchikata:2016qku,Adam:2022nlq} in the inspiral and a post-merger
signal that can be different from a standard GR black hole ringdown \citep[see, e.g.,][]{Siemonsen:2023age}.

We consider two specific waveforms, both from the CoRe catalog~\citep{Dietrich:2018phi, Gonzalez:2022mgo}, and including all spherical harmonic modes through $\ell = 4$:
\begin{itemize}
	\item{The numerical relativity (NR) simulation of a precessing BNS with masses of $1.35M_{\odot}$ and $1.11M_{\odot}$ and dimensionless spins of
	$(-0.077,-0.077,-0.077), (-0.090, -0.090, -0.090)$ at a dimensionless frequency of $G Mf/c^3 = 4.76\times 10^{-3}$ from~\citet[CoRe ID BAM:0110]{dietric_scaled_bns}.
	This simulation uses the SLy~\citep{douchin_eos} equation of state (EOS), so the stars have dimensionless tidal deformabilities of $392$ and $1291$, giving an effective
	dimensionless tidal deformability, $\tilde{\Lambda}$ \citep{Wade:2014vqa}, of $724$.}
	\item{The NR simulation from~\cite{ujevic_scaled_bns} of a non-spinning BNS with masses of $1.72M_{\odot}$ and $0.98M_{\odot}$ (CoRe ID BAM:0131). This
	simulation also uses the SLy EOS, so the stars have dimensionless tidal deformabilities of $66$ and $2569$, giving an effective dimensionless tidal deformability of $509$.
	This case is chosen since it has a more BBH-like post-merger signal than the precessing binary does (see the plots in Appendix~\ref{subsec:app_wf_illustration}).}
\end{itemize}
An approach used to look for possible exotic compact objects is to look for echoes, repeated signals due to signal
reflecting multiple times off radial potential barriers. So far, no such echoes have been 
detected~\citep{gwtc3_tgr}. However, echoes will only be present for some binaries of black hole mimickers, and are
not present for, e.g., mergers of binary boson stars or in the scaled BNS waveforms. We also note that for some particular
events, some more in-depth analyses with exotic
objects have been performed. For example, GW190521, an event with exceptionally high 
masses~\citep{2020PhRvL.125j1102A,2020ApJ...900L..13A}, 
has been analyzed with waveforms for head-on Proca star collisions, finding a small preference for such waveforms compared to
BBH waveforms~\citep{190521_proca_star,CalderonBustillo:2022cja}. The latter reference also analyzes some other high-mass events with the Proca star collision waveforms, but does not find these waveforms to be favored for any firm detection.

\section{Results}\label{sec:results}
In general, if one analyses a signal that is not described by the
underlying waveform model, then that analysis will attempt to use any freedom
available in the model to best fit the model to the provided data. Depending
on the specific ways in which parameters affect the model, this can lead to a
spurious apparent match with the analysis model. In this context, to
investigate how each of the deviations from GR or the binary black hole
hypothesis described in Section~\ref{sec:tgr} would impact the single image
lensing searches, a series of 65 simulated signals using as their
basis GW150914-like parameters \citep[cf.][]{gw150914_detection,gwtc2.1}, given
in Table~\ref{tab:GW150914-parameters}, was constructed. This total
number was broken down across the deviation types as follows:

\begin{itemize}
	\item{\textit{Massive Graviton}: $20$ logarithmically spaced values of $A_{0}$ from $5 \times 10^{-45}\,\mathrm{eV}^2$ to $5 \times 10^{-43}\,\mathrm{eV}^2$, corresponding to $m_g \simeq 7.07\times10^{-23}\,\mathrm{eV}/c^2$ and $m_g \simeq 7.07\times10^{-22}\,\mathrm{eV}/c^2$, respectively.}
	\item{\textit{Modified Energy Flux}: The following values of $a_{2}$ with waveform scaling: $2, 5, 10, 20, 50$ and the following values without waveform scaling: $20, 50, 75, 100, 125, 150, 175, 200$.}
	\item{\textit{Modified QNM Spectrum}: Values of the final dimensionless charge, $Q_{f}$, from $0.50$ to $0.75$ in steps of $0.05$.}
	\item{\textit{Scalar Polarization}: Values of $\mathcal{A}$ from $0.05$ to $0.50$ in steps of $0.05$.}
	\item{\textit{Precessing Black Hole Mimicker following~\citet{dietric_scaled_bns} (BHM-P)}: Scaling the total mass from $90M_{\odot}$ to $150M_{\odot}$ in steps of $10M_{\odot}$.}
	\item{\textit{Non-Spinning Black Hole Mimicker following~\citet{ujevic_scaled_bns} (BHM-NS)}: Scaling the total mass from $80M_{\odot}$ to $160M_{\odot}$ in steps of $10M_{\odot}$.}
\end{itemize}

\begin{center}
\begin{deluxetable}{cc}\label{tab:GW150914-parameters}
	\tablehead{Source Parameter & Value}

	\tablecaption{Source parameter value used in the GW150914-like injections.}

	\startdata
	Total Mass, $M_{\textrm{tot}}$ & $72 M_{\odot}$ \\
	Mass Ratio, $q$ & 0.8 \\
	Luminosity Distance, $d_{L}$ & 452 Mpc \\
	Inclination, $\iota$ & 1.59 rad \\
	Right Ascension, RA & 1.68 rad \\
	Declination, dec & -1.27 rad \\
	Polarisation Angle, $\psi$ & 3.93 rad \\
	Coalescence Time, $t_{c}$ & 1126259462 s \\
	Primary Dimensionless Spin Vector, $\bm{\chi}_{1}$ & $(-0.1, -0.1, -0.1)$\\
	Secondary Dimensionless Spin Vector, $\bm{\chi}_{2}$ & $(-0.1, 0.0, 0.1)$\\
	\enddata

\end{deluxetable}
\end{center}

These choices for the non-GR parameters were designed to give cases where one expects the GR deviation to be easily detectable by at least some analyses
as well as cases where it will be less easily detectable. In particular, the graviton masses are all well above the constraints set by the LVK, since those come from combining
together the constraints from many events, while we only analyze single events here. In the black
hole mimicker cases, the lower bound on the total mass is chosen to ensure sufficient higher mode
content is within the LVK-band (i.e.,\ from $20$~Hz)---here the lowest mass ensures that the $|m| \leq 3$ modes are within band. The upper bound is chosen to retain a sufficient number of inspiral cycles in band for analysis.
Each simulated signal was 
then subjected to the \textsc{Gravelamps} micro- and millilensing analysis pipeline and the \textsc{GOLUM} 
type II analysis, relying on the \textsc{bilby} inference framework~\citep{Ashton_2019} and using the \textsc{dynesty}~\citep{dynesty} nested sampler and the
\texttt{IMRPhenomXPHM}~\citep{imrphenomxphm} model
for precessing binary black hole waveforms implemented in \textsc{LALSuite}~\citep{lalsuite}.
\texttt{IMRPhenomXPHM} includes the same higher modes in the coprecessing frame as the \texttt{TEOBResumS-v3-GIOTTO}
model, except for the $(5,\pm 5)$ modes. The injected frames were made without noise, and were
analysed under the assumption of the predicted O4 noise curves for the LIGO-Virgo network, using the LIGO noise curve with a $190$ Mpc horizon~\citep{Aasi:2013wya}. 
This yielded SNRs
for the individual injections between 21.81 and 65.88 which is sufficiently high for obvious
detection. The configurations for the lensing analyses conform
to those that were deployed in~\cite{o3_lensing} and~\cite{o3_lensing_followup}.

We stress here that a false positive from the following analyses would not by
itself consistute a claim of detection in a realistic scenario---this would
require additional analyses as well as a background study to fully calibrate the
significance of the results of these investigations. False positives from the
following analyses, however, would be the impetus for additional investigations
such as those carried out in~\cite{o3_lensing_followup}.

\subsection{General Relativity Baseline}\label{subsec:gr-baseline}

As a step before the examination of non-GR waveforms, we first examine the results of
our investigations on an injection of the GW150914-like parameters given in
Table~\ref{tab:GW150914-parameters} to set a baseline from which the non-GR results may be compared.
Here, we briefly summarize the results of investigations of this waveform, with the more detailed
descriptions of how to interpret these results provided in the following respective analysis
sections.

\begin{itemize}
  \item{An investigation using the isolated point mass microlensing model did not find support for
    the microlensing hypothesis, with a $\log_{10}\mathcal{B}^{\textrm{ML}}_{\textrm{U}}$---the Bayes
    factor comparing the evidence for the microlensed and unlensed cases---of $-2.5$.}
    \item{A similar investigation with the singular isothermal sphere microlensing model found this
        model more closely aligned with the unlensed case, with a
        $\log_{10}\mathcal{B}^{\textrm{SIS}}_{\textrm{U}}$ of $0.0$. An examination of the
        posteriors for this case revealed that the lens mass posterior reflects the prior, and the
        source position posterior is entirely above $y=1$, which mirrors the expectations for an
         unlensed event.} 
   \item{An investigation using the phenomenological millilensing model did not find support for the
      existence of any additional millisignals, and preferred the unlensed waveform with
      $\log_{10}\mathcal{B}^{\textrm{MiL}}_{\textrm{U}} = -0.4$.}
   \item{The type II strong lensing analysis using a discrete prior on the Morse index did not
       find support for the type II hypothesis. The
       $\log_{10}\mathcal{B}^{\textrm{II}}_{\textrm{I}+\textrm{III}}$ in this case was $-2.7$, in line
	   with the expectations for an unlensed signal.}
   \item{The type II strong lensing investigation using a continuous prior on the Morse index did
     not find support for the type II case, with a median value of the recovered Morse index at 0.47 and the
     90\% credible interval (C.I.) boundaries at 0.02 and 0.91. In addition to a broad posterior, it is also
	 important to note a significantly lower density of posterior points around the $n=0.5$ region compared
	 to the cases for the type I and III images. The posterior is shown in Figure~\ref{fig:illustration_posterior_no_lens},
	 where we overlay it with a non-GR case which is indistinguishable from the GR case.}
\end{itemize}

The conclusion would therefore be that this specific event would not be
considered as a lensing candidate, as it adheres to the expectations for an
unlensed event. Thus, one would expect that any suggestion of false positivity
in the non-GR waveform analysis based upon this system would be driven by the
deviations from GR rather than the source parameter combination.

\subsection{Point Mass Microlensing Analysis}\label{subsec:pm-microlensing}

The first step of microlensing investigations is an analysis using the isolated point mass 
microlensing model, due to the increased computational tractability of calculating the amplification
 factor compared with other models. 

Table~\ref{tab:pm-microlensing} gives the complete results of this investigation, with the 
superscripted $\beta$ indicating those analyses with a Bayes factor comparing the point mass
microlensing and unlensed cases, $\mathcal{B}^{\textrm{ML}}_{\textrm{U}}$,
indicating a sufficiently high support for the microlensing hypothesis that they would be 
flagged for additional follow-up in the style of~\cite{o3_lensing_followup}. This threshold is set
at $\log_{10} \mathcal{B}^{\textrm{ML}}_{\textrm{U}} > 1$. The choice of this value is based upon
the background microlensing analysis done in~\cite{o3a_lensing}. The errors on the Bayes factor are
expected to be of $\mathcal{O}(0.1)$ in this and the subsequent SIS and millilensing analyses. As can be seen 
within the table, modifications to the QNM spectrum, the addition of a scalar polarization, 
and non-waveform-scaling modifications to the energy flux do not yield any scenarios in the 
space tested in which a false preference for the isolated point mass microlensing model is obtained. 

The remaining cases, however, demonstrate at least one scenario in which not only is a false
preference for the microlensing hypothesis obtained, but that false preference is
strong---particularly in the cases of the injections with a large graviton mass and the
precessing black hole mimicker waveforms based on~\cite{dietric_scaled_bns} (BHM-P). This latter case
has two additional points of note. In the analysis of the BHM-P waveforms, the prior on the
luminosity distance needed to be extended from an upper limit of $2$~Gpc to an upper limit of
$5$~Gpc in order to fully enclose the estimated posterior. Analysis of some of these injections also
revealed a secondary weaker local maximum likelihood point in the parameter space found by some of
the completed chains leading to significant disagreement between the chains. The combination of the
chains for the results given in Table~\ref{tab:pm-microlensing} reflects those with the highest
preference for the investigation. This behavior is seen in both the lensed and unlensed analyses of
these waveforms.

A possible reason that the scaled BNS waveforms display such strong
preference for the microlensing model may lie in their morphology. These
waveforms display dephasing and amplitude modification in their mergers as well
as an oscillatory post-merger. These are similar to the kinds of effects produced by
the oscillatory wave optics regime amplification factor in
microlensing and thus these additional effects may be relatively well fit using
these parameters as compared with the available fits from the unlensed waveform
model.

One of the first-stage checks of potential microlensing candidates performed 
in~\cite{o3_lensing_followup} was the checking of the lens parameter posteriors.
In the case of an unlensed event, one would not expect to see a strong preference for a non-zero
lens mass or a source position less than the upper bound of the prior. The reasoning for the latter
instance is that, in all cases, the amplification factor of the point mass model is oscillatory but
higher values of source position have decreased amplitude and so more closely align with unlensed
waveforms. Seeing deviation from either of these expectations would therefore be a sign that further
investigation is required, and that the signal may be identified as microlensed.

\begin{figure*}[ht]
	\centering
	\includegraphics[width=\linewidth]{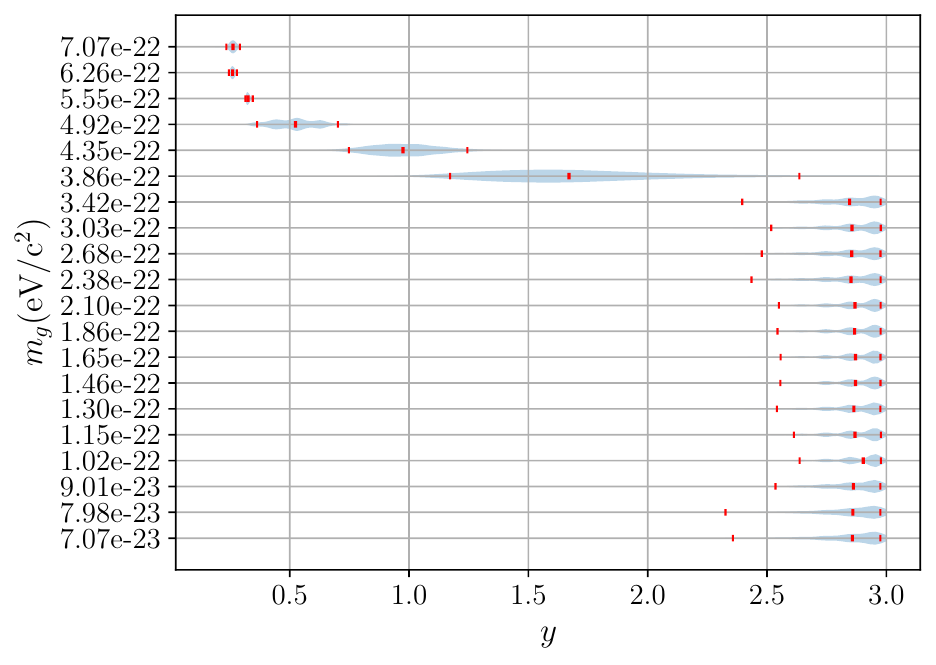}
	\caption{Posteriors on the source position for the massive graviton series of injections analyzed
  using the isolated point mass microlensing model. The red ticks indicate the median and 90\%
C.I. Lower graviton mass injections display a preference for source position values
towards the upper prior boundary and are railing against it, which is the expectation for unlensed
events. However, the higher graviton mass injections display significant constraints towards a
particular value of source position within the prior space, which would be the expectation for a
microlensed event.}\label{fig:massive-graviton-y-violin}
\end{figure*}

Figure~\ref{fig:massive-graviton-y-violin} shows the posteriors on the dimensionless 
source position for the analyses performed on the massive graviton injections. 
For the cases with a smaller graviton mass, one finds the expected result for a non-lensed event of
significant support for values around the upper prior bound on the source position. However, beginning
with the injection with $m_{g} = 3.86 \times 10^{-22}\text{ eV}/c^2$, one instead obtained constrains within
the prior range. As one can see in Table~\ref{tab:pm-microlensing}, the threshold case is the 
last of the injections that retains preference for the non-lensed hypothesis, albeit with 
significantly less support compared to the injections with smaller graviton masses. 
As such, these events would require further investigation as potential microlensing events combined 
with the Bayes factors shown. However, here, it is worth noting that the lowest deviations
leading false positives already correspond to graviton masses above the limit set by current analyses~\citep{gwtc3_tgr}.

The remaining posterior plots are shown within Appendix~\ref{subsec:pm-violins}. These results are
similar to those shown for the massive graviton injections. The injections that show 
significant support for the microlensing hypothesis also demonstrate apparent false constraints 
on the lensing parameters with those with high Bayes factor support demonstrating significant 
such false constraints, which would require further investigation.

\begin{center}
\begin{deluxetable*}{cccccccc}\label{tab:pm-microlensing}
	\tablecaption{Results of performing isolated point mass microlensing analysis using \textsc{Gravelamps} on 
	the simulated modified GR events. Cells with a $\beta$ superscript indicate the cases with $\log_{10} \mathcal{B}^{\textrm{ML}}_{\textrm{U}} > 1$ that would be flagged for 
	further investigation as microlensing candidates. As can be seen, aside from the modified energy flux with no waveform scaling, the modified QNM spectrum,
	and the addition of a scalar polarization, each GR deviation has at least some cases in which false positives may be identified.}
	\tablehead{
		\multicolumn{2}{c}{Massive Graviton} & \multicolumn{2}{c}{Modified Energy Flux} & \multicolumn{2}{c}{Modified QNM Spectrum} & \multicolumn{2}{c}{BHM-P} \\
		$m_{g}$ ($\text{eV}/c^2$) & $\log_{10}\mathcal{B}^{\textrm{ML}}_{\textrm{U}}$ & $a_{2}$ & $\log_{10}\mathcal{B}^{\textrm{ML}}_{\textrm{U}}$ & $Q_{f}$ & $\log_{10}\mathcal{B}^{\textrm{ML}}_{\textrm{U}}$ & $M$ $(M_{\odot})$ & $\log_{10}\mathcal{B}^{\textrm{ML}}_{\textrm{U}}$
	}
	\tablecolumns{8}

	\startdata
	$7.07 \times 10^{-23}$ & $-2.3$ & 2 & $-2.4$ & 0.50 & $-2.4$ & $80$&  $15.1^{\beta}$\\
	$7.98 \times 10^{-23}$ & $-2.0$ & 5 & $-2.2$ & 0.55 & $-2.4$ & $90$&  $18.8^{\beta}$\\
	$9.01 \times 10^{-23}$ & $-2.1$ & 10 & $-1.6$ & 0.60 & $-2.3$ & $100$ & $26.8^{\beta}$\\
	$1.02 \times 10^{-22}$ & $-2.3$ & 20 & $0.9$ & 0.65 & $-2.3$ & $110$&  $31.2^{\beta}$\\
	$1.15 \times 10^{-22}$ & $-2.4$ & 50 & $8.6^{\beta}$ & 0.70 & $-2.0$ & $120$ & $24.3^{\beta}$\\
	$1.30 \times 10^{-22}$ & $-2.3$ & 20$^{\dag}$ & $-2.4$ & 0.75 & $-2.1$ & $130$ & $41.2^{\beta}$\\
	$1.46 \times 10^{-22}$ & $-2.3$ & 50$^{\dag}$ & $-2.2$ & & & $140$ & $47.4^{\beta}$\\
	$1.65 \times 10^{-22}$ & $-2.2$ & 75$^{\dag}$ & $-2.1$ & & & $150$ & $34.3^{\beta}$\\
	\cmidrule[\heavyrulewidth]{5-8} $1.86 \times 10^{-22}$ & $-2.3$ & 100$^{\dag}$ & $-2.2$ & \multicolumn{2}{c}{Scalar Polarization} & \multicolumn{2}{c}{BHM-NS}\\
	$2.10 \times 10^{-22}$ & $-2.2$ & 125$^{\dag}$ & $-2.3$ & $\mathcal{A}$ & $\log_{10}\mathcal{B}^{\textrm{ML}}_{\textrm{U}}$ & $M$ $(M_{\odot})$ & $\log_{10}\mathcal{B}^{\textrm{ML}}_{\textrm{U}}$\\
	\cmidrule{5-8} $2.38 \times 10^{-22}$ & $-2.2$ & 150$^{\dag}$ & $-2.2$ & 0.05 & $-2.6$ & 80 & $-0.7$\\
	$2.68 \times 10^{-22}$ & $-2.0$ & 175$^{\dag}$ & $-2.1$ & 0.10 & $-2.5$ & 90 & 0.1\\
	$3.03 \times 10^{-22}$ & $-2.2$ & 200$^{\dag}$ & $-1.8$ & 0.15 & $-2.7$ & $100$ & $1.3^{\beta}$\\
	$3.42 \times 10^{-22}$ & $-2.2$ & & & 0.20 & $-2.4$ & $110$ & $3.2^{\beta}$ \\
	$3.86 \times 10^{-22}$ & $-0.5$ & & & 0.25 & $-2.6$ & $120$ & $5.9^{\beta}$ \\
	$4.35 \times 10^{-22}$ &  $9.6^{\beta}$ & & & 0.30 & $-3.1$ & 130 & $7.3^{\beta}$ \\
	$4.92 \times 10^{-22}$ &  $27.3^{\beta}$ & & & 0.35 & $-3.0$ & 140 & $7.4^{\beta}$ \\
	$5.55 \times 10^{-22}$ &  $28.1^{\beta}$ & & & 0.40 & $-3.1$ & 150 & $9.2^{\beta}$ \\
	$6.26 \times 10^{-22}$ &  $15.6^{\beta}$ & & & 0.45 & $-3.2$ & 160 & $11.3^{\beta}$ \\
	$7.07 \times 10^{-22}$ &  $24.6^{\beta}$ & & & 0.50 & $-3.1$\\
	\enddata

	\tablecomments{\small \, $^{\dag}$ indicates the injections have been done without waveform scaling.}
\end{deluxetable*}
\end{center}

\subsection{SIS Microlensing Analysis}\label{subsec:sis-microlensing}

The follow-up analysis performed in~\cite{o3_lensing_followup} on microlensing candidates was to 
investigate an additional microlensing mass density profile; specifically that of the SIS profile. 
It was noted in that work that the microlensing candidate event displayed increased preference for 
the SIS model compared to the point mass model. However, the primary purpose of that follow-up 
would be to gain information about the lensing object from an actual lensed event. Nonetheless, 
we perform this analysis here on these injections to ascertain whether the false positive status 
shown in the point mass model is restricted to that model or is retained/increased by this additional 
model. We also deploy this investigation on all of the injections as opposed to only the subset that 
would be flagged for additional follow-up to check if additional false positives could occur. 

Table~\ref{tab:sis-microlensing} shows the results of performing this analysis, comparing the SIS model with the point mass model. Cells with a superscripted $\beta$ are those that displayed a $\log_{10}$ Bayes factor preference for the point mass model over the unlensed case and for the SIS model over the point mass model. This represents a minority of cases, specifically the highest mass massive graviton injections and some of the non-spinning black hole mimicker cases---though many of these latter cases would also be compatible with indistinguishability. Cells with a superscripted $\gamma$ are those that displayed a $\log_{10}$ Bayes factor preference for the point mass model over the unlensed model and over the SIS model. 

In all but one case where the point mass model is preferred over the SIS, the SIS would remain
preferred over the unlensed case---though in the case of the $80M_{\odot}$ precessing black hole
mimicker waveform, this preference would be just below the
$\log_{10}\mathcal{B}^{\textrm{L}}_{\textrm{U}} > 1$ threshold of interest. The exception is the
case of the $140M_{\odot}$ precessing black hole mimicker. In this case, the source parameter
posterior estimates do not reflect the primary or secondary likelihood maxima found in the point
mass and unlensed analyses and are instead in between the source parameter values corresponding to
these, which may have resulted in the substantially weaker support for this
model. Similarly to the point mass case, the precessing black hole mimicker investigations required
extending the luminosity distance prior to $5$~Gpc. 

\begin{figure*}
	\includegraphics[width=\linewidth]{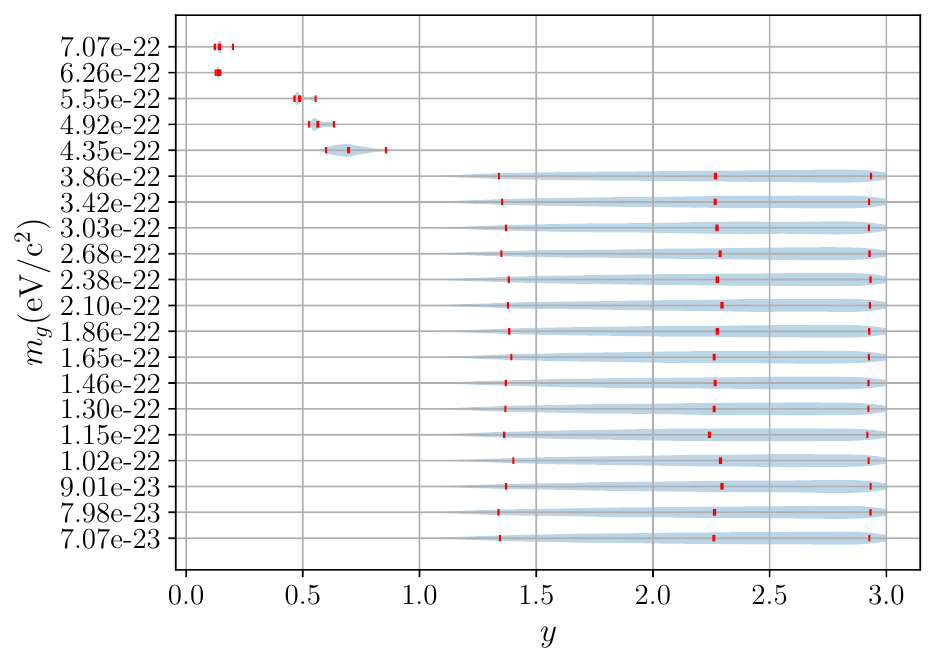}
	\caption{Recovered posteriors on the source position for the massive graviton series of injections
    analyzed using the SIS microlensing model. The red ticks indicate the median and 90\%
	C.I. Lower graviton mass injections display support
    broadly across the $y > 1$ range. This is the expectation for unlensed events as in this region
    of parameter space only a single image is produced resulting in a flat amplification factor that
    resembles an unlensed event. By contrast, the higher mass injections display constraint towards
    a particular value within the prior boundary and specifically below $y = 1$, i.e., within the
    regions which the amplification factor would be oscillatory. This is the expectation for
    microlensing candidates.}\label{fig:massive-graviton-sis-position-violin}
\end{figure*}

Figure~\ref{fig:massive-graviton-sis-position-violin} shows the source position posteriors for the
massive graviton series of injections. Similarly to the results shown in
Figure~\ref{fig:massive-graviton-y-violin}, the posteriors for those injections that display
preference for the microlensed waveform show constraints within the prior boundary. Those events that
display preference for the unlensed waveform show broader posteriors than those for the point mass
case, with all support above $y = 1$. This is the expectation in the SIS case, where above
this point, geometric optics predicts that only one image will be produced by lensing from this
profile. This results in a flat amplification factor within this parameter region which therefore
more closely resembles an unlensed waveform. This may be the reason for the favouring of the SIS
model over the point mass lens model in the cases where the unlensed waveform was preferred over the
point mass model, and why the difference would be sufficient to bring the log Bayes factor between
the SIS and unlensed waveforms into indistinguishability.

Appendix~\ref{subsec:sis-violins} shows the remaining posteriors for the injections, which as is the case with 
the point mass results display similar behavior to the massive graviton results. In those results 
with preference for the microlensing hypothesis sufficient to trigger this follow-up investigation, these posteriors are consistently below the $y=1$ threshold, thus retaining oscillatory behavior of the amplification 
factor even at high frequency. Consequently, the SIS microlensing follow-up would not allow for any 
additional candidates to be ruled out. Some additional steps that could be deployed in order to
further test these cases are discussed in~\cite{o3_lensing_followup}.

\begin{center}
\begin{deluxetable*}{cccccccc}\label{tab:sis-microlensing}
	\tablecaption{Results of performing the SIS microlensing analysis using \textsc{Gravelamps} 
	on the simulated modified GR events. Cells wtih a $\gamma$ superscript are those events that had a $\log_{10}$ 
	Bayes factor in favor of the point mass microlensing case that display a $\log_{10}$ Bayes factor 
	supporting the point mass model over the SIS, though the SIS remains preferred over the non-lensed case. 
	Cells with a $\beta$ superscript are those events that had a $\log_{10}$ Bayes factor in favor of the point 
	mass microlensing case that display a $\log_{10}$ Bayes factor supporting the SIS model over the point mass one.}
	\tablehead{
		\multicolumn{2}{c}{Massive Graviton} & \multicolumn{2}{c}{Modified Energy Flux} & \multicolumn{2}{c}{Modified QNM Spectrum} & \multicolumn{2}{c}{BHM-P} \\
		$m_{g}$ ($\text{eV}/c^2$) & $\log_{10}\mathcal{B}^{\textrm{SIS}}_{\textrm{PM}}$ & $a_{2}$ & $\log_{10}\mathcal{B}^{\textrm{SIS}}_{\textrm{PM}}$ & $Q_{f}$ & $\log_{10}\mathcal{B}^{\textrm{SIS}}_{\textrm{PM}}$ & $M$ $(M_{\odot})$ & $\log_{10}\mathcal{B}^{\textrm{SIS}}_{\textrm{PM}}$
	}
	\tablecolumns{8}

	\startdata
	$7.07 \times 10^{-23}$ & $2.4$ & 2 & $2.5$ & 0.50 & $2.5$ & 80 &  $-14.3^{\gamma}$\\
	$7.98 \times 10^{-23}$ & $2.2$ & 5 & $2.3$ & 0.55 & $2.6$ & 90 &  $-1.3^{\gamma}$\\
	$9.01 \times 10^{-23}$ & $2.3$ & 10 & $1.8$ & 0.60 & $2.4$ & 100 &  $-0.8^{\gamma}$\\
	$1.02 \times 10^{-22}$ & $2.3$ &  20 & $-0.7^{\gamma}$ & 0.65 & $2.4$ & 110 & $-4.3^{\gamma}$\\
	$1.15 \times 10^{-22}$ & $2.4$ &  50 & $-2.3^{\gamma}$ & 0.70 & $2.2$ & 120 & $-1.2^{\gamma}$\\
	$1.30 \times 10^{-22}$ & $2.4$ & 20$^{\dag}$ & $2.5$ & 0.75 & $2.1$ &  130 &  $-7.6^{\gamma}$\\
	$1.46 \times 10^{-22}$ & $2.3$ & 50$^{\dag}$ & $2.4$ & & & 140 & $-85.8$\\
	$1.65 \times 10^{-22}$ & $2.4$ & 75$^{\dag}$ & $2.2$ & & &  150 &  $-5.4^{\gamma}$\\
	\cmidrule[\heavyrulewidth]{5-8} $1.85 \times 10^{-22}$ & $2.4$ & 100$^{\dag}$ & $2.3$ & \multicolumn{2}{c}{Scalar Polarization} & \multicolumn{2}{c}{BHM-NS}\\
	$2.10 \times 10^{-22}$ & $2.4$ & 125$^{\dag}$ & $2.3$ & $\mathcal{A}$ & $\log_{10}\mathcal{B}^{\textrm{SIS}}_{\textrm{PM}}$ & $M$ $(M_{\odot})$ & $\log_{10}\mathcal{B}^{\textrm{SIS}}_{\textrm{PM}}$\\
	\cmidrule{5-8} $2.38 \times 10^{-22}$ & $2.3$ & 150$^{\dag}$ & $2.4$ & 0.05 & $2.7$ & 80 & $-0.1$\\
	$2.68 \times 10^{-22}$ & $2.2$ & 175$^{\dag}$ & $2.3$ & 0.10 & $2.8$ & 90 & 0.0\\
	$3.03 \times 10^{-22}$ & $2.3$ & 200$^{\dag}$ & $2.0$ & 0.15 & $3.0$ &  100 &  $0.7^{\beta}$ \\
	$3.42 \times 10^{-22}$ & $2.3$ & & & 0.20 & $2.6$ & 110 & $-0.3^{\gamma}$ \\
	$3.86 \times 10^{-22}$ & $0.9$ & & & 0.25 & $2.8$ & 120 & $0.2^{\beta}$ \\
    $4.35 \times 10^{-22}$ &  $-1.3^{\gamma}$ & & & 0.30 & $3.4$ &  130 &  $0.2^{\beta}$ \\
    $4.92 \times 10^{-22}$&  $-3.8^{\gamma}$ & & & 0.35 & $3.3$ &  140 &  $0.2^{\beta}$ \\
    $5.55 \times 10^{-22}$&  $-8.0^{\gamma}$ & & & 0.40 & $3.1$ &  150 &  $-0.9^{\gamma}$ \\
    $6.26 \times 10^{-22}$&  $4.0^{\beta}$ & & & 0.45 & $3.4$ & 160 & $-0.2^{\gamma}$ \\
    $7.07 \times 10^{-22}$&  $1.3^{\beta}$ & & & 0.50 & $3.1$\\
	\enddata

	\tablecomments{\small \, $^{\dag}$ indicates the injections have been done without waveform scaling.}
\end{deluxetable*}
\end{center}

\subsection{Millilensing Analysis}\label{subsec:millilesning-analysis}

In addition to the secondary microlensing model, \citet{o3_lensing_followup} investigated the 
potential for the microlensing candidate to be a millilensing candidate. In this scenario, 
the lens is sufficiently massive that geometric optics applies and Equation~\eqref{eq:geometric-amplification} 
describes the amplification. However, the time delays are sufficiently short that the images overlap 
resulting in similar beating patterns to microlensing but without wave optics effects. 
We replicate this analysis using a phenomenological millilensing model introduced 
in~\cite{liu_millilensing} and implemented within~\textsc{Gravelamps}. 
We allow for up to 
four signals to be present within the overall signal which is sufficient to give an indication of whether there is support for
additional images but remains tractable for performing a large number of analyses. 


Table~\ref{tab:millilensing} gives the results of the millilensing analysis
showing the $\log_{10}$ Bayes factor between the millilensing and the
non-lensed waveforms. The superscripted symbol indicates the recovered number
of images from the posterior from one to four. As can be seen, all cases are
present. Table~\ref{tab:millilensing} indicates that in addition to all
candidates that are flagged for investigation by the microlensing analysis, the
millilensing analysis also flags two of the injections with an additional scalar
polarization as false positives with a preference for a single image. Examining
the posterior for the first image for the $\mathcal{A} = 0.20$ case
indicates a preference for a type II image, comprising 98\% of the
total posterior, which is confirmed by the strong lensing discrete type II
analysis shown in Table~\ref{tab:discrete_SL_analyses}. The lowest two mass
non-spinning black hole mimicker injections are additionally now found to have
sufficient support for follow-up consideration in this case, with one and two
image recoveries respectively. This would be in line with the slender support
for the $M=90M_{\odot}$ case in the microlensing analysis, and the slender
disfavoring for the $M=80M_{\odot}$ case. 

\begin{figure*}
	\includegraphics[width=\linewidth]{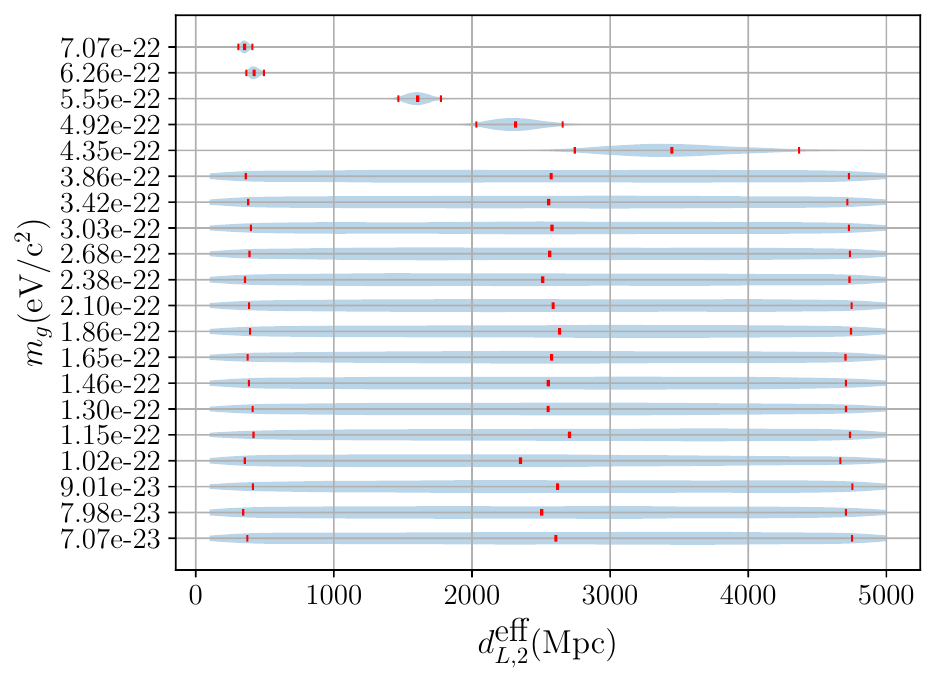}
	\caption{Recovered posteriors on the effective luminosity distance of the proposed second image for the massive graviton series of injections analyzed using the phenomenological millilensing model. The red ticks indicate the median and 90\%
	C.I. Lower graviton mass injections display broad posteriors across the entire space which is an indication of lack of support for a second image. The higher graviton mass injections, which favor the millilensing hypothesis with a number of images greater than two, do show constraints to a particular value within this region as is expected if a millilensed signal is present.}\label{fig:massive-graviton-dl1-violin}
\end{figure*}

We perform a similar posterior consistency check to the one we did for the microlensing analysis. 
For non-lensed events, the posterior space for the lensing observables should reflect the 
respective priors, all of which were uniform. For lensed events, the posteriors should 
show a constraint for each of the parameter sets corresponding to images that are preferred 
to exist and reflect the priors for images which are not. This behavior is illustrated in 
Figure~\ref{fig:massive-graviton-dl1-violin} which shows the posteriors on the effective luminosity 
distance of a proposed second image for the massive graviton series of injections. As can be seen from Table~\ref{tab:millilensing}, 
preference for multiple images begins with $m_{g} = 4.35 \times 10^{-22}
\textrm{ eV}/c^2$, and 
this is the point at which the posteriors cease being uniform and display a constraint 
to a particular value.

Similarly to the SIS analysis, the $140 M_{\odot}$ precessing black hole
mimicker waveform does not yield support for the millilensing hypothesis
despite posterior support for 4 images similar to the $130M_{\odot}$ and
$150M_{\odot}$ cases. In this instance, the source posteriors in all chains
more closely resembled the local maximum found for the point mass and unlensed
analysis, and this was the case with a repeated analysis of this waveform. This
may be an indication that all analyses are untrustworthy in the case of this
waveform. Additionally, as the interference between the images in the
  millilensing case continues to produce an oscillatory amplification factor,
  this may be the reason that similarly to the microlensing case, the scaled
  BNS waveforms yield significant support for the millilensing case. The
increasing number of images would also indicate that at higher masses,
progressively more parameters may be used to gain a better fit to these
waveforms.

Appendix~\ref{app:millilensing} displays the rest of the effective luminosity
distance and time delay posteriors for the second image of the millilensing
analysis. Both of these mirror the trend of
Figure~\ref{fig:massive-graviton-dl1-violin} with broad, uninformative
posteriors when a second image has no support and constrained posteriors when
support exists for that image. This is repeated in the third and fourth image
posteriors. 

The Morse phase posteriors do not display a consistent constraint to a single value for those images
that are supported, varying between support for one value, support for two of the three values, or
are unconstrained between all three. This is not necessarily unexpected for true lensing candidates
and so would not be sufficient to discard the injections as such. 

While the luminosity distance posterior needed to be extended for the microlensing 
investigations into the precessing black hole mimicker waveforms, the millilensing investigations into these 
waveforms additionally required an increase in the maximal effective proposed luminosity 
distance from the baseline $5$~Gpc used. This was allowed to reach $100$~Gpc with the estimates 
for the median effective luminosity distance varying between $5$~Gpc and $12$~Gpc. This reinforces 
the suggestion from the microlensing analyses that these waveforms would demonstrate the 
apparent highest magnifications. While this level of magnification may shed some doubt on 
these events due to it being unlikely, it is difficult to infer the absolute magnification 
of an unknown signal. Rather, it is only possible to infer the relative magnification between the images, which in these 
cases was between $1$ and $10$, and would not be in and of itself disqualifying when compared
to expected values from current models~\citep{More:2021kpb,2023MNRAS.519.2046J}. 

\begin{center}
\begin{deluxetable*}{cccccccc}\label{tab:millilensing}
	\tablecaption{Results of performing the phenomenological millilensing analysis using \textsc{Gravelamps} 
	on the simulated modified GR events. Cells with superscripts indicate those events that had a $\log_{10}$ 
	Bayes factor in favour of the millilensing case over the non-lensing case. 
	The superscripted symbol indicates the number of images favoured with $\xi$
	corresponding to a single image, $\gamma$ corresponding to two images,
  $\delta$ to three, 
	and $\beta$ to four.}
	\tablehead{
		\multicolumn{2}{c}{Massive Graviton} & \multicolumn{2}{c}{Modified Energy Flux} & \multicolumn{2}{c}{Modified QNM Spectrum} & \multicolumn{2}{c}{BHM-P} \\
		$m_{g}$ ($\text{eV}/c^2$) & $\log_{10}\mathcal{B}^{\textrm{MiL}}_{\textrm{U}}$ & $a_{2}$ & $\log_{10}\mathcal{B}^{\textrm{MiL}}_{\textrm{U}}$ & $Q_{f}$ & $\log_{10}\mathcal{B}^{\textrm{MiL}}_{\textrm{U}}$ & $M$ $(M_{\odot})$ & $\log_{10}\mathcal{B}^{\textrm{MiL}}_{\textrm{U}}$
	}
	\tablecolumns{8}

	\startdata
	$7.07 \times 10^{-23}$ & $-0.5$ & 2 & $-0.5$ & 0.50 & $-0.5$ &  80 & $23.1^{\delta}$\\
	$7.98 \times 10^{-23}$ & $-0.1$ & 5 & $-0.3$ & 0.55 & $-0.3$ &  90 & $46.6^{\delta}$\\
	$9.01 \times 10^{-23}$ & $-0.3$ & 10 & $-0.4$ & 0.60 & $-0.4$ &  100 & $69.4^{\delta}$\\
	$1.02 \times 10^{-22}$ & $-0.3$ & 20 &  $1.3^{\delta}$ & 0.65 & $-0.3$ & 110 &  $72.1^{\delta}$\\
	$1.15 \times 10^{-22}$ & $-0.1$ & 50 &  $22.5^{\beta}$ & 0.70 & $0.1$ & 120 &  $28.5^{\gamma}$\\
	$1.30 \times 10^{-22}$ & $-0.4$ & 20$^{\dag}$ & $-0.3$ & 0.75 & $-0.3$ & 130 &  $50.3^{\beta}$\\
	$1.46 \times 10^{-22}$ & $-0.4$ & 50$^{\dag}$ & $-0.3$ & & & 140 & $-4.9$\\
	$1.65 \times 10^{-22}$ & $-0.3$ & 75$^{\dag}$ & $-0.4$ & & &  150 & $60.2^{\beta}$\\
	\cmidrule[\heavyrulewidth]{5-8} $1.86 \times 10^{-22}$ & $-0.3$ & 100$^{\dag}$ & $-0.5$ & \multicolumn{2}{c}{Scalar Polarization} & \multicolumn{2}{c}{BHM-NS}\\
	$2.10 \times 10^{-22}$ & $-0.4$ & 125$^{\dag}$ & $-0.6$ & $\mathcal{A}$ & $\log_{10}\mathcal{B}^{\textrm{MiL}}_{\textrm{U}}$ & $M$ $(M_{\odot})$ & $\log_{10}\mathcal{B}^{\textrm{MiL}}_{\textrm{U}}$\\
	\cmidrule{5-8} $2.38 \times 10^{-22}$ & $-0.1$ & 150$^{\dag}$ & $-0.3$ & 0.05 & $-0.6$ & 80 & $1.1^{\xi}$\\
	$2.68 \times 10^{-22}$ & $-0.2$ & 175$^{\dag}$ & $-0.4$ & 0.10 & $-0.1$ &  90 & $1.0^{\gamma}$\\
	$3.03 \times 10^{-22}$ & $-0.3$ & 200$^{\dag}$ & $0.0$ & 0.15 & $-0.1$ &  100 & $4.1^{\delta}$ \\
	$3.42 \times 10^{-22}$ & $-0.3$ & & &  0.20 &  $1.3^{\xi}$ &  110 & $9.8 ^{\delta}$ \\
	$3.86 \times 10^{-22}$ & $0.0$ & & &  0.25 &  $1.7^{\xi}$ &  120 & $15.9 ^{\beta}$ \\
	$4.35 \times 10^{-22}$ & $11.6^{\gamma}$ & & & 0.30 & $-0.2$ &  130 &  $17.6 ^{\delta}$ \\
	$4.92 \times 10^{-22}$ & $32.0^{\gamma}$ & & & 0.35 & $-0.0$ &  140 &  $18.8 ^{\delta}$ \\
	$5.55 \times 10^{-22}$ & $39.7^{\gamma}$ & & & 0.40 & $-0.3$ &  150 &  $17.7 ^{\delta}$ \\
	$6.26 \times 10^{-22}$ & $74.7^{\beta}$ & & & 0.45 & $-0.2$ &  160 &  $17.9^{\delta}$ \\
	$7.07 \times 10^{-22}$ & $76.6^{\beta}$ & & & 0.50 & $-0.1$\\
	\enddata

	\tablecomments{\small \, $^{\dag}$ indicates the injections have been done without waveform scaling.}
\end{deluxetable*}
\end{center}

\subsection{Type II Strong Lensing Analysis}\label{subsec:typeII_strong_lensing}

When performing strongly-lensed type II image searches, we typically follow a series of steps. 
First, we analyze the data using a discrete and uniform prior on the Morse factor, where we 
allow only for $n = \{0, 0.5, 1\}$, corresponding to the three physical image types. 
From this, we compute two Bayes factors, comparing the type II hypothesis with the 
type I or type III hypothesis. This is done by reweighting the evidence under one of the 
image type hypothesis based on the final samples and the prior. To get the evidence for 
a given image type $X$, $\mathcal{Z}_X$, we compute a series of weights for every Morse
factor sample $w_i^X = p(n_i = \text{type}_X | \text{data})/p(n_i = \text{type}_X)$, where the numerator corresponds to 
the posteriors and the denominator to the prior. The evidence is then computed based on these
weights as $\mathcal{Z}_X = (1/N) \sum_i w_i^X \mathcal{Z}$, where $\mathcal{Z}$ is the evidence 
obtained via the strong lensing run. We then compute the Bayes factor as
$\mathcal{B}^\text{II}_{X} = Z_\text{II}/Z_{X}$, where $X = \{\text{I}, \text{III}\}$ when comparing type I with type I or III, 
respectively. Finally, following the threshold used in~\citet{o3_lensing}, an event is seen as interesting if we find 
$\log_{10}\mathcal{B}^\text{II}_\text{I} > 0.1$ or $\log_{10}\mathcal{B}^\text{II}_\text{III} > 0.1.$\footnote{In
an ideal situation, the two Bayes factors should be the same since type I and 
type III images are degenerate~\citep{ezquiaga_strong_lensing}. However, due to finite 
sampling effects, it can happen that the two have slightly different values. Additionally,
if there are no samples supporting the type II image hypothesis, we get a null Bayes factor, 
hence a $\log_{10}$ value of $-\infty$.} This is a conservative 
choice as one expects an error of $\mathcal{O}(0.1)$ on these Bayes factors,
and, therefore, a value of 0.1 is still only slightly favoring the lensing 
hypothesis. Such a conservative choice makes sense when considering one
does not want to miss a first detection of lensing.

When an event is interesting, we need to assess if the observed evidence for a type II 
image is genuine. Since the discrete prior allows only for the physical values, 
there is no way of knowing if a different value of the Morse factor would be a better fit to the data.
If one finds an unphysical value of the Morse factor to be preferred,
the event is probably not lensed, but something different could be happening, needing 
further inquiries to settle whether the effect has a noise or physical origin. 
Thus, for interesting events, we rerun the lensing analysis, this time using a continuous prior 
on the Morse phase. In this work, we directly perform 
the two analyses, sampling over the Morse factor in a discrete and continuous way regardless
of the result coming from the former analysis, to see if the second step could be 
used as to flag anomalies in other situations than lensed candidates.

For the discrete Morse factor analyses, whether or not one detects possible strong lensing features depends 
on the specific non-GR waveform considered and the value of the deviation parameter. 
For the massive graviton waveforms, no deviation is found. In some cases, the type II hypothesis
is clearly disfavored, in line with the event having higher-order modes (HOMs) and not being a lensed type II image. 
For the scalar polarizations, we find either a strong support for type II images---when 
$\mathcal{A} \in \{0.15, 0.2, 0.25\}$---or a strong disfavoring. 
While this behavior is clarified in the case of a continuous Morse factor, this would mean 
that, in type II image search set ups, some of these deviations would be followed-up 
with more in-depth analyses. In the case of a modified energy flux, none of the cases
leads to a favoring of the type II image. 
For the modified QNM spectrum, none of 
the discrete analyses favor the presence of a type II image, and the disfavoring grows 
stronger as $Q_f$ decreases. Finally, turning to the black hole mimickers, 
the situation is different from one model to the other. For the precessing signals 
following~\citet{dietric_scaled_bns}, we have an alternation between favoring and disfavoring for the 
type II hypothesis, with support for total masses of $120$, $130$, and $150M_{\odot}$.
For the nonspinning~\citet{ujevic_scaled_bns} waveform, no support for type II images is found, 
with only some strong disfavoring in some cases. So, for the exotic object case, 
follow ups would take place or not depending on the specifics of the binary being considered
and how the exotic objects' properties affect the waveform. 
Table~\ref{tab:discrete_SL_analyses} gives a more quantitative overview of the results 
for the different non-GR/non-BBH waveforms and their corresponding parameter values.

\begin{center}
\begin{deluxetable*}{cccccccc}\label{tab:discrete_SL_analyses}
	\tablecaption{Results of performing type II image searches with a discrete prior 
	on the simulated modified GR events. Cells with a $\beta$ superscript indicate those analyses that 
	would be flagged for further investigation as strongly-lensed type II images. 
	As can be seen, follow-up analyses would be performed for some events with an 
	extra scalar polarization, one case of modified energy flux, and some cases of 
	the precessing~\citet{dietric_scaled_bns}-based black hole mimickers (BHM-P).}
	\tablehead{
		\multicolumn{2}{c}{Massive Graviton} & \multicolumn{2}{c}{Modified Energy Flux} & \multicolumn{2}{c}{Modified QNM Spectrum} & \multicolumn{2}{c}{BHM-P} \\
		$m_{g}$ ($\text{eV}/c^2$) & $\log_{10}\mathcal{B}^\text{II}_{\text{I}+\text{III}}$ & $a_{2}$ & $\log_{10}\mathcal{B}^\text{II}_{\text{I}+\text{III}}$ & $Q_{f}$ & $\log_{10}\mathcal{B}^\text{II}_{\text{I}+\text{III}}$ & $M$ $(M_{\odot})$ & $\log_{10}\mathcal{B}^\text{II}_{\text{I}+\text{III}}$
	}
	\tablecolumns{8}

	\startdata
	$7.07 \times 10^{-23}$ & $-1.5$ & 2 & $-2.3$ & 0.50 & $-1.1$ & 80 & $-0.1$ \\
	$7.98 \times 10^{-23}$ & $-1.2$ & 5 & $-3.0$ & 0.55 & $-0.9$ &  90 & $-0.5$ \\
	$9.01 \times 10^{-23}$ & $-0.4$ & 10 & $-\infty$ & 0.60 & $-0.8$ & 100 & $-0.1$ \\
	$1.02 \times 10^{-22}$ & $0.1$ & 20 & $-\infty$ & 0.65 & $-0.1$ & 110 & $-0.1$ \\
	$1.15 \times 10^{-22}$ & $0.1$ & 50 & $-3.6$ & 0.70 & $-0.1$ &  $120$&  $0.3^{\beta}$ \\
	$1.30 \times 10^{-22}$ & $0.1$ & 20$^{\dag}$ & $-1.6$ & 0.75 & $-0.1$ &  130 &  $0.2^{\beta}$\\
	$1.46 \times 10^{-22}$ & $0.0$ & 50$^{\dag}$ & $-0.6$ & & & 140 & $-0.6$ \\
	$1.65 \times 10^{-22}$ & $0.0$ & 75$^{\dag}$ & $-0.5$ & & &  150 &  $0.5^{\beta}$ \\
	\cmidrule[\heavyrulewidth]{5-8}$1.86 \times 10^{-22}$ & $0.0$ & 100$^{\dag}$ & $-1.0$ & \multicolumn{2}{c}{Scalar Polarization} & \multicolumn{2}{c}{BHM-NS} \\
	$2.10 \times 10^{-22}$ & $0.0$ & 125$^{\dag}$ & $-0.9$ & $\mathcal{A}$ & $\log_{10}\mathcal{B}^\text{II}_{\text{I}+\text{III}}$ & $M$ $(M_{\odot})$ & $\log_{10}\mathcal{B}^\text{II}_{\text{I} + \text{III}}$ \\
	\cmidrule{5-8}
	$2.38 \times 10^{-22}$ & $0.1$ & 150$^{\dag}$ & $-0.6$ & 0.05 & $-0.7$ & 80 & $-0.1$\\
	$2.68 \times 10^{-22}$ & $0.0$ &  175$^{\dag}$ &  $0.5^{\beta}$ & 0.10 & $-0.2$ & 90 & $0.0$\\
	$3.03 \times 10^{-22}$ & $0.0$ & 200$^{\dag}$ & $-0.1$ &  0.15 &
   $1.0^{\beta}$ & 100 & $0.0$ \\
	$3.42 \times 10^{-22}$ & $0.1$ & & &  0.20 &  $2.2^{\beta}$ & 110 & $-0.3$ \\
	$3.86 \times 10^{-22}$ & $0.1$ & & &  0.25 &  $3.2^{\beta}$ & 120 & $-0.2$ \\
	$4.35 \times 10^{-22}$ & $0.0$ & & & 0.30 & $-1.0$ & 130 & $-1.4$ \\
	$4.92 \times 10^{-22}$ & $0.0$ & & & 0.35 & $-1.8$ & 140 & $-1.6$ \\
	$5.55 \times 10^{-22}$ & $0.0$ & & & 0.40 & $-1.9$ & 150 & $-1.5$ \\
	$6.26 \times 10^{-22}$ & $0.0$ & & & 0.45 & $-2.3$ & 160 & $-1.3$ \\
	$7.07 \times 10^{-22}$ & $0.0$ & & & 0.50 & $-2.8$\\ 
	\enddata

	\tablecomments{\small \, $^{\dag}$ indicates the injections have been done without waveform scaling.}
\end{deluxetable*}
\end{center}

By comparing Tables~\ref{tab:pm-microlensing} and~\ref{tab:discrete_SL_analyses}, 
one sees that the apparent hierarchy seen in microlensing is not as present in the 
discrete type II strong lensing searches. The reason is clarified by the continuous runs. 
As will be clear in what follows, in some cases, the peak value for the Morse factor posterior---which
resemble a Gaussian with a relatively small width---is between the values attainable from lensing
when analyzing modified GR waveforms. However, the discrete priors 
force the samples to be at specific values. Therefore, when deviations are detectable 
through the Morse phase shift but the Morse factor corresponds to a non-GR value, 
the position of the peak will strongly influence the (dis-)favoring of the type II image. 
As an illustrative example, if the GR deviations can be mimicked with $n = 0.8$ for a 
given value of its deviation parameter, we will most likely disfavor the type II hypothesis. 
However, if for a different deviation parameter value, the mimicking happens for $n = 0.7$, 
then the type II image is favored.
However, both deviations are expected 
to be seen in the continuous search as they would manifest themselves as a Morse factor 
value not compatible with the lensing hypothesis.

In Table~\ref{tab:continuous_SL_analyses}, we show the median values and the 90\% C.I.
found for the different alternative GR theories. Here, more cases display 
evidence for non-GR values. The explanation is the one given in the previous paragraph: 
some deviations correspond to unphysical Morse factors which lead to a type II disfavoring 
in the discrete case. Such a case is indistinguishable from a GR event with strong 
HOM contribution, and therefore would probably not be followed up in usual analyses. 
In particular, we see that for the scalar polarization, all continuous runs would lead 
to questioning the lensing hypothesis based on the posterior. For the modified energy flux, except for a couple 
of cases, most analyses would also lead to a follow up. For these two non-GR waveforms, it is
particularly interesting to note that the smallest deviations already lead to questioning the
Morse factor value, and while the recovered posteriors are not compatible with lensing, they are
away enough from the GR values to warrant further scrutiny.
 For the modified QNMs, only the three lowest values, 
strongly disfavoring lensing in the discrete case, would have enough significance in their posterior to 
entail follow-up analyses. For the precessing black hole mimickers following~\citet{dietric_scaled_bns}, 
the events favoring type II analyses still do here. The other events do not display enough 
evidence to deserve a follow-up, even if a few have a distribution mildly preferring non-GR values. 
Finally, for the massive graviton and the nonspinnning black hole mimickers following~\citet{ujevic_scaled_bns},
none of the continuous runs would lead to follow-up analyses, 
which is in line with the values seen in the discrete case.

\begin{figure}
\centering
\includegraphics[keepaspectratio, width=\linewidth]{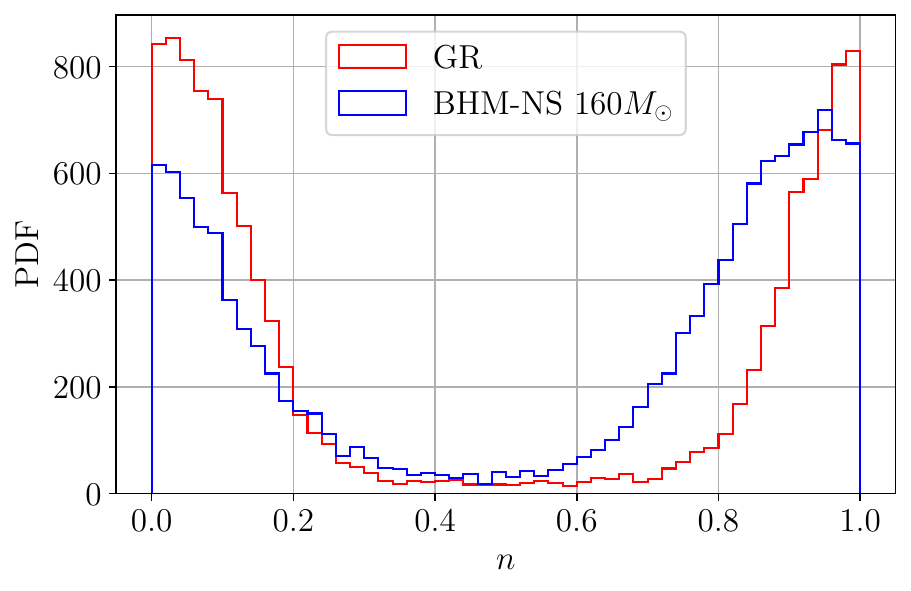}
\caption{Example posterior for the continuous Morse factor analysis 
where one could not distinguish whether the event has deviation or not, since the posterior is
compatible with having a Morse factor of 0 or 1,
 which corresponds to an unlensed GR event. 
 This example is the posterior
 for the nonspinning~\citet{ujevic_scaled_bns}-based black hole mimickers with $M = 160 M_{\odot}$ (in blue).
 Overlayed in red is the results of the GR analysis. One can clearly see
 the two distributions are close and present the same features, meaning 
 we cannot claim deviation from GR in the BHM-NS case.}
\label{fig:illustration_posterior_no_lens}
\end{figure}

\begin{figure*}
\centering
\includegraphics[keepaspectratio, width=\linewidth]{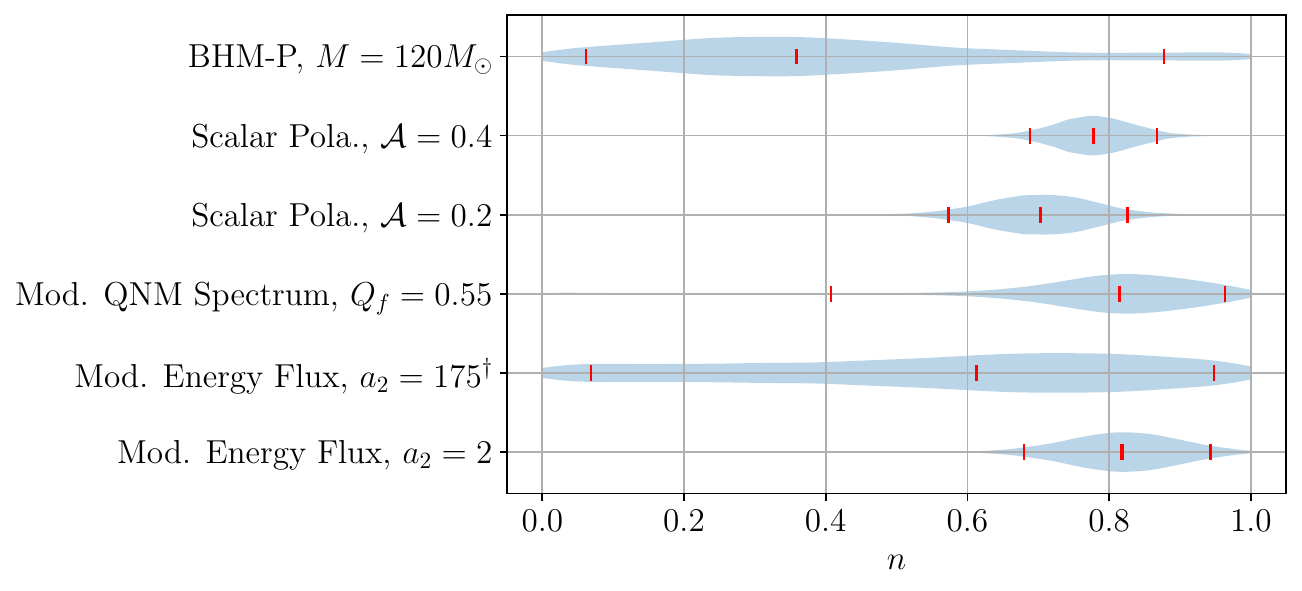}
\caption{Examples of a few posteriors illustrating the continuous Morse factor recovery for cases in 
which extra follow-up would be required. The red ticks indicate the median and 90\%
C.I.
Some of these posteriors simply favor values that are 
not compatible with GR and possibly type II, while other have Morse factor values excluding lensing
or GR.}
\label{fig:violin_intriguing_typeII}
\end{figure*}

Here, it is also important 
to stress that, in some cases, we face a disfavoring of the type II hypothesis in 
the discrete case which is also seen in the continuous case. However, the posterior 
can be such that the favored values for the Morse factor are 0 and 1, which is, again, 
indistinguishable from the unlensed GR case with strong HOM contribution. 
An illustration of such a posterior is given in Figure~\ref{fig:illustration_posterior_no_lens},
where we overlay the GR and a non-GR case for illustration purposes. 
Additionally, we give a few example of intriguing posteriors, incompatible with lensing or GR, 
in Figure~\ref{fig:violin_intriguing_typeII}. One sees that some of these posteriors are still 
compatible with GR but have a shape such that more in-depth studies are needed. 
On the other hand, some posteriors simply do not include values explainable through GR and lensing. 
A more complete overview of the recovered posteriors for all the 
considered GR deviations can be found in Appendix~\ref{app:SL_extra_res}.

\begin{center}
\begin{deluxetable*}{cccccccc}\label{tab:continuous_SL_analyses}
	\tablecaption{Results of performing type II image searches with a continuous prior on the simulated modified 
	GR events. The values given are the median values with the $90\%$ C.I. 
	Cells with a superscripted $\beta$ are those that would favor the type II hypothesis, and therefore lead 
	to follow up analyses. Those with a superscripted $\gamma$ are the cases that show deviation from the GR and lensing values but that 
	are not flagged by the discrete analysis. 
	As can be seen, follow-up analyses would be performed for all the events with a scalar polarization, 
	most of the events with a modified energy flux, a few with modified QNMs, and the same events as for the discrete 
	case for the precessing~\citet{dietric_scaled_bns}-based black hole mimickers.}
	\tablehead{
		\multicolumn{2}{c}{Massive Graviton} & \multicolumn{2}{c}{Modified Energy Flux} & \multicolumn{2}{c}{Modified QNM Spectrum} & \multicolumn{2}{c}{BHM-P} \\
		$m_{g}$ ($\text{eV}/c^2$) & med. and 90\% C.I. & $a_{2}$ & med. and 90\% C.I. & $Q_{f}$ & med. and 90\% C.I. & $M$ $(M_{\odot})$ & med. and 90\% C.I.
	}
	\tablecolumns{8}

	\startdata
    $7.07 \times 10^{-23}$ & $0.17^{+0.81}_{-0.16}$ & 2 & $0.82^{+0.12}_{-0.14} {}^{\gamma}$ & 0.50 & $0.82^{+0.14}_{-0.21} {}^{\gamma}$ & 80 & $0.45^{+0.50}_{-0.40} \hphantom{{}^\beta}$ \\
    $7.98 \times 10^{-23}$ & $0.24^{+0.73}_{-0.22}$ & 5 & $0.82^{+0.11}_{-0.11} {}^{\gamma}$ & 0.55 & $0.81^{+0.15}_{-0.41} {}^{\gamma}$ &  90 & $0.76^{+0.20}_{-0.71} \hphantom{{}^\beta}$ \\
    $9.01 \times 10^{-23}$ & $0.41^{+0.53}_{-0.37}$ & 10 & $0.83^{+0.11}_{-0.12} {}^{\gamma}$ & 0.60 & $0.79^{+0.18}_{-0.70} {}^{\gamma}$ & 100 & $0.49^{+0.46}_{-0.44} \hphantom{{}^\beta}$ \\
    $1.02 \times 10^{-22}$ & $0.47^{+0.47}_{-0.42}$ & 20 & $0.83^{+0.09}_{-0.09} {}^{\gamma}$ & 0.65 & $0.61^{+0.35}_{-0.55} \hphantom{{}^\gamma}$ & 110 & $0.48^{+0.47}_{-0.43} \hphantom{{}^\beta}$ \\
    $1.15 \times 10^{-22}$ & $0.50^{+0.45}_{-0.44}$ & 50 & $0.82^{+0.07}_{-0.07} {}^{\gamma}$ & 0.70 & $0.51^{+0.44}_{-0.46} \hphantom{{}^\gamma}$ & 120 & $0.36^{+0.52}_{-0.30} {}^{\beta}$ \\
    $1.30 \times 10^{-22}$ & $0.50^{+0.45}_{-0.45}$ & 20$^{\dag}$ & $0.81^{+0.14}_{-0.15} {}^{\gamma}$ & 0.75 & $0.49^{+0.45}_{-0.45} \hphantom{{}^\gamma}$ & 130 & $0.65^{+0.29}_{-0.56} {}^{\beta}$ \\
    $1.46 \times 10^{-22}$ & $0.49^{+0.46}_{-0.44}$ & 50$^{\dag}$ & $0.78^{+0.15}_{-0.17} {}^{\gamma}$ & & & 140 & $0.34^{+0.63}_{-0.30} \hphantom{{}^\beta}$ \\
    $1.65 \times 10^{-22}$ & $0.51^{+0.44}_{-0.46}$ & 75$^{\dag}$ & $0.76^{+0.16}_{-0.17} {}^{\gamma}$ & & & 150 & $0.59^{+0.32}_{-0.46} {}^{\beta}$ \\
    \cmidrule[\heavyrulewidth]{5-8}$1.86 \times 10^{-22}$ & $0.51^{+0.44}_{-0.46}$ & 100$^{\dag}$ & $0.83^{+0.13}_{-0.20} {}^{\gamma}$ & \multicolumn{2}{c}{Scalar Polarization} & \multicolumn{2}{c}{BHM-NS} \\
    $2.10 \times 10^{-22}$ & $0.50^{+0.44}_{-0.45}$ & 125$^{\dag}$ & $0.80^{+0.17}_{-0.65} {}^{\gamma}$ & $\mathcal{A}$ & med. and 90\% C.I. & $M$ $(M_{\odot})$ & med. and 90\% C.I. \\
	\cmidrule{5-8}
    $2.38 \times 10^{-22}$ & $0.49^{+0.45}_{-0.44}$ & 150$^{\dag}$ & $0.73^{+0.24}_{-0.67} \hphantom{{}^\beta}$ & 0.05 &  $0.78^{+0.13}_{-0.14} {}^{\gamma}$ & 80 & $0.57^{+0.38}_{-0.52}$\\
    $2.68 \times 10^{-22}$ & $0.50^{+0.45}_{-0.45}$ & 175$^{\dag}$ & $0.77^{+0.17}_{-0.51} {}^{\beta}$ & 0.10 &  $0.76^{+0.13}_{-0.14} {}^{\gamma}$ & 90 & $0.50^{+0.45}_{-0.45}$\\
    $3.03 \times 10^{-22}$ & $0.50^{+0.44}_{-0.45}$ & 200$^{\dag}$ & $0.58^{+0.37}_{-0.53} \hphantom{{}^\beta}$ & 0.15 &  $0.73^{^+0.13}_{-0.14} {}^{\beta}$ & 100 & $0.49^{+0.46}_{-0.44}$ \\
    $3.42 \times 10^{-22}$ & $0.49^{+0.46}_{-0.44}$ & & & 0.20 &  $0.70^{+0.12}_{-0.13} {}^{\beta}$ & 110 & $0.67^{+0.29}_{-0.64}$ \\
    $3.86 \times 10^{-22}$ & $0.50^{+0.45}_{-0.45}$ & & & 0.25 &  $0.68^{+0.12}_{-0.12} {}^{\beta}$ & 120 & $0.47^{+0.49}_{-0.42}$ \\
    $4.35 \times 10^{-22}$ & $0.50^{+0.45}_{-0.44}$ & & & 0.30 &  $0.77^{+0.10}_{-0.10} {}^{\gamma}$ & 130 & $0.58^{+0.37}_{-0.53}$ \\
    $4.92 \times 10^{-22}$ & $0.49^{+0.46}_{-0.44}$ & & & 0.35 &  $0.77^{+0.10}_{-0.09} {}^{\gamma}$ & 140 & $0.81^{+0.17}_{-0.79}$ \\
    $5.55 \times 10^{-22}$ & $0.49^{+0.46}_{-0.44}$ & & & 0.40 &  $0.78^{+0.09}_{-0.09} {}^{\gamma}$ & 150 & $0.81^{+0.16}_{-0.77}$ \\
    $6.26 \times 10^{-22}$ & $0.45^{+0.49}_{-0.41}$ & & & 0.45 &  $0.78^{+0.09}_{-0.09} {}^{\gamma}$ & 160 & $0.77^{+0.21}_{-0.75}$ \\
    $7.07 \times 10^{-22}$ & $0.49^{+0.46}_{-0.44}$ & & & 0.50 &  $0.78^{+0.08}_{-0.08} {}^{\gamma}$\\ 
	\enddata

	\tablecomments{\small \, $^{\dag}$ indicates the injections have been done without waveform scaling.}
\end{deluxetable*}
\end{center}

Once an event has been found with an unusual Morse factor value, standard follow-up would 
be to check for noise artifacts or waveform systematics~\citep{o3_lensing, o3_lensing_followup}
to verify whether the observed deviation has a physical origin. What our result show is that, 
should an unusual Morse factor value be found---favoring or not a type II image, 
it would be important to verify whether GR deviations are observed or not,
 and which of the two hypotheses is favored.

\subsection{Additional Biases on Source Parameters}

The primary focus of this work is whether deviations from GR would impact the single image lensing analyses; in particular, whether any of these deviations would cause a false positive result from the investigations.
If such a false positive were mistakenly identified in a real analysis, one question would also be what additional biases are added to the source parameters due to this misinterpretation of the data. In particular,
one obvious question is whether the parameters estimated by the lensing analyses are significantly degraded 
compared to their GR estimates.

Table~\ref{tab:microlensing-source-parameters} displays the comparison of the median and 90\% C.I.
boundaries between the lensed and unlensed estimates for the chirp mass, mass ratio, and luminosity
distance for the point mass microlensing and unlensed investigations. These are the parameters that
are most likely to be affected in the microlensing analysis. 

The chirp mass and mass ratio estimates in the massive graviton and modified energy flux cases do
not display significant signs of consistent bias as a result of the microlensing investigation. In
the black hole mimicker cases, the lensed investigations are typically larger with one or two
exceptions, particularly in the non-spinning $140M_{\odot}$ case---in line with the peculiarities
noted about the analysis of this event. 

The luminosity distance estimates are significantly more affected, as would be expected, given that
the posteriors demonstrate constraints in the lens parameter posteriors. In all cases, the lensed
estimate is significantly higher than the unlensed case, with the estimates in the precessing black
hole mimicker case comparatively higher than those of the other deviation types, again indicating that these
have the highest apparent magnification. 

The conclusion of this exercise in the microlensing analysis therefore is that in 
addition to the false identification of the signal as a lensed signal would be that
the source parameters under this false model would be worse in all cases than that of 
the already incorrect GR model. 

\begin{center}
	\begin{deluxetable*}{ccccccc}\label{tab:microlensing-source-parameters}
	\tablehead{& \multicolumn{3}{c}{Non-Lensed Estimate} & \multicolumn{3}{c}{Lensed Estimate} \\
		   Injection & $\mathcal{M}_{c}$ $(M_{\odot})$ & $q$ & $d_{L}$ (Mpc) & $\mathcal{M}_{c}$ $(M_{\odot})$ & $q$ & $d_{L}$ (Mpc)}
	\tablecolumns{7}
	\tablecaption{Estimates on the chirp mass, $\mathcal{M}_{c}$, the mass ratio, $q$, 
	and the luminosity distance, $d_{L}$ for the point mass and non-lensed analyses of 
	those events that would be identified as false positives under the point mass 
	microlensing hypothesis. Estimates given are the median of the posterior with the error 
	bars indicating the boundaries of the 90\% C.I. For easy reference, the true values are $\mathcal{M}_c = 31.11M_\odot$, $q = 0.8$ for the massive graviton and modified energy injections;
	$\mathcal{M}_c = 0.4328M$, $q = 0.82$ for the BHM-P injections; and $\mathcal{M}_c = 0.4154M$, $q
= 0.57$ for the BHM-NS injections. The true value of $d_L$ is $452$~Mpc for all injections.}

	\startdata
	Massive Graviton \\
	\cmidrule{1-1}$m_{g} = 4.35 \times 10^{-22} \text{ eV}/c^2$ & ${26.31}_{-0.23}^{+0.23}$ &
  ${0.166}_{-0.003}^{+0.004}$ & ${1118}_{-51}^{+59}$ & ${26.29}_{-0.20}^{+0.20}$ &
  ${0.167}_{-0.003}^{+0.007}$ & ${1338}_{-101}^{+137}$\\
	$m_{g} = 4.92 \times 10^{-22} \text{ eV}/c^2$ & ${23.81}_{-0.12}^{+0.49}$ &
  ${0.182}_{-0.038}^{+0.008}$ & ${1134}_{-67}^{+64}$ & ${24.25}_{-0.14}^{+0.12}$ &
  ${0.144}_{-0.002}^{+0.004}$ & ${1644}_{-194}^{+279}$ \\
	$m_{g} = 5.55 \times 10^{-22} \text{ eV}/c^2$ & ${21.74}_{-0.13}^{+0.12}$ &
  ${0.139}_{-0.005}^{+0.008}$ & ${1101}_{-62}^{+77}$ & ${21.65}_{-0.10}^{+0.09}$ &
  ${0.137}_{-0.002}^{+0.004}$ & ${1930}_{-104}^{+62}$\\
	$m_{g} = 6.26 \times 10^{-22} \text{ eV}/c^2$ & ${16.97}_{-0.19}^{+0.16}$ &
  ${0.25}_{-0.01}^{+0.02}$ & ${279}_{-31}^{+37}$ & ${16.23}_{-0.25}^{+0.32}$ &
  ${0.83}_{-0.01}^{+0.01}$ & ${544}_{-69}^{+84}$\\
	$m_{g} = 7.07 \times 10^{-22} \text{ eV}/c^2$ & ${14.10}_{-0.17}^{+0.17}$ &
  ${0.85}_{-0.02}^{+0.02}$ & ${255}_{-33}^{+43}$ & ${14.08}_{-0.18}^{+0.17}$ &
  ${0.83}_{-0.02}^{+0.02}$ & ${463}_{-63}^{+79}$\\
	\cmidrule[\heavyrulewidth]{1-1} Modified Energy Flux\\
	\cmidrule{1-1} $a_{2} = 50$ & ${31.25}_{-0.88}^{+0.76}$ & ${0.51}_{-0.06}^{+0.07}$ &
  ${502}_{-56}^{+53}$ & ${31.89}_{-0.90}^{+0.73}$ & ${0.50}_{-0.05}^{+0.06}$ &
  ${536}_{-48}^{+49}$\\ 
	\cmidrule[\heavyrulewidth]{1-1} BHM-P\\
	\cmidrule{1-1} $M = 80M_{\odot}$ & ${35.62}_{-1.64}^{+2.53}$ & ${0.31}_{-0.04}^{+0.09}$ &
  ${923}_{-136}^{+135}$ & ${35.46}_{-1.61}^{+2.44}$ & ${0.31}_{-0.04}^{+0.08}$ &
  ${926}_{-167}^{+734}$\\
	$M = 90M_{\odot}$ & ${41.91}_{-2.40}^{+2.99}$ & ${0.40}_{-0.05}^{+0.08}$ &
  ${945}_{-136}^{+140}$ & ${41.02}_{-2.21}^{+3.05}$ & ${0.39}_{-0.05}^{+0.07}$ &
  ${1342}_{-487}^{+850}$\\
	$M = 100M_{\odot}$ & ${47.73}_{-2.34}^{+3.02}$ & ${0.50}_{-0.05}^{+0.07}$ &
  ${927}_{-151}^{+149}$ & ${50.62}_{-5.06}^{+7.25}$ & ${0.57}_{-0.11}^{+0.19}$ &
  ${2152}_{-918}^{+796}$\\
	$M = 110M_{\odot}$ & ${52.63}_{-2.71}^{+4.05}$ & ${0.60}_{-0.06}^{+0.10}$ &
  ${841}_{-163}^{+170}$ & ${60.22}_{-7.89}^{+6.61}$ & ${0.79}_{-0.20}^{+0.18}$ &
  ${2342}_{-661}^{+903}$\\
	$M = 120M_{\odot}$ & ${57.37}_{-4.72}^{+11.40}$ & ${0.70}_{-0.11}^{+0.29}$ &
  ${863}_{-176}^{+283}$ & ${65.67}_{-5.89}^{+3.11}$ & ${0.93}_{-0.17}^{+0.07}$ &
  ${2566}_{-796}^{+949}$\\
	$M = 130M_{\odot}$ & ${55.94}_{-4.04}^{+6.92}$ & ${0.68}_{-0.10}^{+0.17}$ &
  ${994}_{-253}^{+223}$ & ${66.85}_{-3.81}^{+2.15}$ & ${0.96}_{-0.09}^{+0.03}$ &
  ${2537}_{-713}^{+757}$\\
	$M = 140M_{\odot}$ & ${54.80}_{-3.41}^{+4.86}$ & ${0.65}_{-0.08}^{+0.10}$ &
  ${1073}_{-218}^{+226}$ & ${67.07}_{-9.86}^{+2.00}$ & ${0.97}_{-0.20}^{+0.03}$ &
  ${2400}_{-918}^{+598}$\\
	$M = 150M_{\odot}$ & ${53.72}_{-3.52}^{+4.03}$ & ${0.62}_{-0.08}^{+0.10}$ &
  ${1100}_{-237}^{+277}$ & ${59.47}_{-5.05}^{+6.16}$ & ${0.86}_{-0.19}^{+0.13}$ &
  ${2347}_{-650}^{+561}$\\
	$M = 160M_{\odot}$ & ${50.52}_{-3.53}^{+4.63}$ & ${0.55}_{-0.08}^{+0.11}$ &
  ${1053}_{-215}^{+287}$ & ${60.24}_{-6.34}^{+6.29}$ & ${0.85}_{-0.20}^{+0.13}$ &
  ${2585}_{-561}^{+649}$\\
	\cmidrule[\heavyrulewidth]{1-1} BHM-NS\\
	\cmidrule{1-1} $M = 100M_{\odot}$ & ${48.02}_{-1.76}^{+1.97}$ & ${0.50}_{-0.03}^{+0.04}$ &
  ${1021}_{-77}^{+90}$ & ${48.78}_{-2.52}^{+2.41}$ & ${0.54}_{-0.06}^{+0.07}$ &
  ${1735}_{-331}^{+158}$\\ 
	$M = 110M_{\odot}$ & ${47.74}_{-1.72}^{+1.77}$ & ${0.50}_{-0.03}^{+0.04}$ & ${950}_{-72}^{+80}$
  & ${48.78}_{-2.06}^{+1.89}$ & ${0.54}_{-0.05}^{+0.06}$ &
  ${1612}_{-164}^{+136}$\\
  $M = 120M_{\odot}$ & ${49.76}_{-1.70}^{+1.80}$ & ${0.54}_{-0.04}^{+0.04}$ & ${944}_{-74}^{+85}$
                     & ${51.79}_{-2.06}^{+2.20}$ & ${0.60}_{-0.05}^{+0.08}$ &
                     ${1658}_{-165}^{+154}$\\
  $M = 130M_{\odot}$ & ${49.63}_{-1.55}^{+1.60}$ & ${0.53}_{-0.03}^{+0.04}$ & ${799}_{-60}^{+70}$
                     & ${51.58}_{-1.69}^{+1.80}$ & ${0.58}_{-0.04}^{+0.05}$ &
                     ${1395}_{-102}^{+112}$\\
  $M = 140M_{\odot}$ & ${33.54}_{-0.46}^{+0.18}$ & ${0.255}_{-0.004}^{+0.003}$ & ${294}_{-23}^{+26}$
                     & ${33.45}_{-0.20}^{+0.13}$ & ${0.253}_{-0.003}^{+0.001}$ &
                     ${397}_{-29}^{+33}$\\
  $M = 150M_{\odot}$ & ${35.10}_{-0.78}^{+0.62}$ & ${0.28}_{-0.01}^{+0.01}$ & ${376}_{-38}^{+46}$
                     & ${45.67}_{-1.46}^{+1.84}$ & ${0.45}_{-0.03}^{+0.04}$ &
                     ${761}_{-94}^{+108}$\\
  \enddata{}
\end{deluxetable*}
\end{center}

For the type II image searches, extra bias could also be present when including lensing. However, it
is expected to be less important than for the microlensing case since the extra Morse factor can only 
lead to some additional dephasing while microlensing gives rise to frequency-dependent effects, tunable 
to match the effect of non-GR effects.
This is confirmed by our results. For all the cases leading
to follow-up analyses in the continuous Morse factor scenario except for the precessing black hole mimickers 
following~\citet{dietric_scaled_bns}, the type II results present no significant bias compared to
the unlensed analysis. In the worst case scenario, some slight shift in the peak value can be
observed, but the posterior still covers the same region of the parameter space for all parameters.

\begin{figure*}
	\centering
	\includegraphics[keepaspectratio, width=0.8\textwidth]{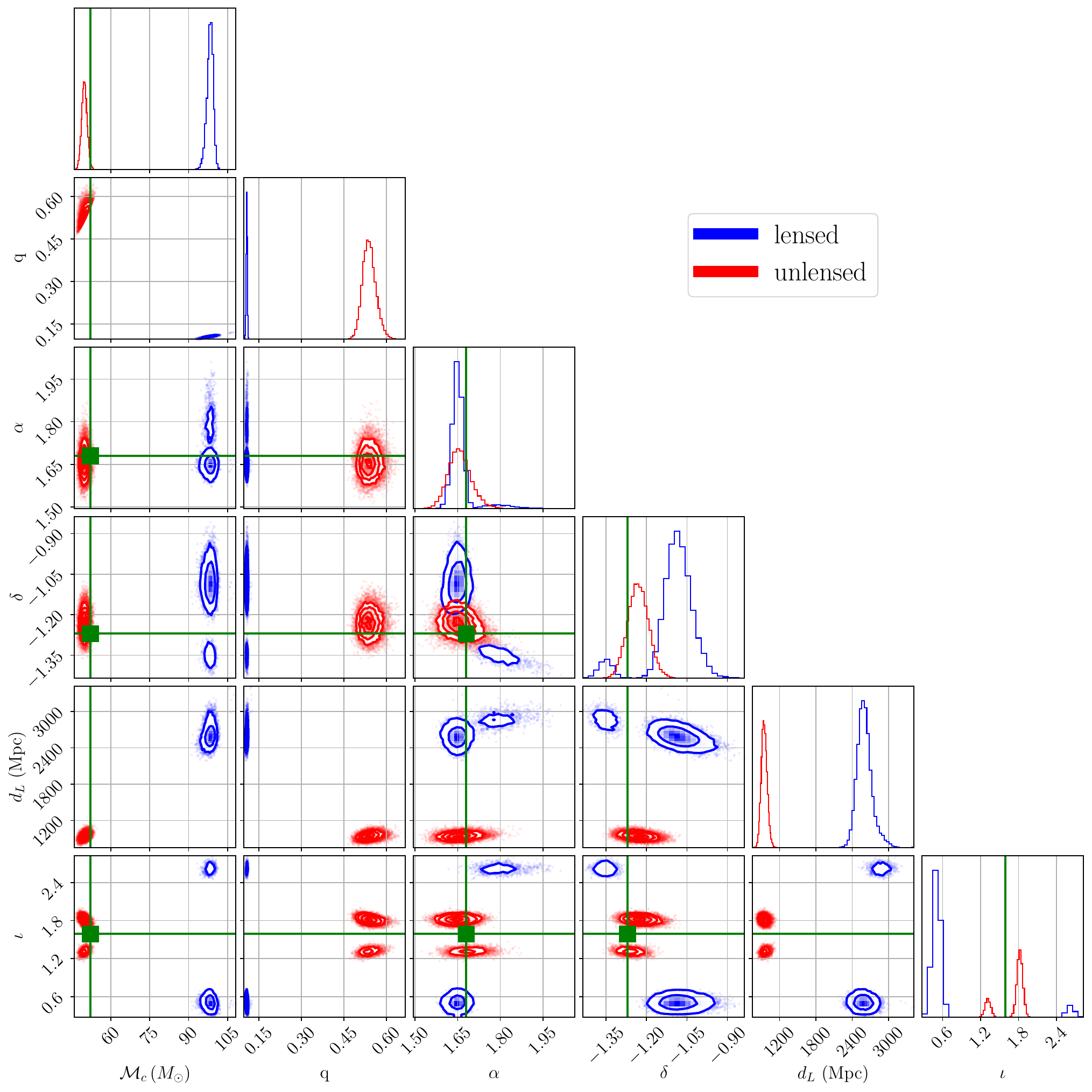}
	\caption{Representative corner plot for the parameter recovery when analyzing the $M = 120M_\odot$
	precessing black hole mimickers waveform following~\citet{dietric_scaled_bns} and searching for type II images with continuous 
	Morse factor values. The parameters shown, from left to right, are the chirp mass, the mass ratio, 
	the right ascension, the declination, the luminosity distance, and the inclination.
	Both the lensed and the unlensed do not recover the true
	parameters (in green), though the unlensed analysis is quite close for the sky location, but they are biased differently. (The true values of $q = 0.82$ and $d_L = 452$~Mpc are off of the plotted regions.)}
	\label{fig:illustration_bias_Dietrich}
\end{figure*}

\begin{figure}
	\centering
	\includegraphics[keepaspectratio, width=0.49\textwidth]{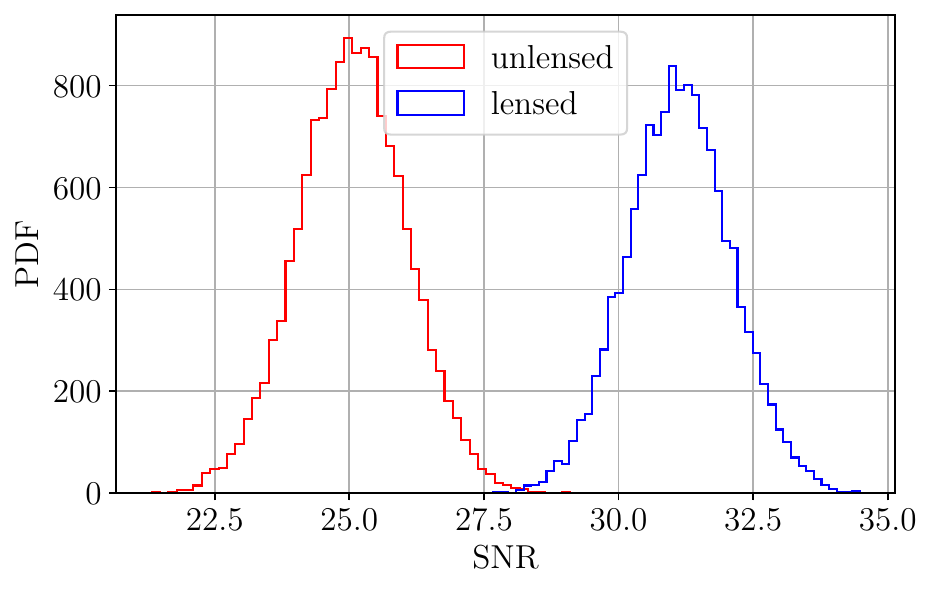}
	\caption{Representation of the recovered SNR distribution for the lensed and unlensed recovery
	of the $M = 120M_\odot$ BHM-P case. One sees that the SNR is higher for the samples
	corresponding to the lensed recovery, meaning that the extra phase shift offered 
	in that case, combined with the biasing of the other parameters, better fits the data.}
	\label{fig:snr_difference_lens_unlens}
\end{figure}

Looking more into the three triggers related to black hole mimickers, we find that their posteriors
are strongly biased for all the parameters. In particular, the lensed case leads to higher and more
asymmetric masses, a higher spin for the first object, and a larger luminosity distance when compared
to the unlensed analysis. Moreover, the sky location that is found is completely changed. 
An illustrative corner plot for the $M = 120M_\odot$ case is given in 
Figure~\ref{fig:illustration_bias_Dietrich}, where one also sees the true values. Because these black hole mimicker
waveforms are significantly different from BBH waveforms, particularly in the post-inspiral region
and changes in the Morse factor can impact the posterior quite a bit in some 
cases~\citep{2023PhRvD.108d3036V, Janquart:2021nus, Wang:2021kzt}.  These changes lead
to a result better matching the injected waveforms, 
as confirmed by the higher SNR recovered for the posterior samples
corresponding to the lensed case compared to the unlensed case, represented
in Fig.~\ref{fig:snr_difference_lens_unlens}. We see that the lensed case 
recover an SNR of $31.1^{+1.6}_{-1.6}$, while the unlensed case finds $25.1^{+1.7}_{-1.7}$.

\section{Conclusion}\label{sec:conclusion}
In the coming years, with continued improvements to the sensitivity of the current 
ground-based detector network \citep{Aasi:2013wya}, there is a reasonable likehood that a lensed GW signal 
will be observed~\citep[see e.g.][]{li_rate}. Should candidates for such a detection be identified, it will be 
important to identify any potential sources of false positivity that may mimic the 
signatures of lensing. In this work, we have analyzed phenomenological deviations from General Relativity 
in order to understand if any such deviations would in and of themselves cause such 
false positivity in those searches for lensing which examine one signal at a
 time---the microlensing, millilensing, and type II strong lensing searches.

To perform this investigation, we have taken several possible 
deviations from GR---specifically a massive graviton, modifications 
to the energy flux and QNM spectrum, and an additional scalar polarization mode--- 
as well as the possibility of exotic compact objects---and performed both the 
microlensing and strong lensing type II image search as found in~\cite{o3_lensing} 
as well as the follow-up analyses performed in~\cite{o3_lensing_followup} to determine 
to what extent any false positive candidates would be identified, and if any could 
be ruled out by the follow-up analyses. Here we have used the predicted noise curves for
O4. Our results indicate that in the microlensing, 
millilensing, and type II image search cases, there are a number of potential scenarios 
that could result in a false positive. 

In the initial isolated point mass microlensing analysis, a graviton with mass
$\geq 4.36 \times 10^{-22}\,\textrm{eV}/c^2$, would cause a sufficient preference for 
the microlensing hypothesis to warrant further follow-up investigation. 
This further investigation in the form of the SIS microlensing analysis and the millilensing analysis would similarly 
find significant support for their respective models meaning that these would be identified as lensing candidates. 
However, gravitational wave observations place constraints on the mass of the graviton to well below this threshold
($< 1.27 \times 10^{-23}\,\textrm{eV}/c^2$ at $90\%$ credibility). 
Beneath this threshold, support for the unlensed waveform was consistent.
On the contrary, for type II strongly-lensed images, dispersion due to a massive graviton 
does not lead to support
for lensing, and no false positivity is expected in this case.

Modifications to the energy flux with $a_{2} = 50$ were able to produce a false positive in both the 
initial point mass microlensing search and the follow-up SIS microlensing analysis. 
The follow-up millilensing analysis would also identify the modification 
with $a_{2} = 20$ as a slenderly favored candidate. Such a deviation also lead to strong disfavoring
of type II images in the discrete type II images searches. This is indistinguishable from the results for a GR signal
with significant higher-order mode content and would not be followed up. Still, when investigating the continuous
results, we find a deviating value for the Morse factors in nearly all cases, and in particular for the
smallest deviation parameters in this work. Those are not compatible
with lensing Morse factors but would be sufficiently odd to warrant extra scrutiny.

Neither modifications to the QNM spectrum nor the addition of an extra scalar polarisation mode 
yielded any situations in which the initial microlensing investigation would yield a sufficiently 
confident false positive to warrant additional investigation, with the unlensed model 
confidently preferred in all cases. 
However, the $\mathcal{A} = 0.20$ and $\mathcal{A}=0.25$ 
additional scalar polarisation cases would be slenderly preferred in a millilensing 
analysis as single image type II candidates which is consistent with the findings of 
the strong lensing type II analysis in these cases. Additionally, the modified QNM spectrum leads to Morse factor
values incompatible with GR or lensing but sufficiently constrained to require additional investigations.
For the scalar polarization, values not found as type II images in the discrete analyses lead to a strong
disfavoring, and their continuous equivalents show Morse factor posteriors that are incompatible with lensing. 
Again, the posteriors are sufficiently well constrained to entail extra investigations. We stress here 
that $\mathcal{A} = 0.15, 0.20, 0.25$ lead to false positives in the discrete and continuous
case and also in the millilensiing analyses. Such events could lead to false identifications of lensing.

Binaries of exotic compact objects, for which we used the proxy of scaling BNS waveforms to BBH masses, 
resulted in significant false positivity in both the precessing and non-spinning cases. 
These waveforms displayed preference for all three of the point mass, SIS, and millilensing 
models compared with a standard unlensed case in the majority of the analyses performed. 
As such objects have not been conclusively ruled out, microlensing and millilensing 
candidates would need to be examined carefully for any signatures of being such objects. 
One particular indicator in the case of the non-spinning analysis may be significant chain 
disagreement when parameter estimation was performed on these waveforms. 
For strongly-lensed type II images, the picture is somewhat different as only
a few cases lead to a preference of type II images, corresponding to the precessing case
using the~\citet{Dietrich:2018phi} waveform with $M \in \{120, 130, 150 \} M_{\odot}$. Still, those events would require extra
investigations as both the discrete and continuous runs lead to a preference for the type II 
image hypothesis.

We also examined the recovered posteriors of the lensed waveforms compared with the 
unlensed waveforms to note whether false identification as a lensed waveform would further
degrade the source parameter estimates compared with the already incorrect GR waveform. 
In the case of the microlensing analysis, this would have significant consequences on the 
luminosity distance estimates, which with very few exceptions were considerably 
worse than the unlensed estimates. Similarly, for the type II image analyses,
significant differences between the lensed and the unlensed recovery can happen, 
although large deviations are only present in a few scenarios.

The conclusions of this work therefore are that there are scenarios in which a deviation from 
GR would be the source of a false positive result in the single event lensing searches and therefore 
any candidates for detection should be investigated for potential deviations from GR as part of 
the identification process for lensed events. Should a misidentification of a GR deviation as 
a lensed event take place, the source parameter estimates would display further degradation 
from the truth as compared with the unlensed GR analysis. Additionally, the continuous Morse factor
lensing analysis \citep[which has the same phase modification as one of the modified dispersion testing
GR analyses;][]{Ezquiaga:2022nak} is quite sensitive to certain GR deviations.

It will, therefore, be important to know if the tests of GR being applied to GW data would be able
to identify the GR deviations that can be misidentified as lensed events, and whether the ultimate
preference would be for the GR deviation rather than the lensing hypothesis. On the first question,
it is likely that the GR deviation would be identified in at least some cases, since a smaller graviton mass 
than the minimum that gives significant evidence for lensing was identified
as such with very high confidence in a similar injection study
in~\cite{johnson_mcdaniel_modified_energy_flux}. However, this is not guaranteed to be the case for all deviations,
so it will be important to study of the response of the lensing and testing GR analyses to waveforms in various alternative theories,
once these can be calculated sufficiently accurately, and to first consider phenomenological GR deviations of the type considered here.
Indeed, the LVK testing GR group is also studying the
response of the tests of GR the group applies to the types of waveforms considered here.

\section*{Acknowledgements}\label{sec:acknowledgements}
The authors thank the Testing General Relativity group of the LIGO-Virgo-KAGRA Collaborations for
useful discussions and in particular N.~V.~Krishnendu for internal review of the manuscript. 

We also thank the authors of \cite{dietric_scaled_bns} and \cite{ujevic_scaled_bns} for making 
their waveforms available and Archisman Ghosh for developing the code used to create the injections.

M.W.\ is supported by the Science and Technology Facilities Council (STFC) of the United Kingdom Doctoral Training Grant ST/W507477/1.
J.J.\ is supported by the research program of the Netherlands Organization for Scientific
Research (NWO). N.K.J.-M.\ is supported by NSF Grant AST-2205920.

The authors are grateful for computational resources provided by the LIGO Laboratory and supported
by NSF Grants PHY-0757058 and PHY-0823459. This material is based upon work supported by NSF's LIGO
Laboratory which is a major facility fully funded by the National Science Foundation.

The authors acknowledge the Python community~\citep{van1995python} and the core set of tools
that enabled this work including \textsc{Numpy}~\citep{Harris:2020xlr}, \textsc{SciPy}~\citep{scipy},
\textsc{matplotlib}~\citep{Hunter:2007ouj},
\textsc{corner}~\citep{ForemanMackey2016cornerpySM},
\textsc{bilby\_pipe}~\citep{Romero_Shaw_2020}, and
\textsc{asimov}~\citep{Williams2023}.

\appendix

\section{Details of the Construction of Modified Waveforms and Illustrations of the Waveforms}\label{app:non-GR_constr}
\subsection{Scalings to obtain other Kerr-Newman QNMs}\label{subsec:app_KN_QNM_scaling}

We now describe how we obtain approximate (complex) frequencies for the Kerr-Newman $(2,1,n)$, $(3,2,n)$,
$(4,4,n)$, and $(5,5,n)$ ($n \in \{0,1\}$) QNMs from the results for the $(2,2,0)$ and $(3,3,0)$ QNMs for which we have
tabulated results from \cite{dias_qnm}. They also provide results for the $(2,2,1)$ mode, but we will not end up
using this, except as a cross-check. For $n = 1$, we only need the imaginary part (equivalently the damping time),
since this is all that is used in \texttt{TEOBResumS-v3-GIOTTO}.

\begin{figure}[t]
\centering
\includegraphics[width=0.49\columnwidth]{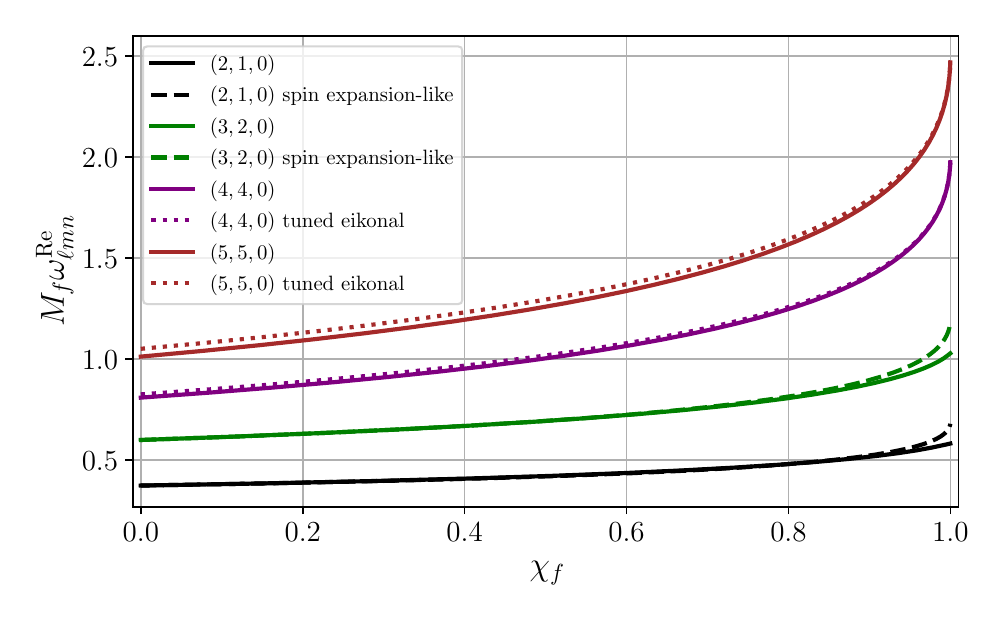}
\includegraphics[width=0.49\columnwidth]{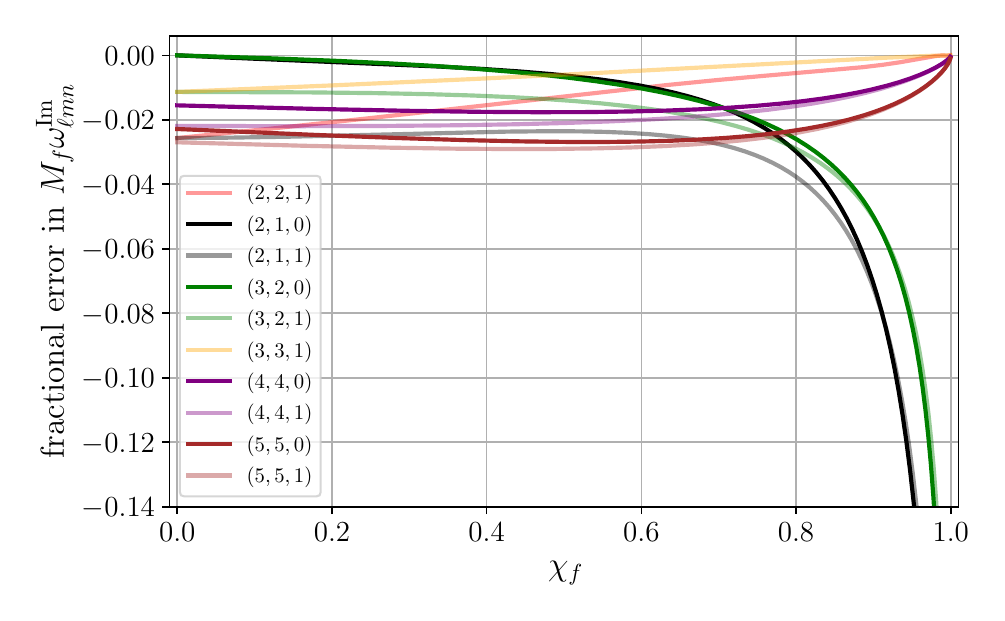}
\caption{\label{fig:QNM_scaling} Illustrations for Kerr of the approximations we use for the (complex) QNM frequencies to obtain our scalings for the modes for which we do not have
Kerr-Newman results: \emph{(Left panel)} The spin expansion-like expression in Equation~\eqref{eq:QNM_llm1} we use for the $(2,1,0)$ and $(3,2,0)$ QNMs and the tuned eikonal expression from Equations~\eqref{eq:QNM_eik} and~\eqref{eq:QNM_eik_tuned} we use for the $(4,4,0)$ and $(5,5,0)$ QNMs. \emph{(Right panel)} The fractional errors in the tuned eikonal expression for the imaginary QNM frequencies. Of course, as discussed in the text, we apply all these corrections to the Kerr
frequencies in such a way that they reduce to Kerr exactly for zero charge. We just show their predictions for Kerr as an indication of how well we expect them to reproduce the
Kerr-Newman results.}
\end{figure}

We start from the eikonal approximation for the complex frequencies, which is valid for large $\ell$, using the version from \cite{yang_qnm,Li:2021zct}:
\begin{equation}
\label{eq:QNM_eik}
\omega^\text{eik}_{\ell m n} = \left(\ell + \frac{1}{2}\right)\omega_\text{orb} + m\omega_\text{prec} - i\left(n + \frac{1}{2}\right)\gamma_\text{L}.
\end{equation}
Here $\omega_\text{orb}$ and $\omega_\text{prec}$ are frequencies corresponding to the spherical photon orbit and $\gamma_\text{L}$ is the Lyapunov exponent of that orbit. We take
all of these to be free parameters, so we will refer to our final result as a \emph{tuned eikonal} expression. Specifically, we determine $\omega_\text{orb}$ and $\omega_\text{prec}$ using
the real parts of the $(2,2,0)$ and $(3,3,0)$ QNMs (denoted with a superscript ``Re''), giving
\begin{subequations}
\label{eq:QNM_eik_tuned}
\begin{align}
\omega_\text{orb} &= 6\omega^\text{Re}_{220} - 4\omega^\text{Re}_{330}\\
\omega_\text{prec} &= -7\omega^\text{Re}_{220} + 5\omega^\text{Re}_{330}.
\end{align}
\end{subequations}
We apply this expression to obtain the real parts of the $(4,4,0)$ and $(5,5,0)$ QNMs, where we find that it gives good accuracy for Kerr, particularly for larger spins; see
Figure~\ref{fig:QNM_scaling}. In the application to Kerr-Newman, we compute the ratio between the Kerr-Newman and Kerr expressions and using this to scale the Kerr QNM frequencies
in \texttt{TEOBResumS-v3-GIOTTO}, so that this reduces to the Kerr expressions used in \texttt{TEOBResumS-v3-GIOTTO} for zero charge.

For the real parts of the $(2,1,0)$ and $(3,2,0)$ QNMs, we find that the tuned eikonal expression is not as accurate as writing
\begin{equation}
\label{eq:QNM_llm1}
\omega^\text{Re, spinning}_{\ell,\ell - 1,0} \simeq \omega^\text{Re, nonspinning}_{\ell\ell0} + \frac{\ell - 1}{\ell}\left(\omega^\text{Re, spinning}_{\ell\ell0} - \omega^\text{Re, nonspinning}_{\ell\ell0}\right).
\end{equation}
We obtained this expression by noting that there is only a difference between the QNM frequencies for different $m$ modes due to spin, and then using the scaling of the frequency with
$m$ from the eikonal expression, which is the same as that for the linear-in-spin corrections \citep[see, e.g.,][]{Pani:2013ija}, so we refer to this as a \emph{spin expansion-like}
expression. As before, we scale the Kerr frequencies by the ratio of this expression for Kerr-Newman and Kerr to recover the Kerr frequencies exactly in the limit of zero charge. We
illustrate the accuracies of this expression for Kerr in Figure~\ref{fig:QNM_scaling}, using the data from \cite{Berti:2005ys,Berti:2009kk}.

For the imaginary parts of all the modes, we take $\gamma_\text{L}/2$ to be given by the $(2,2,0)$ inverse damping time for the $\ell = 2$ modes and by the $(3,3,0)$ inverse damping
time for all other modes. We find that for the final spins and charges we consider, there is a small difference ($< 2\%$) between the values of $\gamma_\text{L}$ that are inferred from
the $(2,2,0)$ and $(3,3,0)$ modes and an even smaller difference ($< 0.6\%$) between the values that would be inferred from the $(2,2,0)$ and $(2,2,1)$ modes [i.e., if one takes the
$(2,2,1)$ damping time to give $3\gamma_\text{L}/2$]. These expressions give accuracies of better than $3\%$ for Kerr for the final spin values we consider, as illustrated in  Figure~\ref{fig:QNM_scaling}. (However, recall that since we are using these expressions to obtain a phenomenological non-GR waveform, the accuracies are not that important, since the use of the Kerr-Newman values is purely a way to obtain reasonable scalings for all the necessary QNM frequencies.)
We then scale the Kerr values of the damping times used in \texttt{TEOBResumS-v3-GIOTTO} by the ratio of the Kerr-Newman value to the Kerr value for $\gamma_\text{L}$.

\subsection{Illustrations of modified EOB final mass and spin}\label{subsec:app_Mfaf}

\begin{figure}[t]
\centering
\includegraphics[width=0.49\columnwidth]{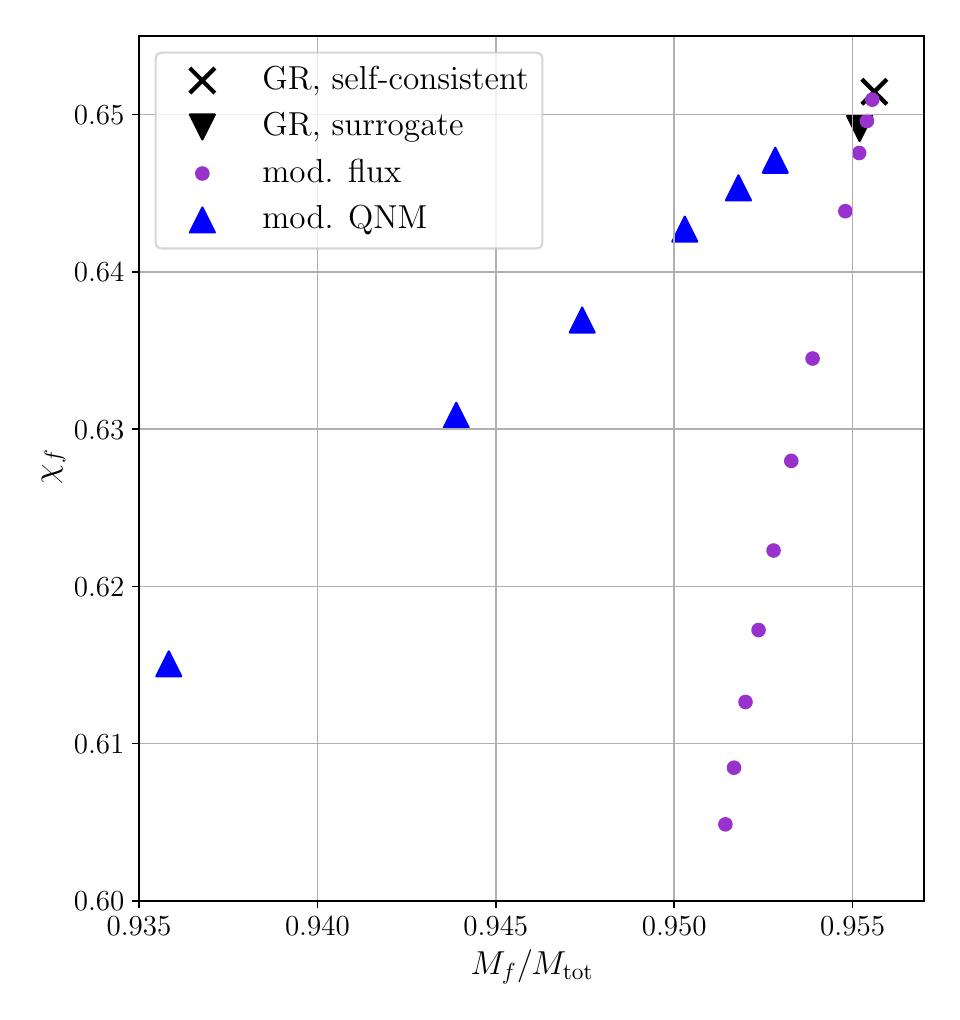}
\caption{\label{fig:Mf_af_modGR} The self-consistent final mass and spin used for the waveforms with the modified energy flux and QNM spectrum, as well as the self-consistent results obtained for the underlying GR \texttt{TEOBResumS-v3-GIOTTO} model and the numerical relativity surrogate prediction. The modified flux points correspond to $a_2 \in \{2, 5, 10, 20, 50 , 75, 100, 125, 150, 175, 200\}$ and the modified QNM points correspond to $Q_f \in \{0.5, 0.55, 0.6, 0.65, 0.7, 0.75\}$. In both cases the points move away from the GR value with increasing values of the deviation.}
\end{figure}

See Figure~\ref{fig:Mf_af_modGR} for the final mass and spin one obtains self-consistently with the changes to the QNM spectrum described in the previous subsection as well as the modifications to the flux. We also compare with the final mass and spin given by the precessing numerical relativity surrogate model \texttt{NRSur7dq4} \citep{Varma:2019csw}. The small differences between the surrogate values and the self-consistent GR value are likely an indication of the accuracy of the \texttt{TEOBResumS-v3-GIOTTO} model, since the surrogate model's errors are smaller than these differences for most cases. Note that while the final spin decreases with increasing charge, the final black hole is always closer to extremal (and thus the QNMs are less damped) as the charge increases.

\subsection{Expressions for scalar modes}\label{subsec:app_ST_scaling}

Here we give the explicit expressions we use to obtain the scalar modes we add to obtain the scalar-tensor waveforms. Recall that we take the sensitivities to be equal, so $\mathcal{S}_- = 0$ in the notation of~\cite{bernard_scalar_polarisation} and the amplitude of the scalar waveform depends on a single combination of scalar-tensor theory parameters, 
\begin{equation}
	a_{\textrm{scalar}} \coloneqq \frac{\sqrt{\alpha}\zeta\mathcal{S}_{+}}{1 - \zeta}.
	\label{eq:scalar-amplitude}
\end{equation}
Here the parameters on the right-hand side are defined in Table~I of~\cite{bernard_scalar_polarisation}. Using just the leading-order expression,
Equation~\eqref{eq:scalar-amplitude} then allows the scalar waveforms to be written in terms of their tensor waveform counterparts. Absorbing all constant quantities into 
$\bar{a}_{\textrm{scalar}}$, this may be written as:
\begin{equation}
	h^{s}_{\ell{}m} = 
	\begin{cases}
		-\bar{a}_{\textrm{scalar}}
	        \sqrt{\frac{\ell{}(\ell{}-1)}{(\ell{}+1)(\ell{}+2)}}
		h_{\ell{}m} & \textrm{for } \ell + m \textrm{ even},\\
		0 & \textrm{otherwise}.
	\end{cases}
	\label{eq:scalar-waveforms}
\end{equation}

The overall scalar contribution to the detector strain is therefore
\begin{equation}
	h^{\textrm{scalar}} = F_{S} \sum_{\ell,m} h^{s}_{\ell{}m} {}^{0}Y_{\ell{}m}(\iota, \phi_{0}),
\end{equation}
where $F_{S} = -\frac{1}{2} \sin^{2} \theta \cos 2\phi$ is the antenna response pattern for the scalar breathing mode~\citep{lab_testing_gr} ($\theta$ and $\phi$ are the spherical
coordinates of the source in the detector frame), and ${}^{0}Y_{\ell{}m}$ are the standard zero spin spherical harmonics, where $\iota$ and $\phi_{0}$ are the binary inclination and
phase at the time of coalescence, respectively. We use the implementation of $F_{S}$ in the \textsc{PyCBC} package~\citep{nitz_pycbc}. 

We set the value of the $\bar{a}_{\textrm{scalar}}$ quantity in terms of the relative amplitude of the $(2,\pm2)$ scalar and tensor polarizations' contribution to the response of the detector with the largest SNR for the tensor signal, here H1 (so thus includes the spherical harmonics and the detector's antenna pattern). Specifically, we fix the amplitude ratio given by
\begin{equation}
	\mathcal{A} = \frac{\bar{a}_{\textrm{scalar}}}{\sqrt{6}}
	                       \frac{|F_{S}|}{\sqrt{F_{+}^{2} + F_{\times}^{2}}}
			       \frac{2 \cdot {}^{0}Y_{22}(\iota, 0)}{{}^{-2}Y_{2,2}(\iota, 0) + {}^{-2}Y_{2,-2}(\iota,0)},
	\label{eq:scalar-amp-ratio}
\end{equation}
where $F_{+}$ and $F_{\times}$ are the antenna responses for the plus and cross tensor polarizations, 
respectively, and ${}^{-2}Y_{2,\pm2}$ are the $(2,\pm2)$ spin-$(-2)$-weighted spherical harmonics.

\subsection{Illustrations of waveforms}\label{subsec:app_wf_illustration}

\begin{figure}[t]
\centering
\includegraphics[width=0.49\columnwidth]{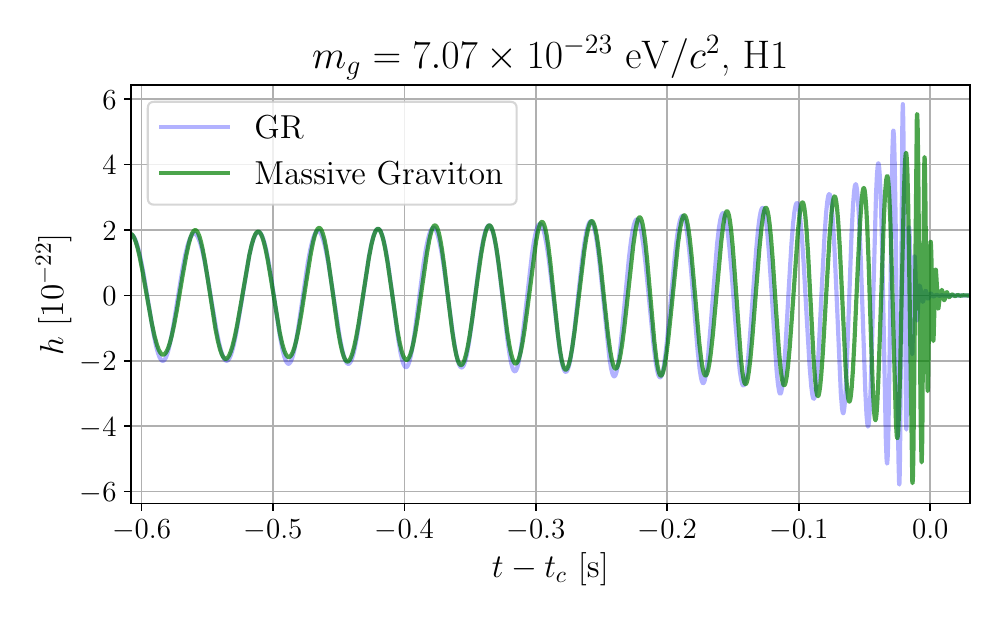}
\includegraphics[width=0.49\columnwidth]{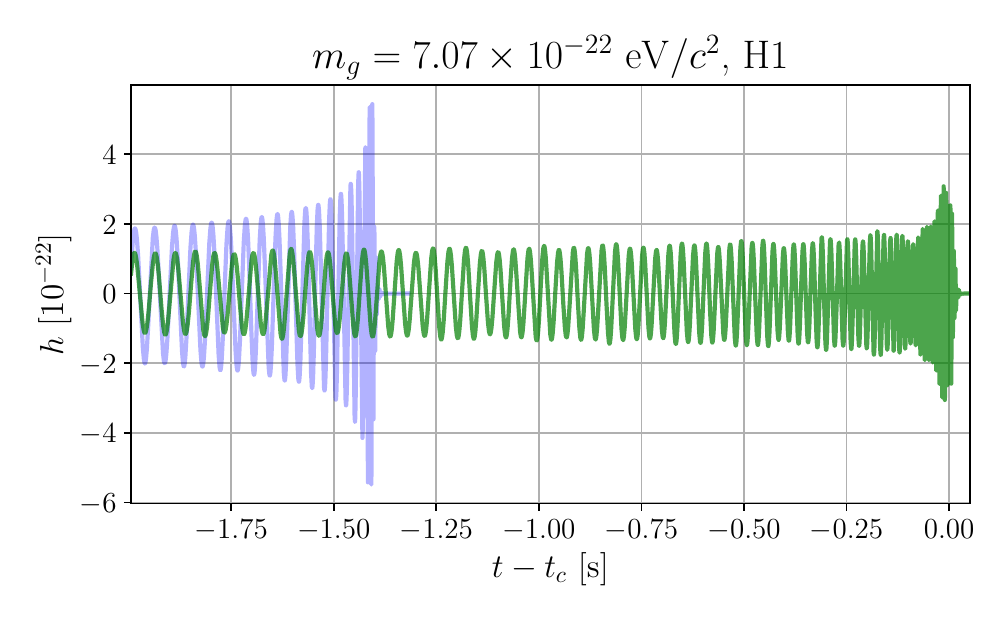}
\caption{\label{fig:mg_wfs} Illustrations of the massive graviton waveforms modified EOB waveforms with modified flux, projected onto the H1 detector and compared with the GR waveform, aligned at $20$~Hz. We show the smallest and largest graviton masses considered in this paper.}
\end{figure}

\begin{figure}[t]
\centering
\includegraphics[width=0.49\columnwidth]{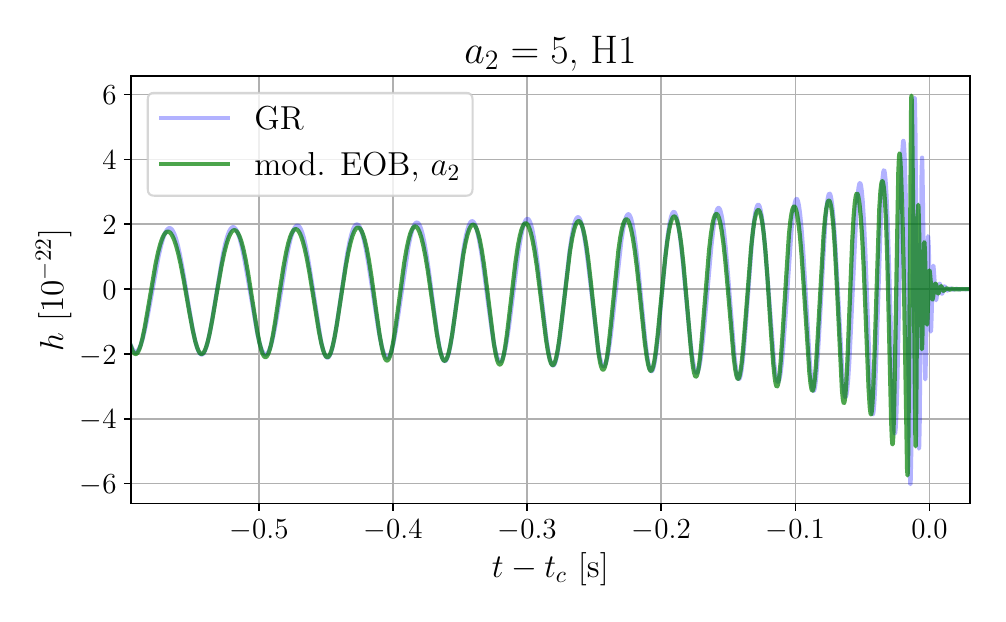}
\includegraphics[width=0.49\columnwidth]{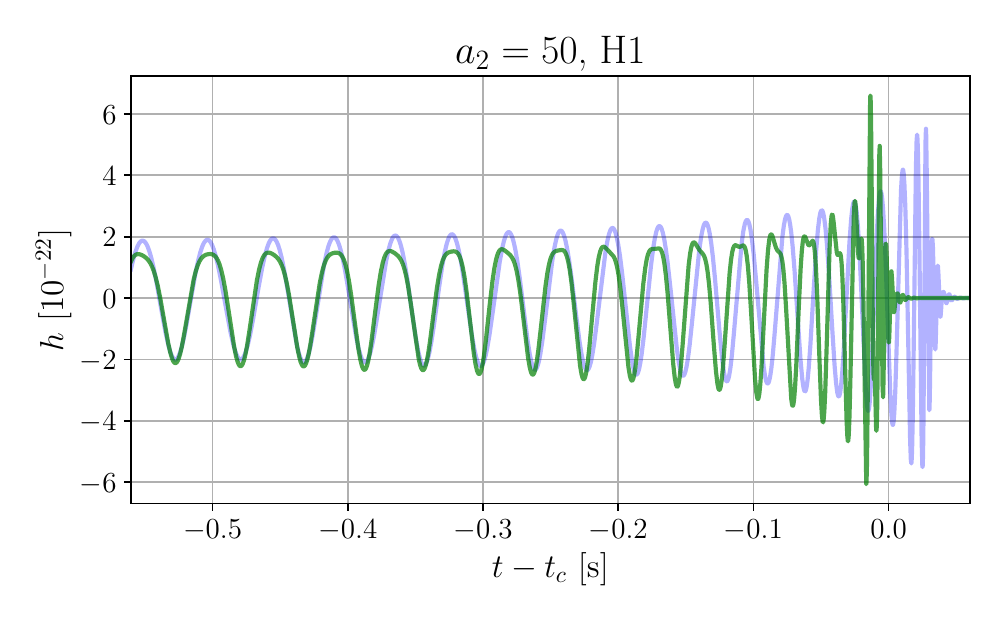}\\
\includegraphics[width=0.49\columnwidth]{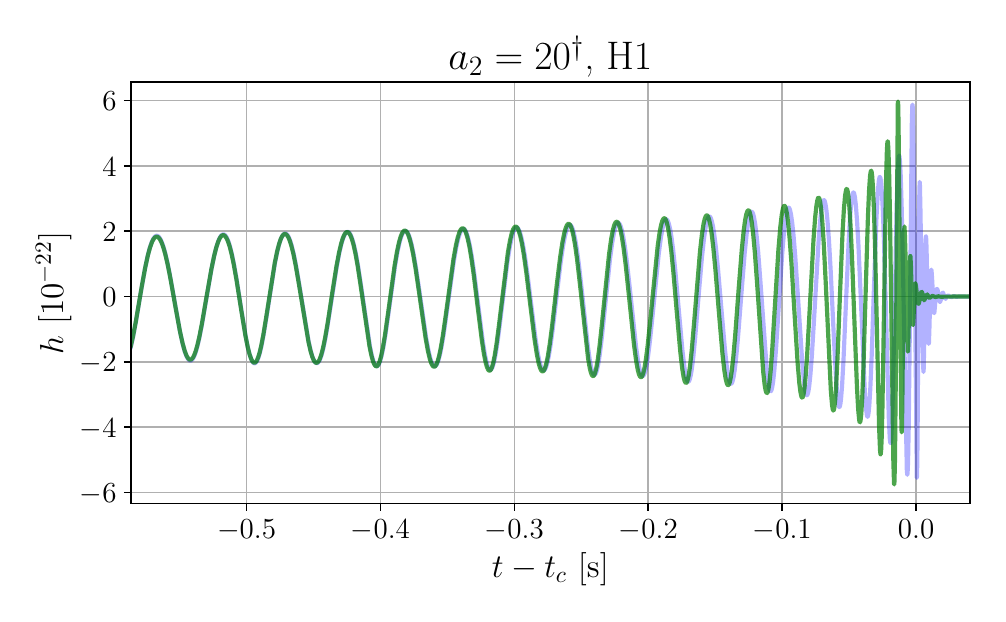}
\includegraphics[width=0.49\columnwidth]{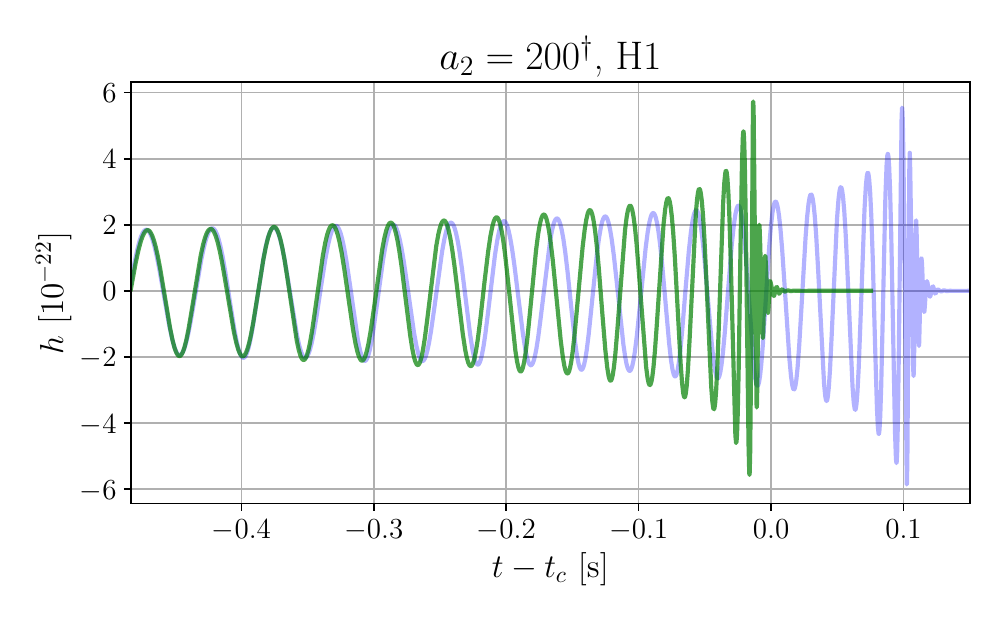}
\caption{\label{fig:alpha2_wfs} Illustrations of the modified EOB waveforms with modified flux, both with (upper panels) and without (lower panels) waveform scaling, compared with the GW150914-like GR waveform. The waveforms are aligned at $20$~Hz and projected onto the H1 detector. We show the smallest and largest GR deviations considered in the case without waveform scaling. In the case with waveform scaling, we show the second-smallest and largest GR deviations, since the modified EOB waveform is almost indistinguishable from the GR waveform for the smallest deviation we consider in this paper.}
\end{figure}

\begin{figure}[t]
\centering
\includegraphics[width=0.49\columnwidth]{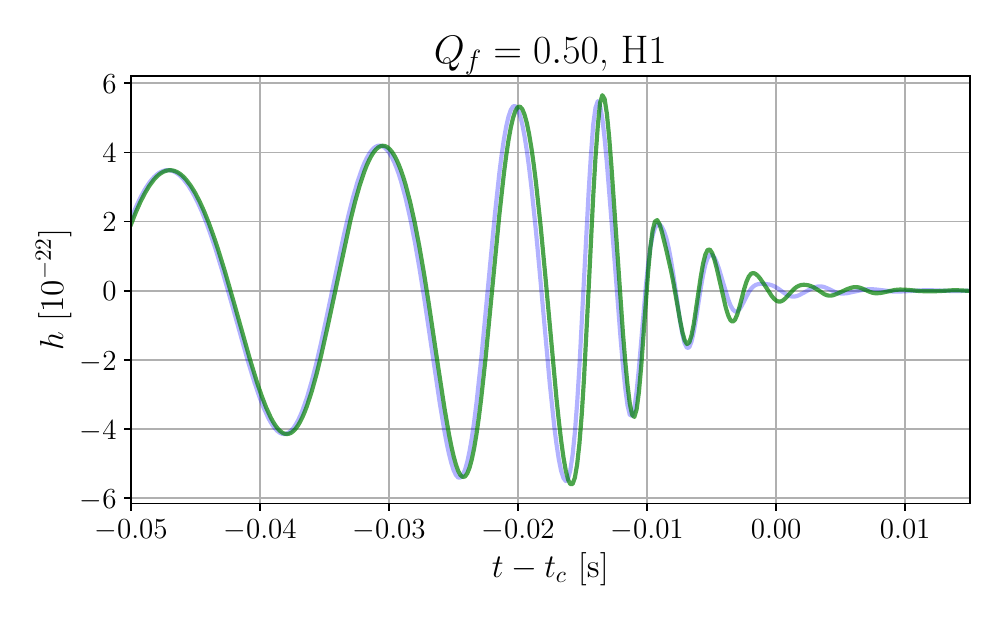}
\includegraphics[width=0.49\columnwidth]{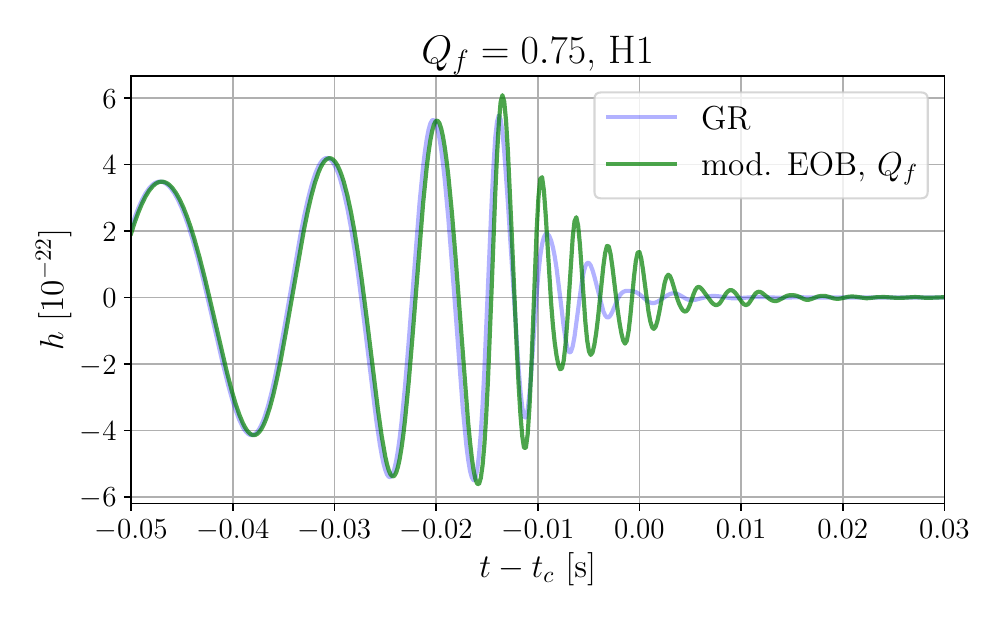}
\caption{\label{fig:qnm_wfs} Illustrations of the modified EOB waveforms with modified QNM spectrum, compared with the GW150914-like GR waveform. The waveforms are aligned at $20$~Hz and projected onto the H1 detector, but are zoomed in to show the merger-ringdown phase, which is the only portion of the waveform where there are any differences from GR. We show the smallest and largest GR deviations we consider in this paper.}
\end{figure}

\begin{figure}[t]
\centering
\includegraphics[width=0.9\columnwidth]{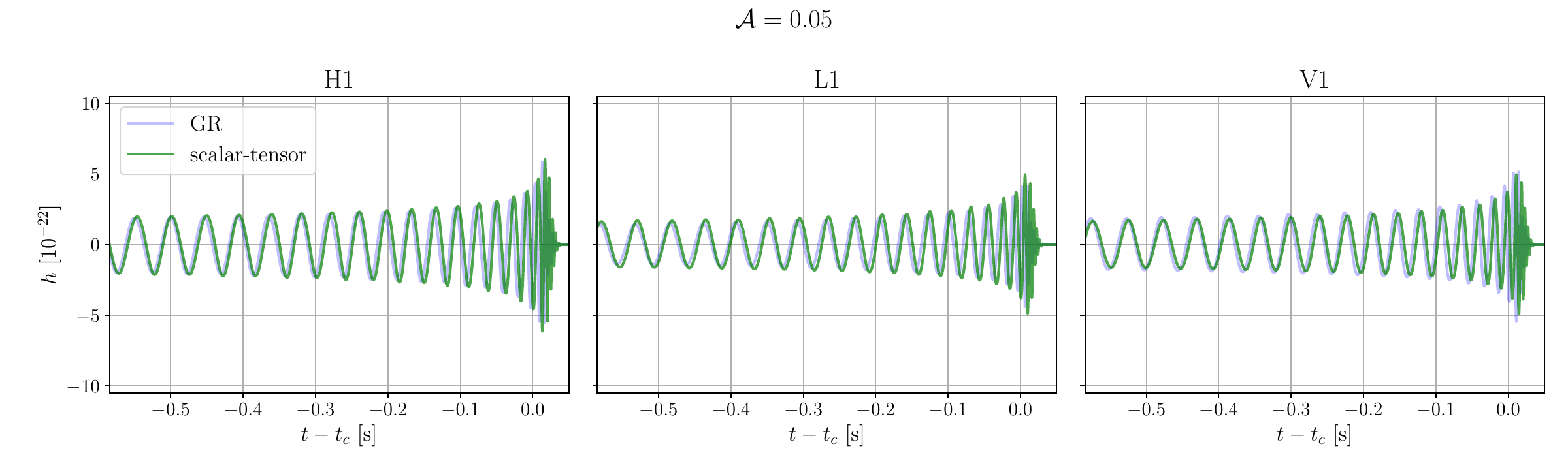}\\
\includegraphics[width=0.9\columnwidth]{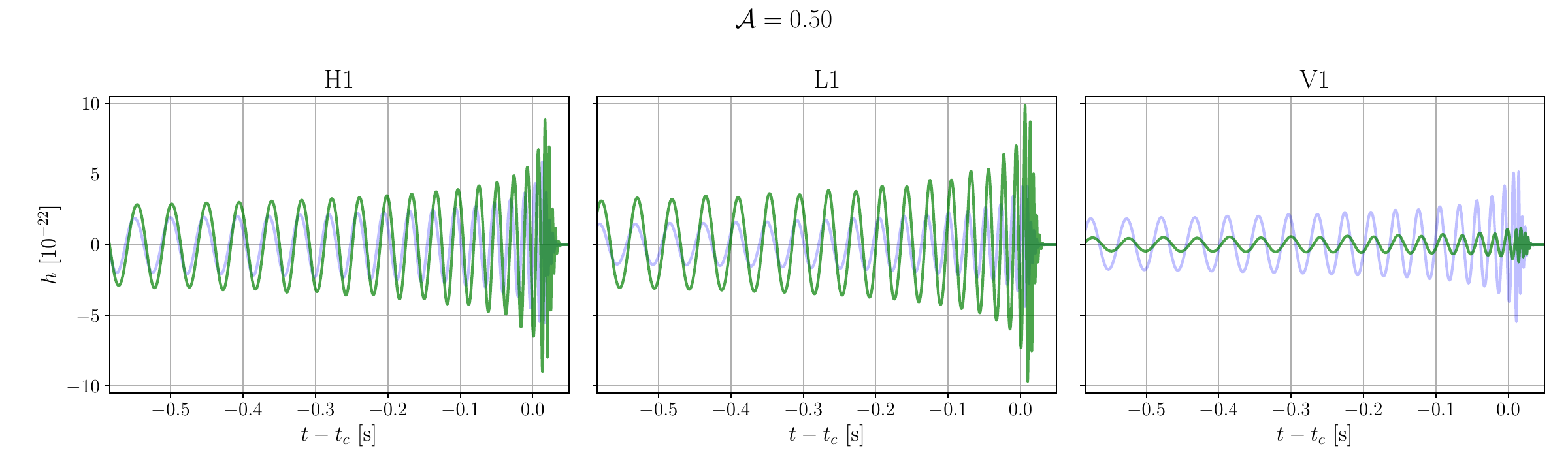}
\caption{\label{fig:ST_wfs} Illustrations of the scalar-tensor waveforms projected onto the three detectors, compared with the GW150914-like GR waveform and aligned at $20$~Hz. These show the smallest and largest $\mathcal{A}$ values considered in this paper.}
\end{figure}

\begin{figure}[t]
\centering
\includegraphics[width=0.45\columnwidth]{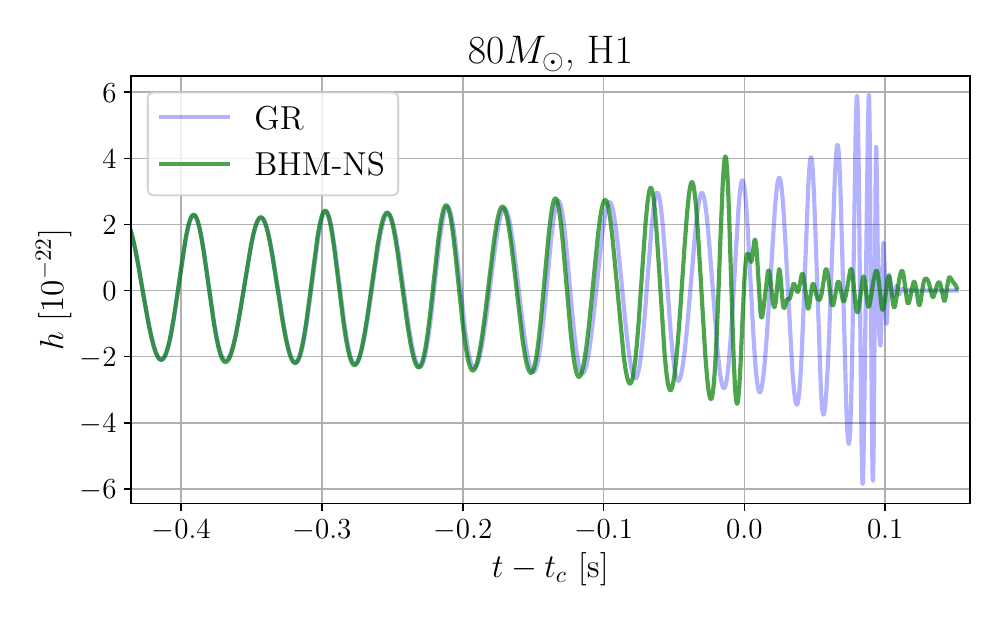}
\includegraphics[width=0.45\columnwidth]{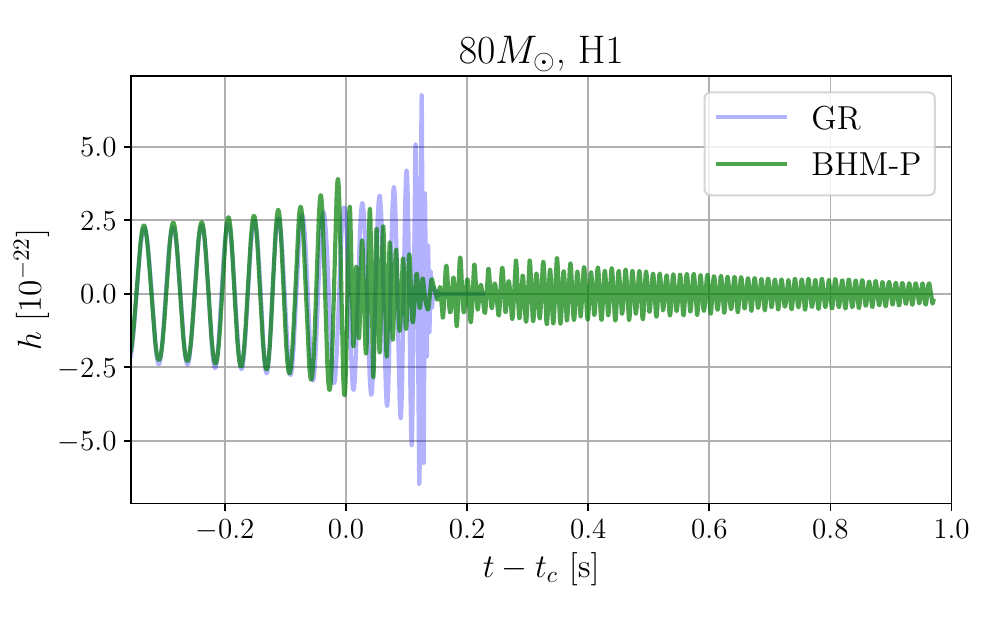}
\caption{\label{fig:BHM_wfs} Illustrations of the black hole mimicker waveforms projected onto the H1 detector for a total mass of $80M_\odot$, compared to the GR waveforms with the same masses and spins, aligned at $20$~Hz.}
\end{figure}

Here we show representative illustrations of the non-GR and BHM waveforms we consider, in all cases comparing with the GR BBH equivalent with the same masses and spins (computed using \texttt{TEOBResumS-v3-GIOTTO}). We align the waveforms at $20$~Hz using the inertial frame $(2,2)$ mode. In all but one of the non-GR cases, we show the smallest and largest GR deviation considered. We show the representative waveforms for the massive graviton, modified flux, modified QNM spectrum, and scalar-tensor waveforms in Figures~\ref{fig:mg_wfs}, \ref{fig:alpha2_wfs}, \ref{fig:qnm_wfs}, and~\ref{fig:ST_wfs}, in all cases comparing with the GW150914-like GR waveforms. We show the BHM waveforms in Figure~\ref{fig:BHM_wfs}, where we use the smallest total mass considered in this paper, in order to show the largest amount of the waveform (since the plot starts at $20$~Hz). The post-merger portions of the BHM waveforms are truncated (as they are in the injections) before they start to exhibit drifts due to numerical inaccuracies.

\section{Prior Choices for the Parameter Estimation}\label{app:prior}
In this appendix, we present the full set of priors that were used for the
parameter estimation investigations. These are the defaults for all analyses. In
the case of the BHM-P investigations, the luminosity distance prior maximum was
increased to $5000$~Mpc and the effective luminosity distance priors for the
millilensing investigation was increased to $10^5$~Mpc. For definitions of each
of the source parameters, see Appendix E of~\cite{Romero_Shaw_2020}.

Additionally, we note that the chirp mass prior range is not a typo. The erroneously
large width of this prior was accidentally placed in the settings. Whilst this will
have made the sampling process less efficient, it should not have impacted the
final results. 

Finally, there is a small inconsistency between the specific \textsc{Planck15}~\citep{Ade:2015xua}
cosmology used by \textsc{bilby}---which corresponds to the one in
\textsc{Astropy}~\citep{2022ApJ...935..167A}---and the one in \textsc{LAL} used for the
generation of the massive graviton waveforms. The differences between these
implementations are minor and not expected to have impacted the final results.
In particular, the cosmology is used in the generation of the injections only to determine the amount of
dephasing from the graviton mass and luminosity distance.

\begin{center}
  \begin{deluxetable}{cccc}\label{tab:source-parameter-priors}
    \tablehead{Parameter & Prior Distribution & Minimum & Maximum}
    \tablecaption{Base prior values used in the injection set}
    \startdata
    \multicolumn{4}{c}{Source Parameters} \\
    \cmidrule{1-4}$\mathcal{M}_{c}$ $[M_{\odot}]$ & Uniform in Component Masses & 5 & 300 \\
    $q$ & Uniform in Component Masses & 0.1 & 1.0 \\
    $m_{1}$ $[M_{\odot}]$ & Constraint & $10$ & $80$ \\
    $m_{2}$ $[M_{\odot}]$ & Constraint & $10$ & $80$ \\
    $d_{L}$ $[\textrm{Mpc}]$ & Uniform in Comoving Volume and Source Time & 100 & 2000 \\
    $a_{1}$ & Uniform & $0$ & $0.99$ \\
    $a_{2}$ & Uniform & $0$ & $0.99$ \\
    $\theta_{1}$ & Sine \\
    $\theta_{2}$ & Sine \\
    $\theta_{\textrm{JN}}$ & Sine \\
    $\phi_{12}$ & Uniform & $0$ & $2\pi$ \\
    $\phi_{\textrm{JL}}$ & Uniform & $0$ & $2\pi$ \\
    RA & Uniform & $0$ & $2\pi$ \\
    dec & Cosine \\
    $\psi$ & Uniform & $0$ & $\pi$ \\
    Phase & Uniform & 0 & $2\pi$ \\
    \cmidrule[\heavyrulewidth]{1-4}\multicolumn{4}{c}{Microlensing Parameters} \\
    \cmidrule{1-4}$M_{L}$ $[M_{\odot}]$ & Uniform & $1$ & $10^{4}$ \\
    $y$ & Power Law ($\alpha = 1$) & $0.1$ & $3.0$ \\
    $d_{\textrm{Lens}}/d_{L}$ & Delta ($0.5$) \\
    \cmidrule[\heavyrulewidth]{1-4}\multicolumn{4}{c}{Millilensing Parameters} \\
    \cmidrule{1-4}Number of Images & Uniform (Discrete) & $1$ & $4$ \\
    $n_{i}$ & Uniform (Discrete) & $0$ & $1$ \\
    $dt_{i}$ $[s]$ & Uniform & $0.001$ & $0.1$ \\
    $dL^{\textrm{eff}}_{i}$ $[M_{\odot}]$ & Uniform & $100$ & $5000$ \\
    \enddata 
  \end{deluxetable}
\end{center}

\section{Additional Results from the Microlensing Investigation}\label{app:microlensing}
\subsection{Point Mass Investigation}\label{subsec:pm-violins}

\begin{figure*}
	\centering
	\includegraphics{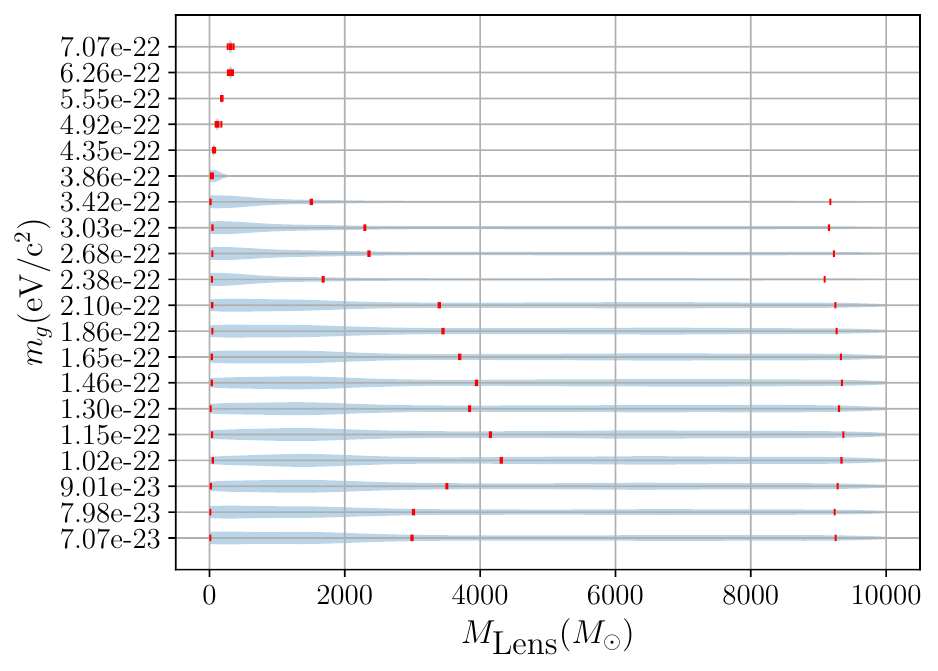}
	\caption{Recovered posteriors on the lens mass (taken to be halfway between the source and
  observer) from the massive graviton series of injections. The red ticks indicate the median and 90\%
  C.I. The middle red tick indicates the median
  of the posterior with the other two representing the boundaries of the 90\% C.I. The lower mass results are broad with
greater support for very low mass lenses, in line with the expectations for unlensed events. The
higher mass graviton injections, however, display support for a specific value of lens mass within
the prior region which is in line with the expectation for microlensed events.}\label{fig:massive-graviton-mass-point-violin}
\end{figure*}

\begin{figure*}
	\centering
	\includegraphics[width=0.49\linewidth]{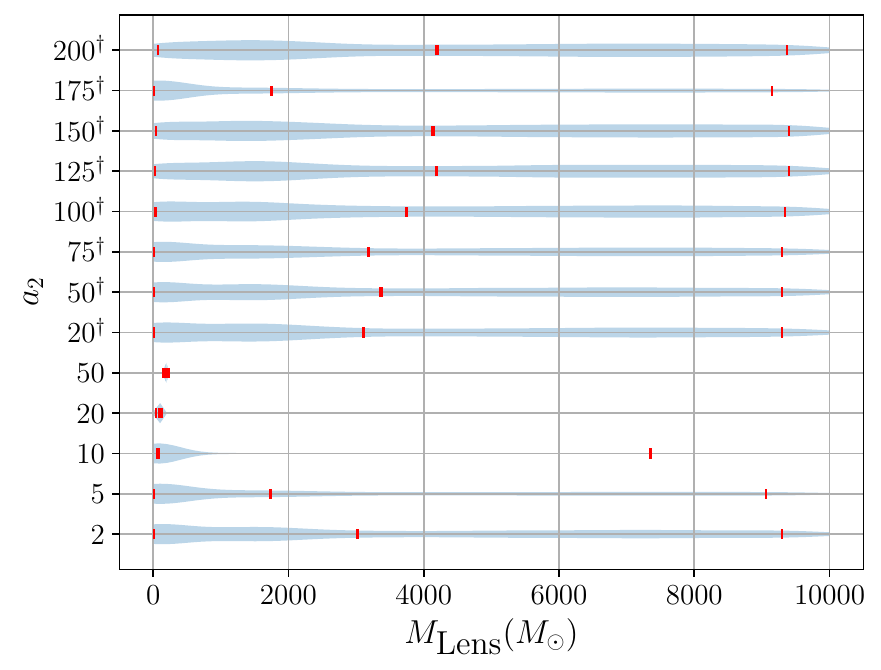}
	\includegraphics[width=0.49\linewidth]{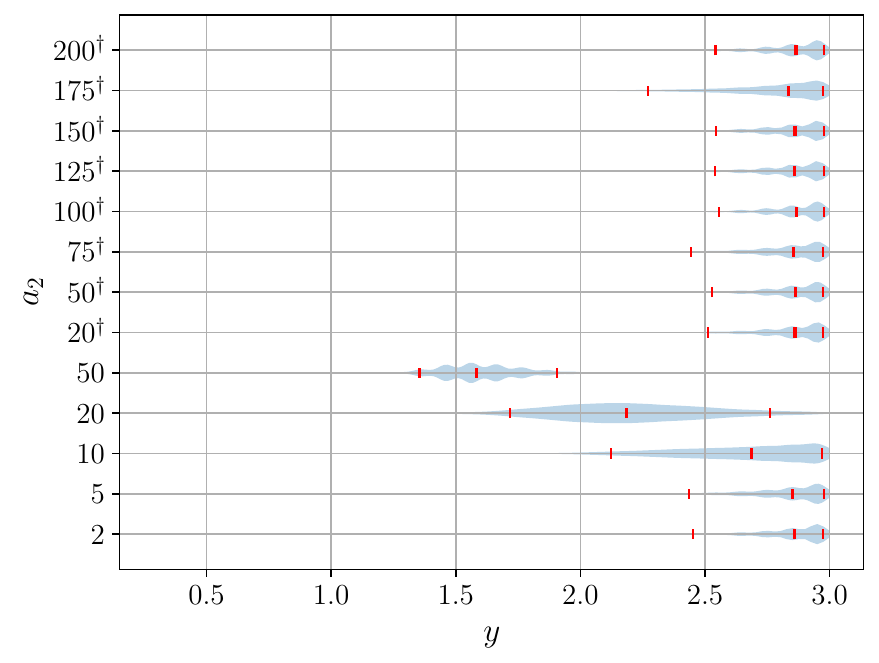}
	\caption{Recovered posteriors on the lens mass (left) and source position (right) from the isolated point mass analysis of the modified energy flux series of injections, where the dagger denotes the injections without waveform scaling. 
	The red ticks indicate the median and 90\%
C.I. As is the case with the massive graviton injections, only those injections that result in preference for the microlensing hypothesis display constraints towards specific values within the prior bounds.}\label{fig:modified-energy-flux-point-violin}
\end{figure*}

\begin{figure*}
	\centering
	\includegraphics[width=0.49\linewidth]{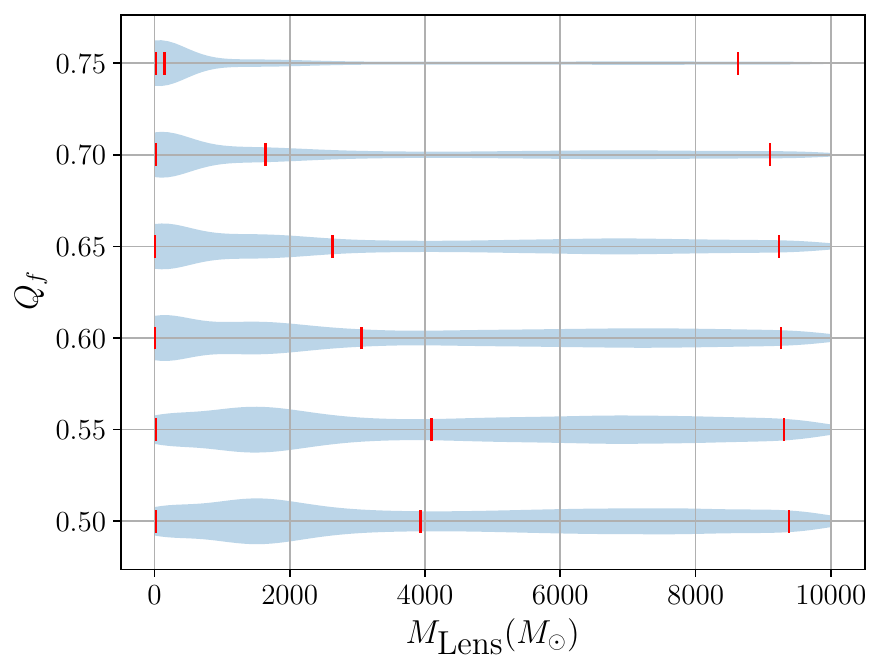}
	\includegraphics[width=0.49\linewidth]{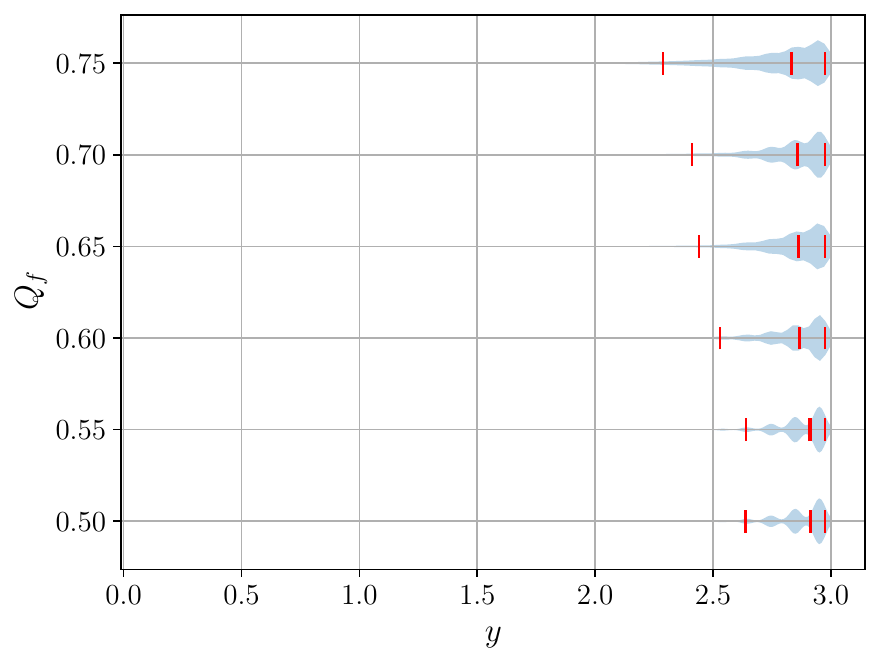}
	\caption{Recovered posteriors on the lens mass (left) and source position (right) from the isolated point mass analysis of the modified QNM spectra series of injections. 
	The red ticks indicate the median and 90\%
C.I. In agreement with the lack of any false positive preference for the microlensing hypothesis for these injections, these posteriors are in line with the expectations for unlensed events.}\label{fig:modified-qnm-point-violin}
\end{figure*}

\begin{figure*}
	\centering
	\includegraphics[width=0.49\linewidth]{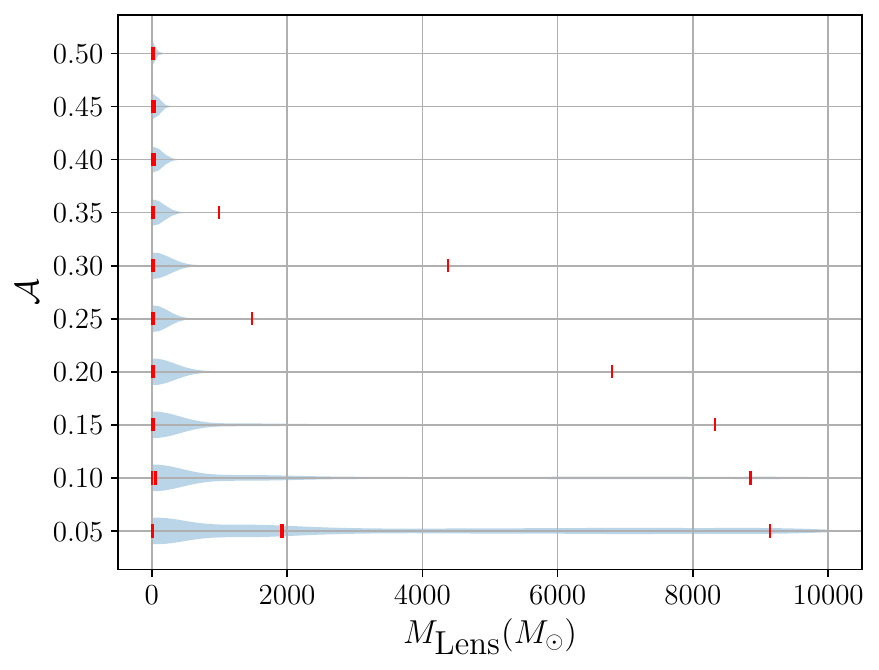}
	\includegraphics[width=0.49\linewidth]{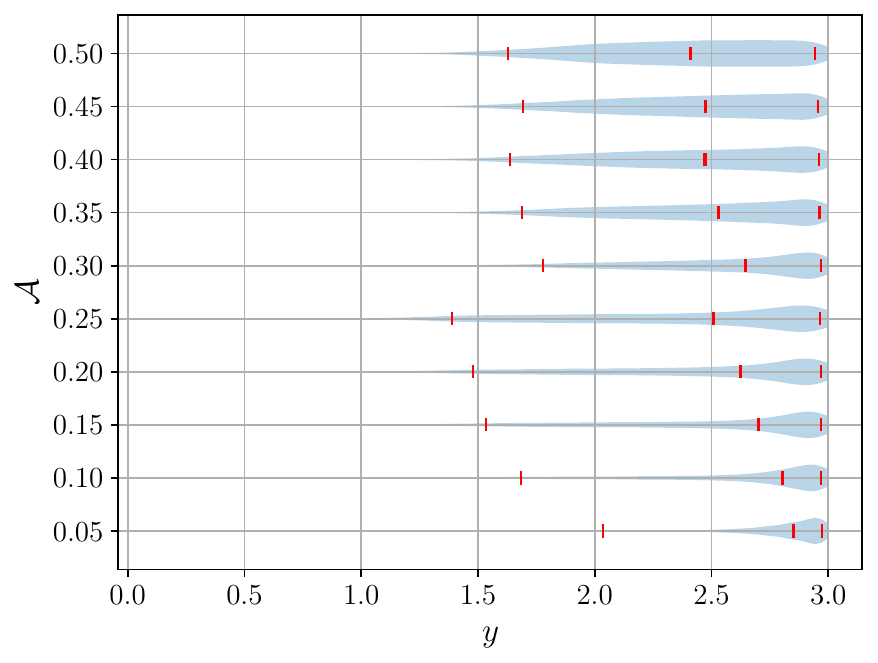}
	\caption{Recovered posteriors on the lens mass (left) and source position (right) from the isolated point mass analysis of the additional scalar polarisation series of injections. 
	The red ticks indicate the median and 90\%
C.I. As for the modified QNM injections, these injections have no false positive preference for the microlensing hypothesis, and the posteriors are within expectations for unlensed events.}\label{fig:scalar-amp-point-violin}
\end{figure*}

\begin{figure*}
	\centering
	\includegraphics[width=0.49\linewidth]{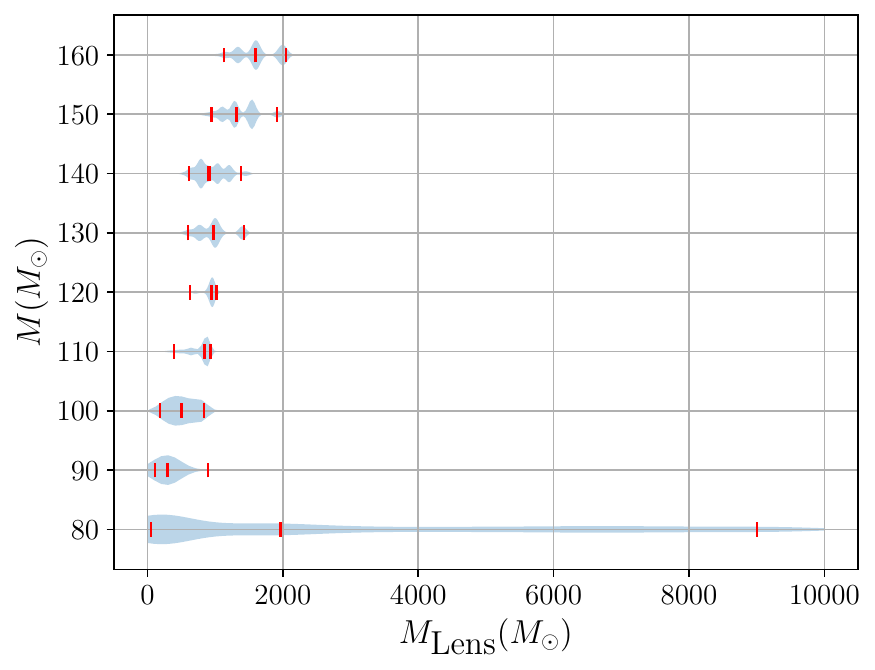}
	\includegraphics[width=0.49\linewidth]{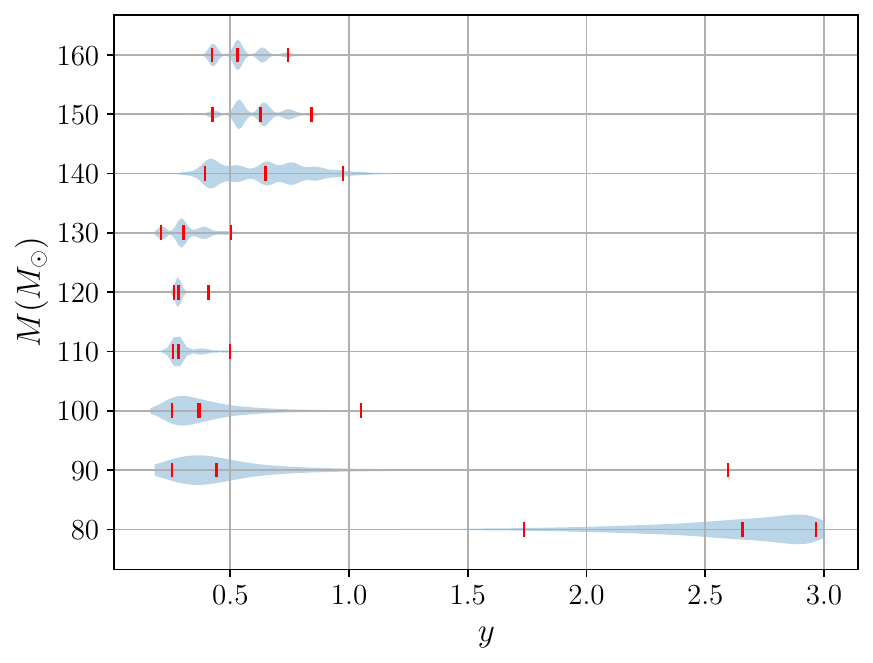}
	\caption{Recovered posteriors on the lens mass (left) and source position (right) from the
    isolated point mass analysis of the black hole mimicker waveforms based on the precessing BNS
    waveform from~\cite{dietric_scaled_bns}. 
	The red ticks indicate the median and 90\%
C.I.
All cases here demonstrate a preference for the
    microlensing hypothesis and this is reflected in the consistently constrained lens parameter
    posteriors. Of particular note is the apparent multimodality which is not expected for true
    microlensing events.}\label{fig:bhm-p-point-violin}
\end{figure*}

\begin{figure*}
	\centering
	\includegraphics[width=0.49\linewidth]{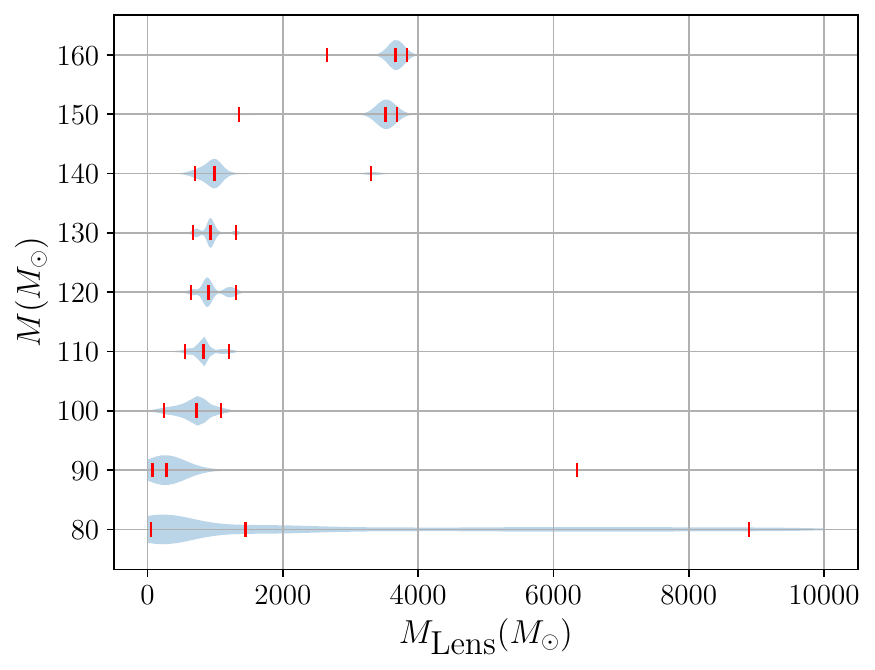}
	\includegraphics[width=0.49\linewidth]{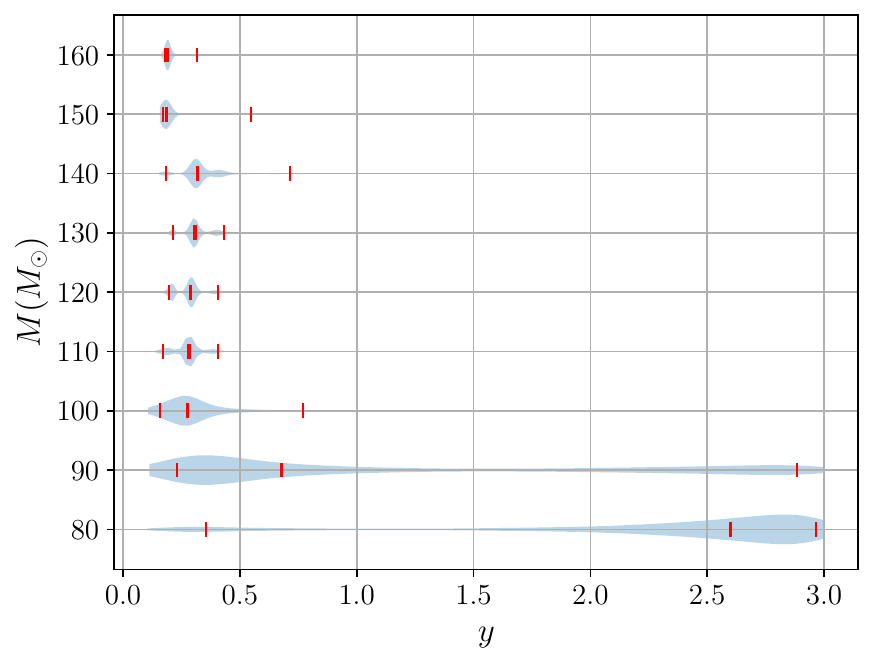}
	\caption{Recovered posteriors on the lens mass (left) and source position (right) from the isolated point mass analysis of the black hole mimicker waveforms based on the non-spinning BNS waveform from~\cite{ujevic_scaled_bns}. 
	The red ticks indicate the median and 90\%
C.I. Of note is that the $90M_{\odot}$ case displays some constraint within the prior boundary despite the microlensed case being insufficiently preferred by the Bayes factor that this event would be selected for further analysis. Some of the results display multimodality which is not expected in true microlensing analysis.}\label{fig:bhm-ns-point-violin}
\end{figure*}

In this appendix, we present the complete set of lens posteriors for the isolated point mass
microlensing analysis in
Figures~\ref{fig:massive-graviton-mass-point-violin}--\ref{fig:bhm-ns-point-violin}. These results
follow the pattern discussed in detail for the source position posteriors from the massive graviton
series of injections in Section~\ref{subsec:pm-microlensing}. For the source position posteriors,
only those injections that favor the microlensing hypothesis yield any constraint within the prior
boundaries with the remainder showing significant railing towards larger source positions. These
behaviors are consistent with the expectations for lensed and unlensed events, respectively. Of note
are the results for the black hole mimicker injections, both the precessing and non-spinning sets
have cases displaying multimodality in the posteriors. This is not expected for true microlensing
candidates and would indicate that further investigation would be needed on these events. However,
given the Bayes factor support this would be insufficient to rule these out as microlensing
candidates.

Similarly, the lens mass posteriors (assuming the lens is placed halfway between the source and
observer) reflect the expectations of unlensed events for those events that yielded no support for
the microlensing hypothesis---broad posteriors with some support for low mass values that rail
against the lower prior boundary. This combination results in the smallest amplification factor
oscillation, resulting in the closest match to an unlensed waveform. Those candidates that did favor
the microlensing hypothesis again display constraints towards specific values within the posterior.
The black hole mimicker waveforms display a preference for significantly higher mass lenses than
those for the massive graviton or modified energy flux cases. The lens mass posteriors in the black
hole mimicker cases also display some multimodality, with the same implications as the multimodality
in the source position posteriors discussed above. 

\subsection{SIS Investigation}\label{subsec:sis-violins}

\begin{figure*}
	\includegraphics{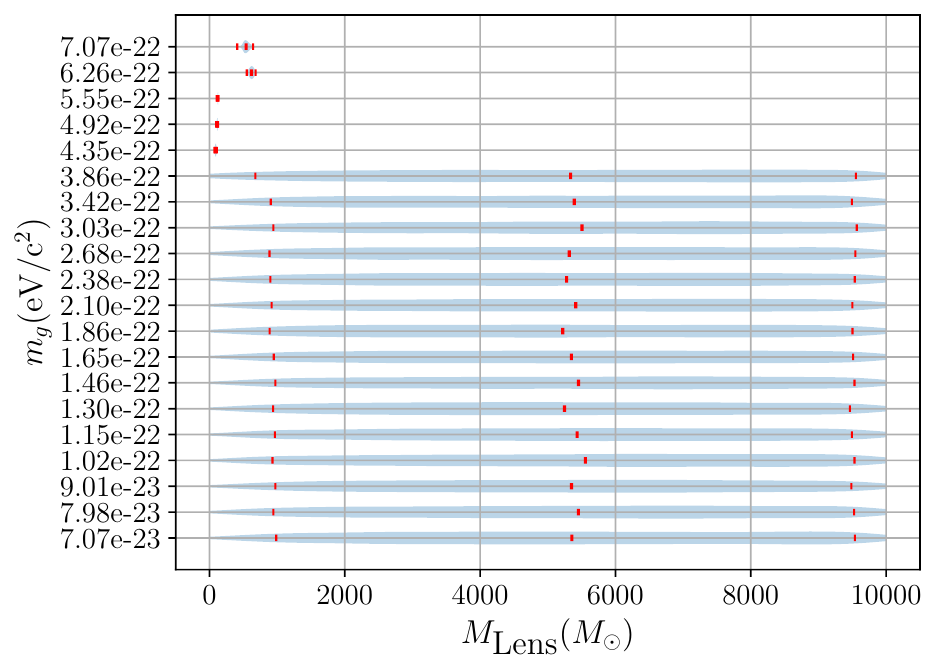}
	\caption{Recovered posteriors on the lens mass from the SIS analysis of the massive graviton series of injections. 
	The red ticks indicate the median and 90\%
C.I. The lower mass injections display broad uninformative posteriors which is in line with the expectations for non-lensed events. In combination with the source position posteriors, these would result in flat amplification factors that more closely  resemeble unlensed signals. The higher mass injections, as was the case with the point mass analysis, display constraints towards specific values which is indicative of their false-positive status.}\label{fig:massive-graviton-mass-sis-violin}
\end{figure*}

\begin{figure*}
	\includegraphics[width=0.49\linewidth]{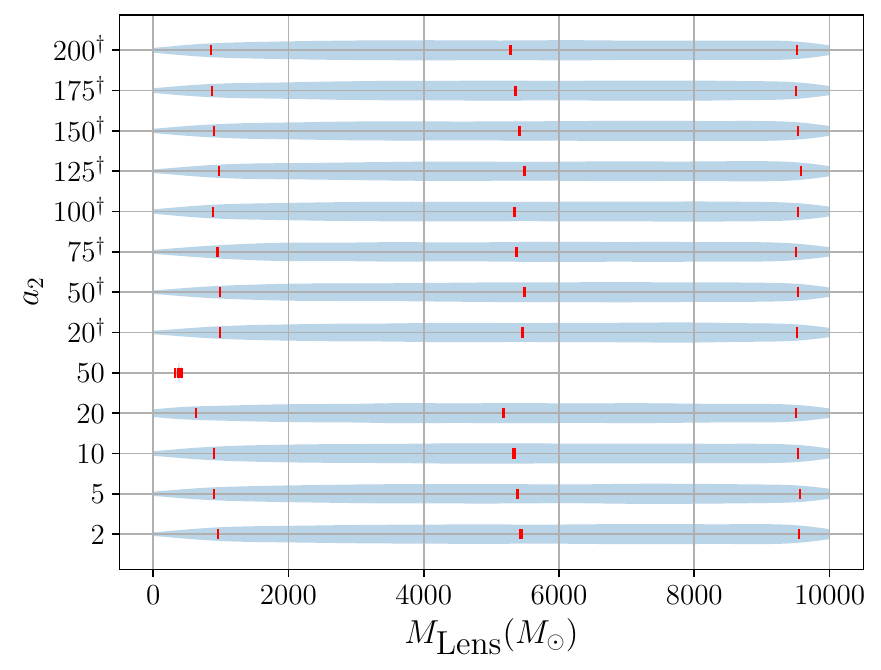}
	\includegraphics[width=0.49\linewidth]{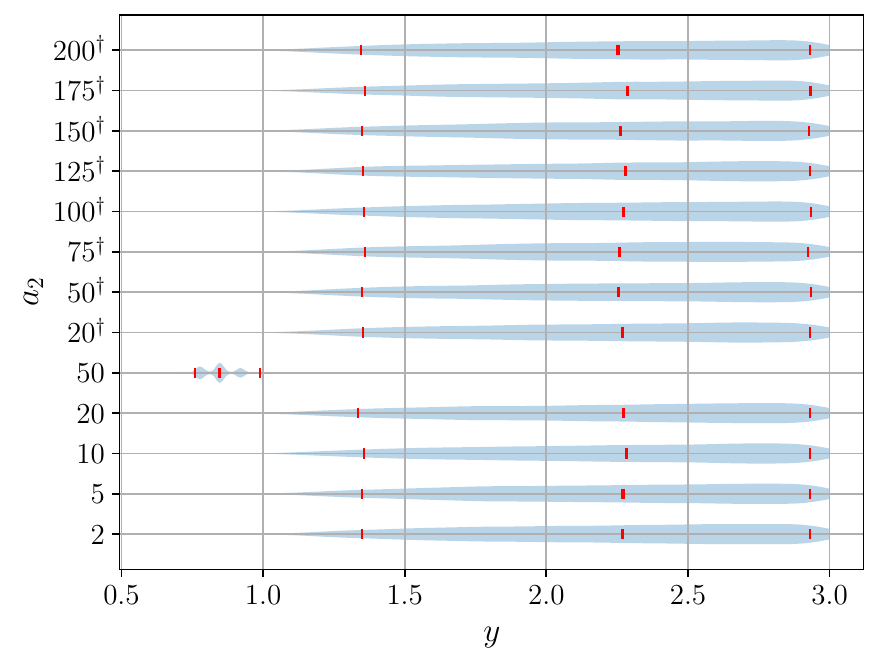}
	\caption{Recovered posteriors on the lens mass (left) and source position (right) from the SIS analysis of the modified energy flux series of injections, where the dagger denotes the injections without waveform scaling. The red ticks indicate the median and 90\%
	C.I. In all but the $a_{2} = 50$ case, these display the same broad posteriors with $y > 1$ that is expected of unlensed events---this includes the $a_{2} = 20$ case which has some support for the point mass model.}\label{fig:modified-energy-flux-sis-violin}
\end{figure*}

\begin{figure*}
	\includegraphics[width=0.49\linewidth]{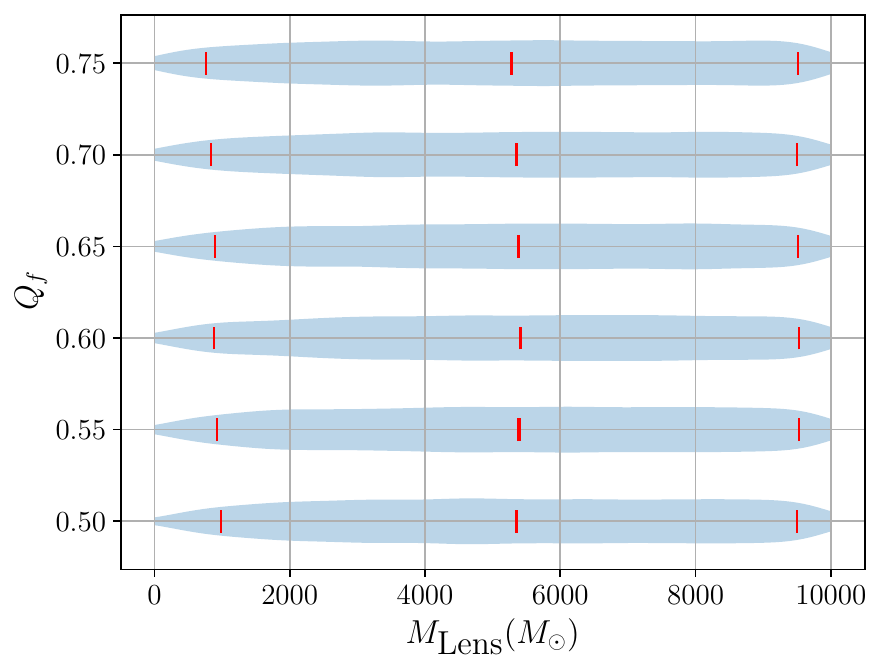}
	\includegraphics[width=0.49\linewidth]{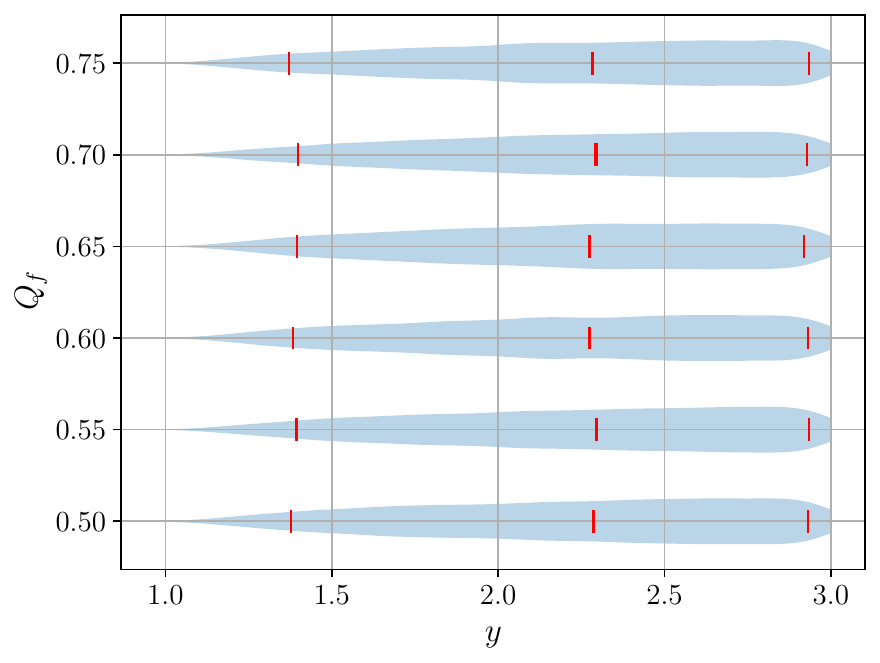}
	\caption{Recovered posteriors on the lens mass (left) and source position (right) from the SIS analysis of the modified QNM spectra series of injections. 
	The red ticks indicate the median and 90\%
C.I. In all cases, these display the same broad posteriors with $y > 1$ that is expected of unlensed events.}\label{fig:modified-qnm-sis-violin}
\end{figure*}

\begin{figure*}
	\includegraphics[width=0.49\linewidth]{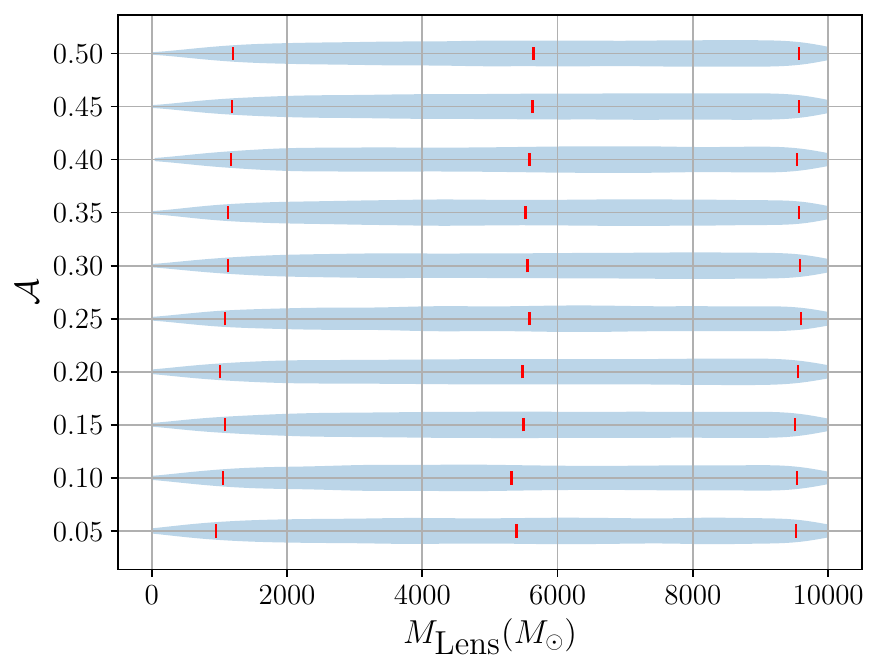}
	\includegraphics[width=0.49\linewidth]{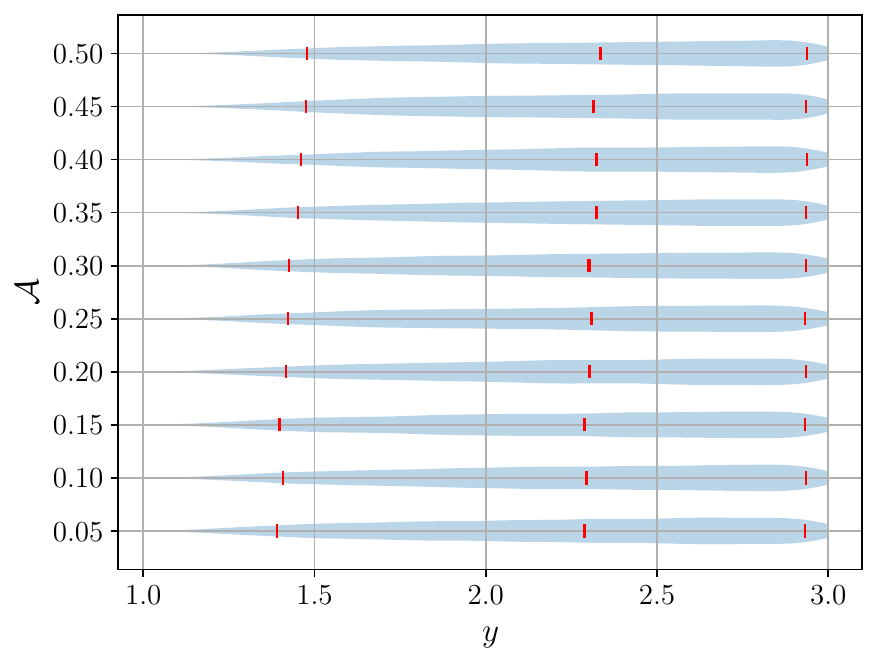}
	\caption{Recovered posteriors on the lens mass (left) and source position (right) from the SIS analysis of the additional scalar polarization series of injections. 
	The red ticks indicate the median and 90\%
C.I. In all cases, these display the same broad posteriors with $y > 1$ that is expected of unlensed events.}\label{fig:scalar-amp-sis-violin}
\end{figure*}

\begin{figure*}
	\includegraphics[width=0.49\linewidth]{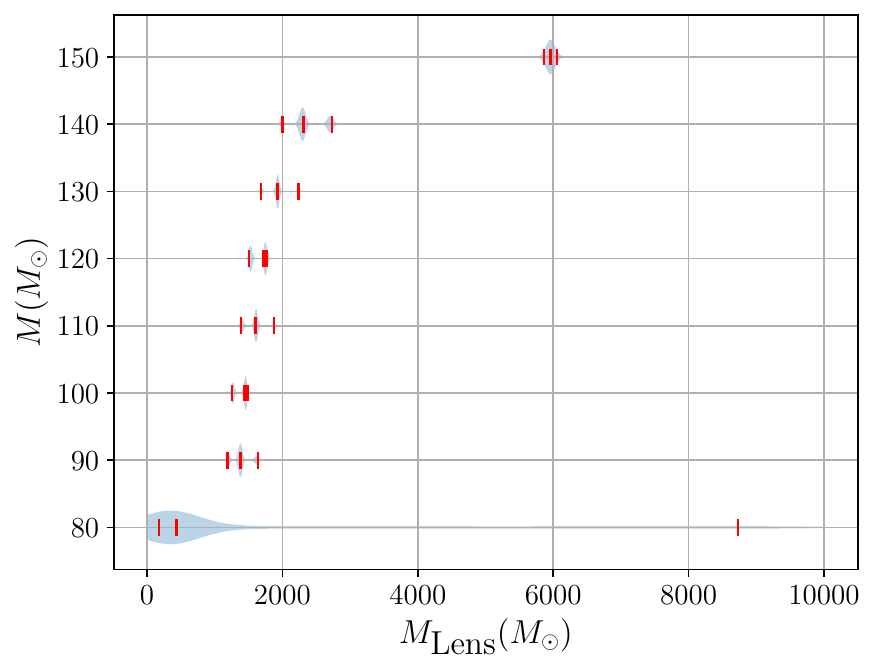}
	\includegraphics[width=0.49\linewidth]{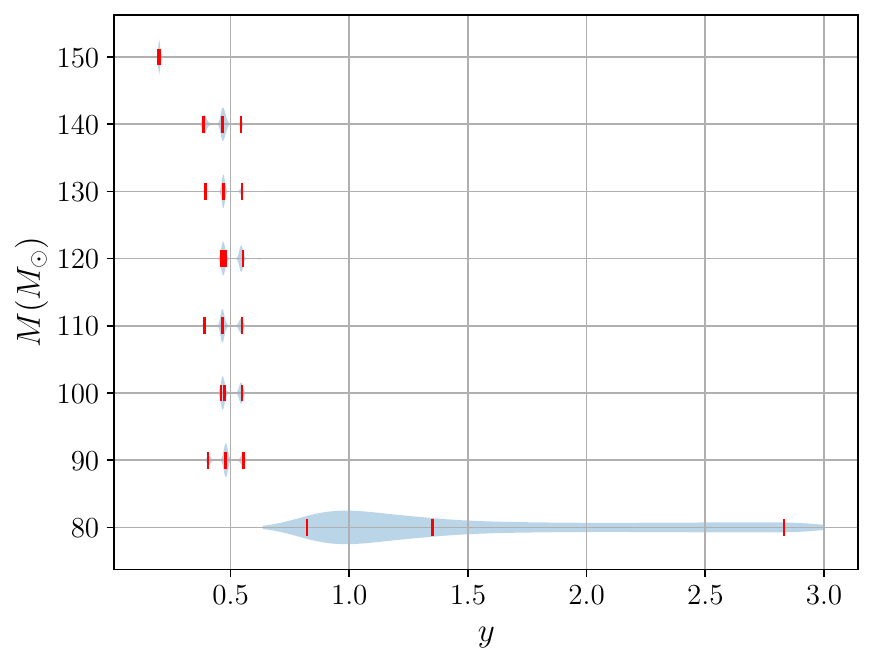}
	\caption{Recovered posteriors on the lens mass (left) and source position (right) from the SIS analysis of the precessing black hole mimicker series of injections. 
	The red ticks indicate the median and 90\%
C.I. In this instance, the $M = 80M_{\odot}$ displays posteriors that have some support for values inside of the region of parameter space that would yield an oscillatory amplification factor, alongisde support for the single image case. The other cases all yield constraint to specific values as is expected of candidate lensing events, including the $M = 140M_{\odot}$ case, despite the significant disfavouring in this case which may be due to finding source parameters between the local and global maximum found in the point mass and unlensed analyses.}\label{fig:bhm-p-sis-violin}
\end{figure*}

\begin{figure*}
	\includegraphics[width=0.49\linewidth]{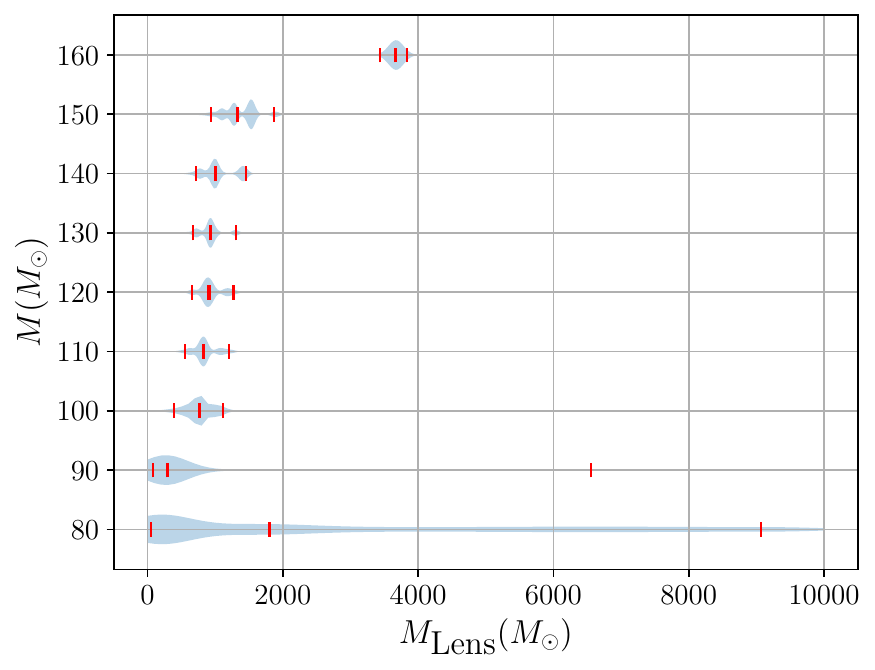}
	\includegraphics[width=0.49\linewidth]{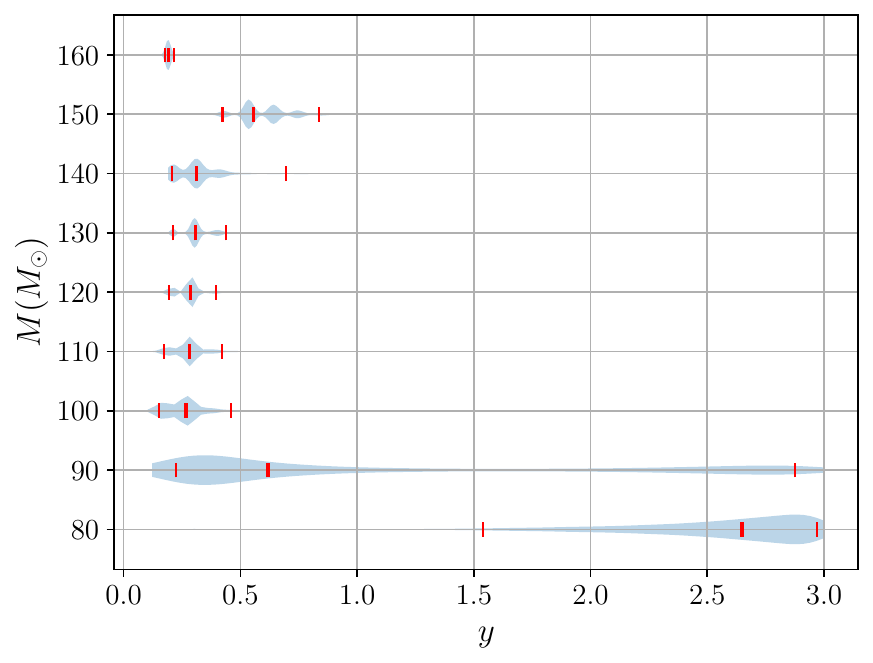}
	\caption{Recovered posteriors on the lens mass (left) and source position (right) from the SIS analysis of the non-spinning black hole mimicker series of injections. 
	The red ticks indicate the median and 90\%
C.I. In this instance, only the $M = 80M_{\odot}$ and $M = 90M_{\odot}$ cases match the expectations for unlensed events, with a more distinct constraint in the remaining cases. Similarly to the point mass results for these, there is however, some unexpected multimodality in a few of the recovered posteriors.}\label{fig:bhm-ns-sis-violin}
\end{figure*}

Similarly to~\ref{subsec:pm-violins}, in this appendix we present the complete set of recovered lens
posteriors for the SIS microlensing analysis in
Figures~\ref{fig:massive-graviton-mass-sis-violin}--\ref{fig:bhm-ns-sis-violin}. The results largely
follow the pattern discussed in Sections~\ref{subsec:pm-microlensing}
and~\ref{subsec:sis-microlensing} in that those candidates that were considered false positives for microlensing largely follow the expectations for microlensed candidates and vice versa.

In particular, as is the case for unlensed events, the source position posteriors are largely uninformative beyond supporting only the $y > 1$ region in which in geometric optics only a single image will be produced and the resultant amplification factor will be flat and thus more closely resemble an unlensed waveform. This leads to an expectation of broader lens mass posteriors compared with the point mass case, and this is observed in the resultant posteriors. One result worth noting is the modified energy flux, $a_{2} = 20$ case. This has some support, though insufficient for further investigation, for the point mass hypothesis. Both the Bayes factor and the posteriors in the SIS case, however, do cast more significant doubt on this as the posteriors match much more closely with the unlensed expectation for this model.

Similarly to the point mass analysis, both the precessing and non-spinning black hole mimicker
injections have cases displaying multimodality in the posteriors. This would not be the expected
behavior of true microlensing candidates. However, it would be insufficient to rule these candidates
out given the Bayes factor support. These candidates would therefore require additional
investigation to give a final determination of their status.

A result of particular note is the $140M_{\odot}$ precessing black hole mimicker waveform, which displays recovered posteriors similar to the candidate lensing events despite the significant disfavoring for this model compared to both the point mass and unlensed models. This may indicate that the source parameters which were found between the local and global likelihood maxima in these analyses are themselves less supported than those found in the maximum for the point and unlensed analyses. 

\section{Additional Results from the Millilensing Investigation}\label{app:millilensing}
\begin{figure*}
	\includegraphics{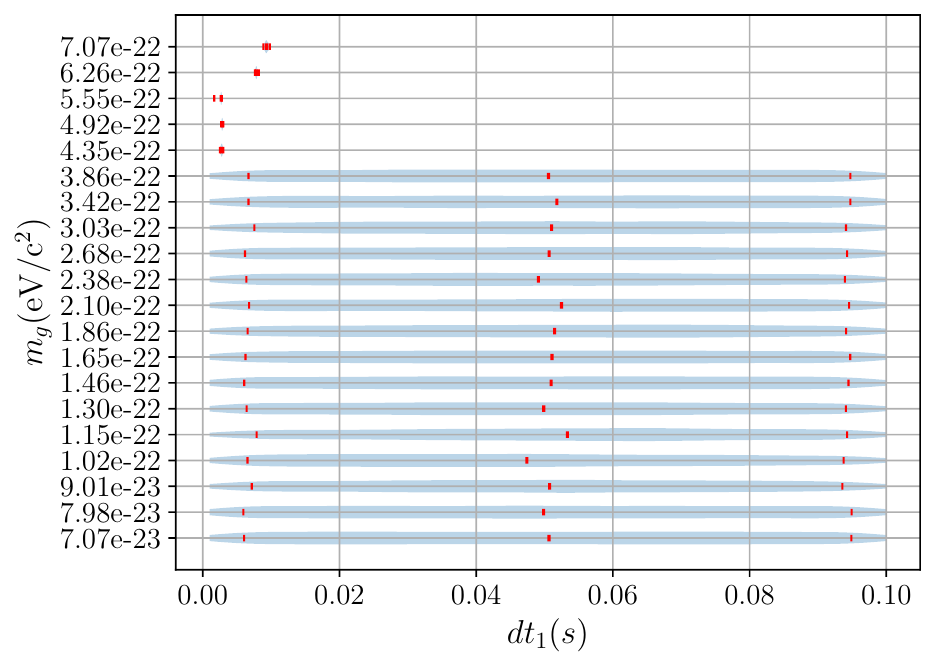}
	\caption{Recovered posteriors on the time delay between the first and second image for the massive graviton series of injections. 
	The red ticks indicate the median and 90\%
C.I. The lower mass injections reflect the expectation for unlensed events, with broad, uniformative posteriors indicative of a lack of support for the second image. The higher mass injections display constraints towards a particular value which is in line with the expectations for a true millilensing event.}\label{fig:massive-graviton-dt-violin}
\end{figure*}

\begin{figure*}
	\includegraphics[width=0.49\linewidth]{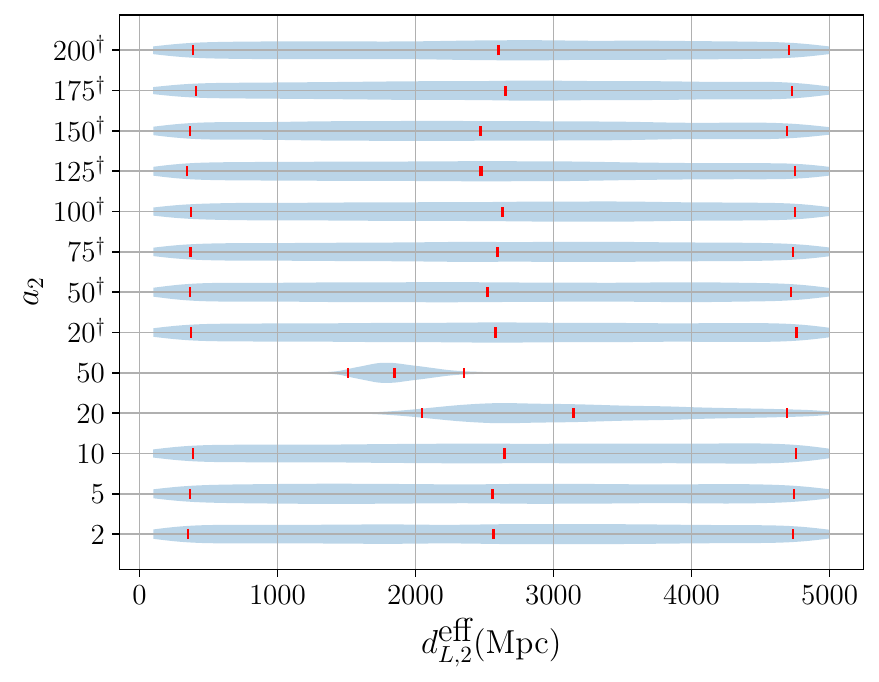}
	\includegraphics[width=0.49\linewidth]{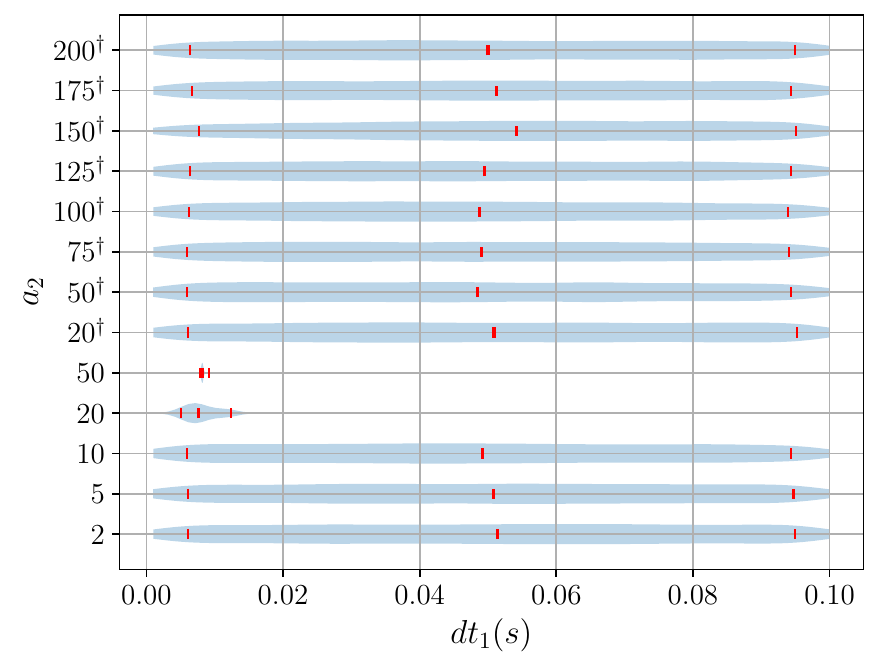}
	\caption{Recovered posteriors on the effective luminosity distance (left) and time delay (right)
  of the second proposed image in the millilensing analysis of the modified energy flux series of
injections, where the dagger denotes the injections without waveform scaling. 
The red ticks indicate the median and 90\%
C.I. In these, only the
$a_{2} = 20$ and $a_{2} = 50$ cases display constraints indicative of support for a second image,
reflected in their support for the millilensing model.}\label{fig:modified-energy-flux-millilensing-violin}
\end{figure*}

\begin{figure*}
	\includegraphics[width=0.49\linewidth]{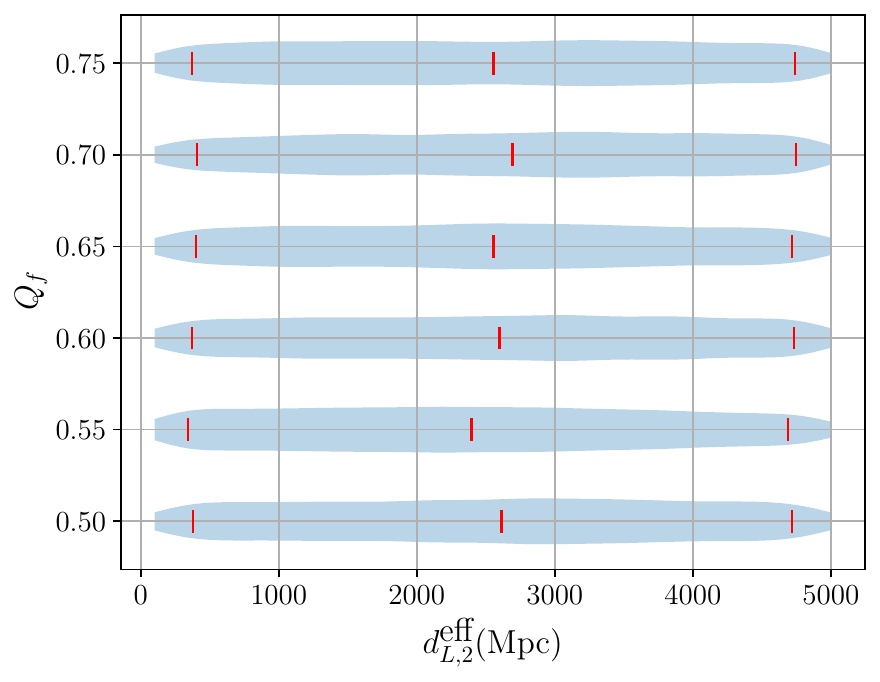}
	\includegraphics[width=0.49\linewidth]{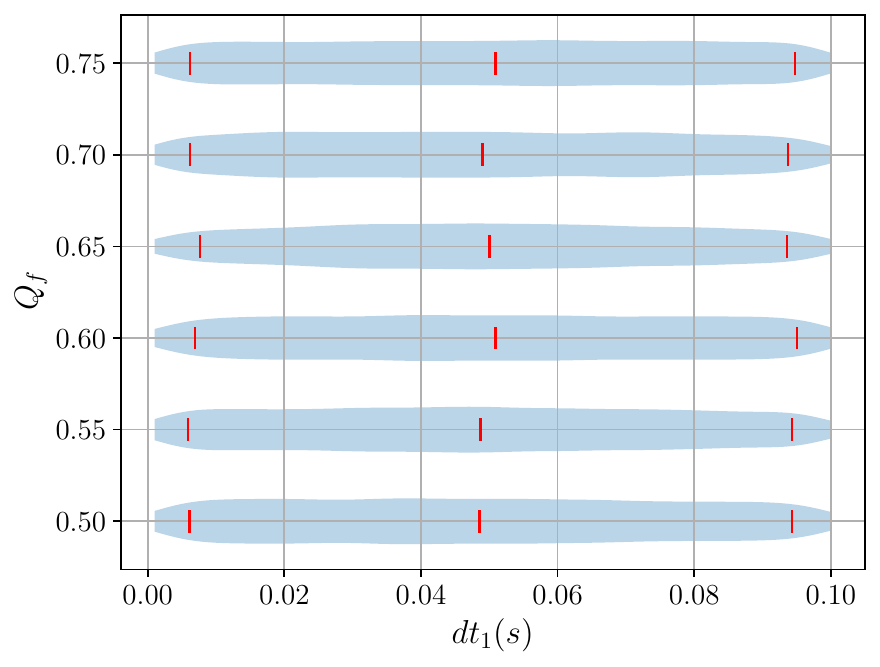}
	\caption{Recovered posteriors on the effective luminosity distance (left) and time delay (right) of the second proposed image in the millilensing analysis of the modified QNM spectra series of injections. 
	The red ticks indicate the median and 90\%
C.I. In these, all cases display the broad, uninformative posteriors indicating no support for a second image.}\label{fig:modified-qnm-millilensing-violin}
\end{figure*}

\begin{figure*}
	\includegraphics[width=0.49\linewidth]{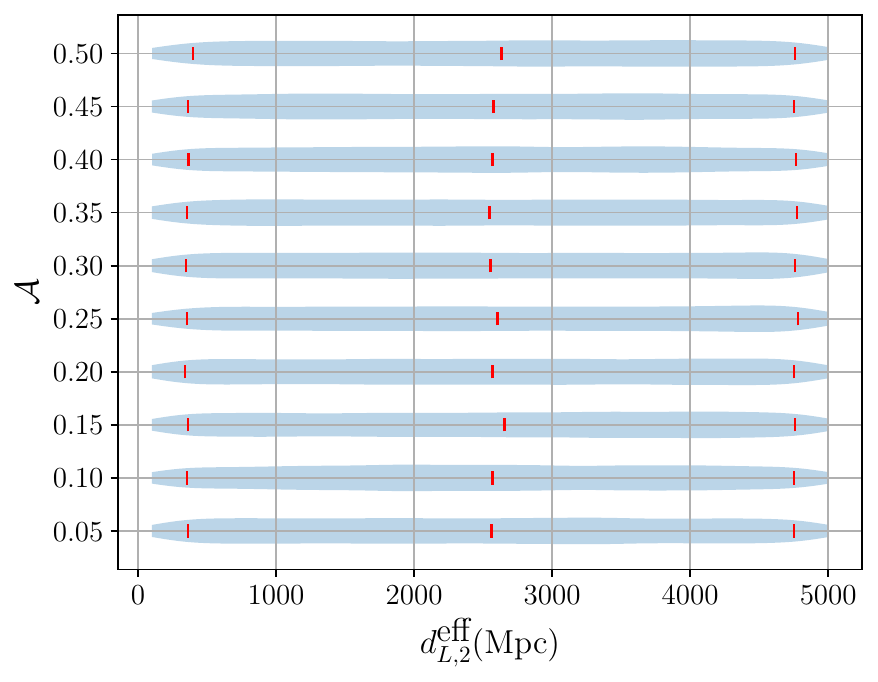}
	\includegraphics[width=0.49\linewidth]{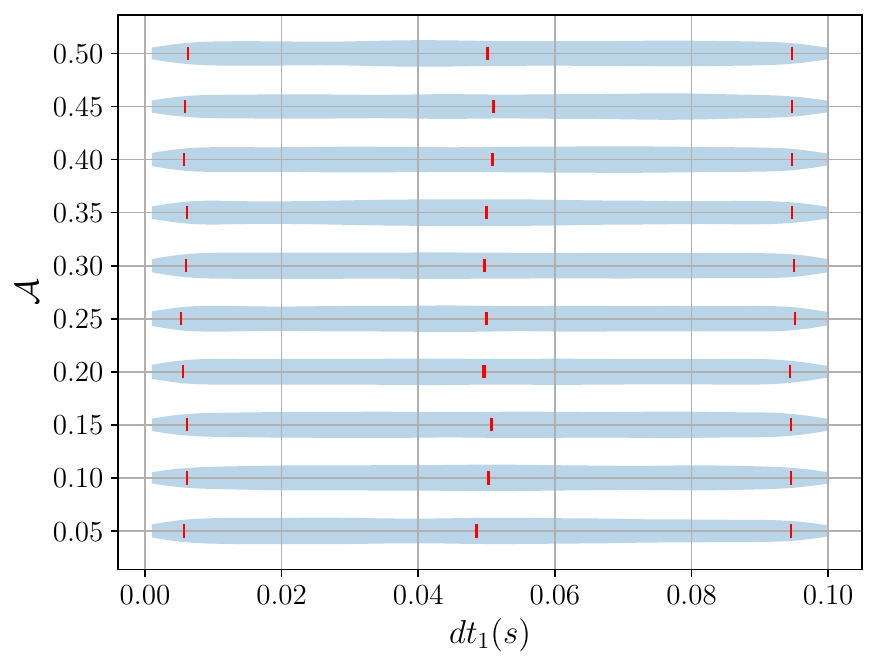}
	\caption{Recovered posteriors on the effective luminosity distance (left) and time delay (right) of the second proposed image in the millilensing analysis of the additional scalar polarization series of injections. 
	The red ticks indicate the median and 90\%
C.I. In all cases, the recovered posteriors are indicative of a lack of support for a second image including in the $\mathcal{A} = 0.20, 0.25$ cases that display Bayes factor support for the millilensing model, which would indicate that support for the model is coming from the Morse factor of the first image, in-line with this resulting in a false positive for the Type II image search as well.}\label{fig:scalar-amp-millilensing-violin}
\end{figure*}

\begin{figure*}
	\includegraphics[width=0.49\linewidth]{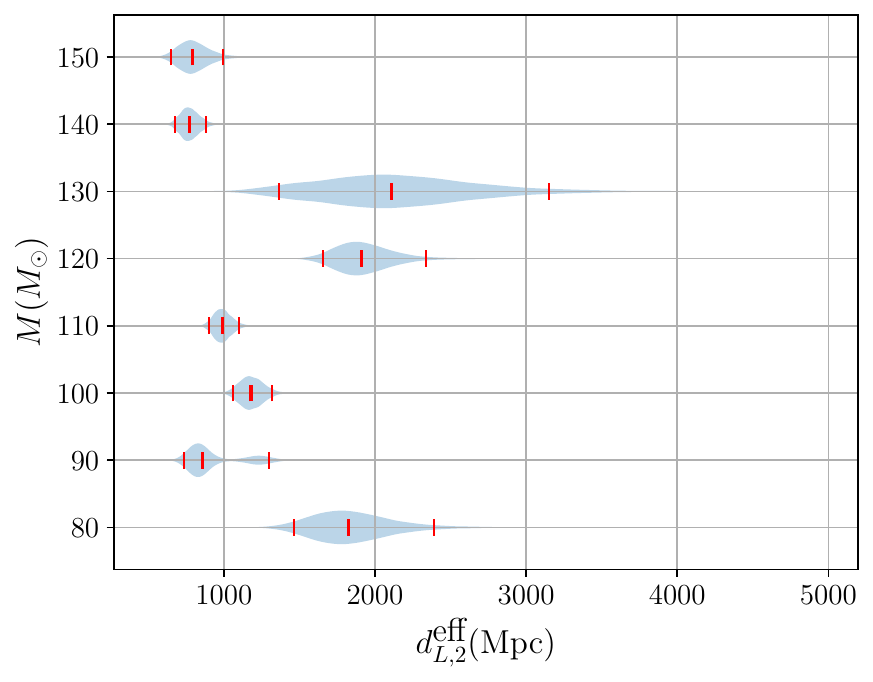}
	\includegraphics[width=0.49\linewidth]{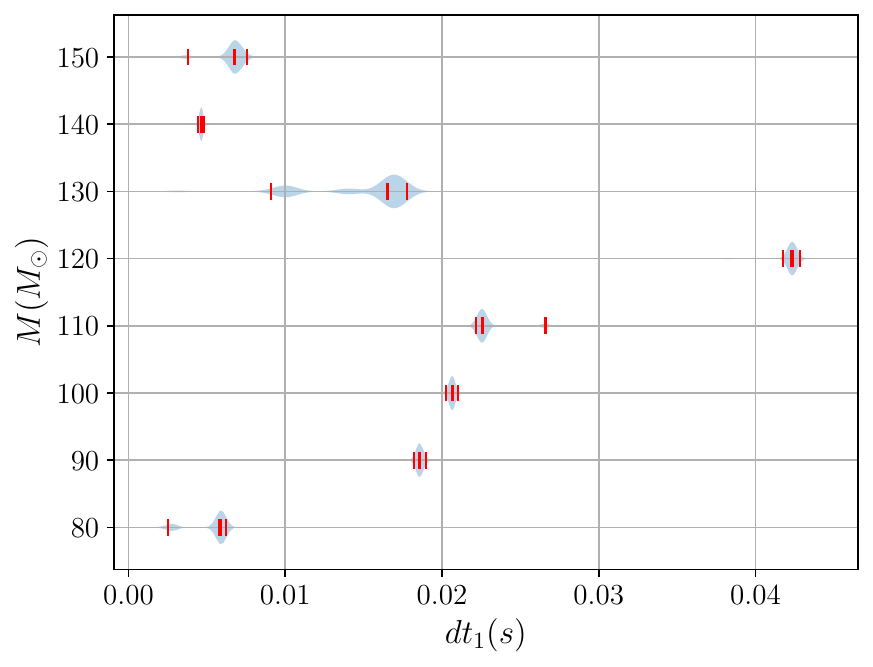}
	\caption{Recovered posteriors on the effective luminosity distance (left) and time delay (right) of the second proposed image in the millilensing analysis of the precessing black hole mimicker series of injections. 
	The red ticks indicate the median and 90\%
C.I. All cases display constraints towards specific values indicating support for a second image in this case, including the $140M_{\odot}$ case which has a Bayes factor preference for the unlensed case.}\label{fig:bhm-p-millilensing-violin}
\end{figure*}

\begin{figure*}
	\includegraphics[width=0.49\linewidth]{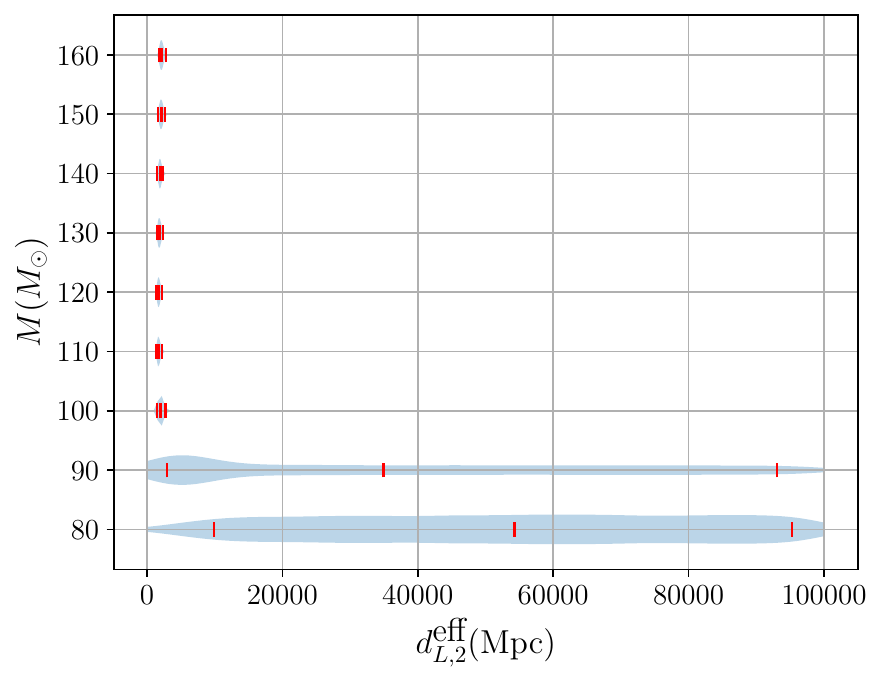}
	\includegraphics[width=0.49\linewidth]{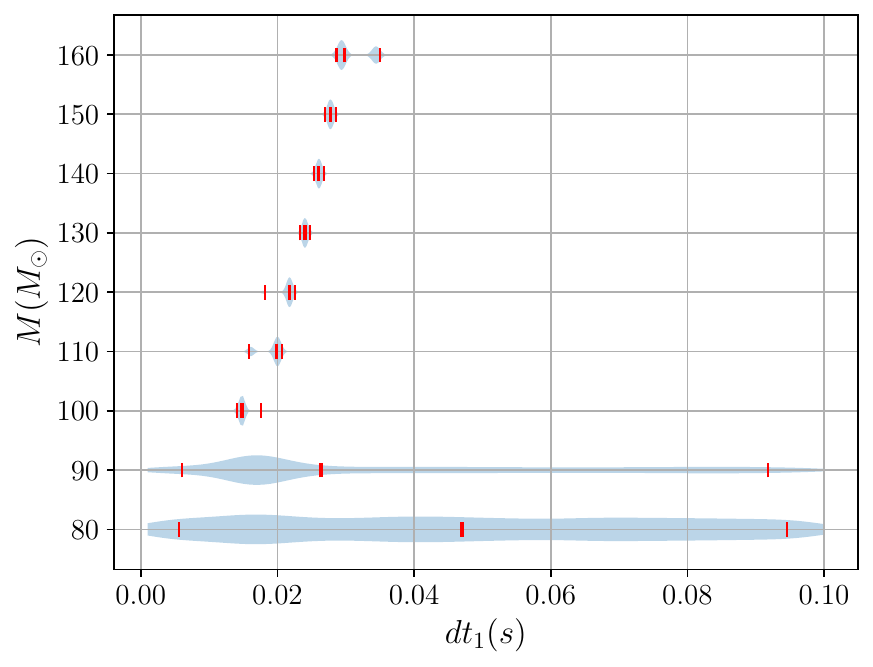}
	\caption{Recovered posteriors on the effective luminosity distance (left) and time delay (right) of the second proposed image in the millilensing analysis of the non-spinning black hole mimicker series of injections. 
	The red ticks indicate the median and 90\%
C.I. In this instance, the $80M_{\odot}$ case displays broad posteriors that do not support a second image. The $90M_{\odot}$ case displays broad posteriors but with a notable clustering towards a particular value, in line with the slender support for two images seen in the Bayes factor. The remainder of the cases which all display strong support for the millilensing model with additional images have tight constraints on the posteriors.}\label{fig:bhm-ns-millilensing-violin}
\end{figure*}

In this appendix, we present the set of recovered posteriors on the effective luminosity distance and time delay of the second proposed image in the millilensing analysis for the injections. These are shown in Figures~\ref{fig:massive-graviton-dt-violin}--\ref{fig:bhm-ns-millilensing-violin}. The expectations for unlensed events would be to see broad, uninformative posteriors spanning the prior space of both parameters, with the expectations for lensing candidates is to obtain constraints on these quantities for as many images as are supported. These expectations are repeated for the third and fourth proposed images. 

In all cases, this is borne out in the recovered posteriors, with the exception of the $140 M_{\odot}$ precessing black hole mimicker case, which similarly to the SIS analysis of this case, displays posteriors in line with the expectation for lensed candidates. This pattern is repeated in the third and fourth proposed image posteriors which are not shown here.

\section{Additional Results from the Type II Strong Lensing Investigation}\label{app:SL_extra_res}
In this appendix, we present the results for all the analyses done for the continuous type II 
lensing searches. These results are complementary with the ones presented in 
Sec.~\ref{subsec:typeII_strong_lensing} and,
more precisely, the numbers shown in Table~\ref{tab:continuous_SL_analyses}.

\begin{figure}
    \centering
    \includegraphics[keepaspectratio, width=0.7\textwidth]{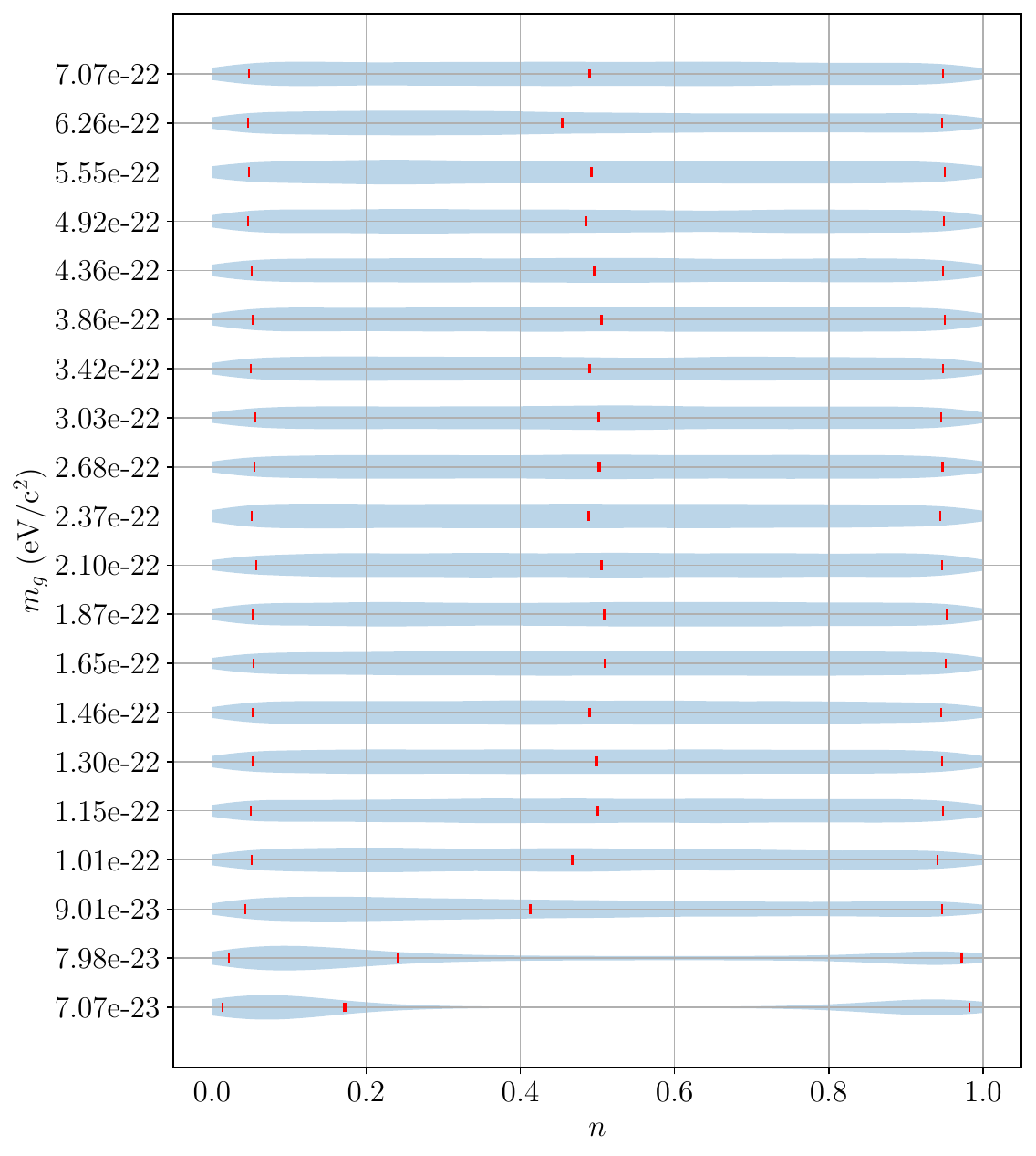}
    \caption{Representation of the recovered posterior for the Morse factor when doing the type II
    strong lensing search using a continuous prior for the massive graviton case. 
    The red ticks indicate the median and 90\%
C.I. Most of the posteriors
    are agnostic to the image type, and the ones for the two lowest masses even disfavor type II images.}
    \label{fig:SL_continuous_massive_graviton}
\end{figure}

First, the results for the massive graviton are given in Figure~\ref{fig:SL_continuous_massive_graviton}.
One sees that the posteriors are flat for most graviton masses considered. For the two lowest masses, 
the posteriors even disfavour the type II value, which is in line with an unlensed event containing
HOMs. Therefore, none of the posteriors would lead to extra inquiries when a massive graviton is present.

\begin{figure}
    \centering
    \includegraphics[keepaspectratio, width = 0.5\textwidth]{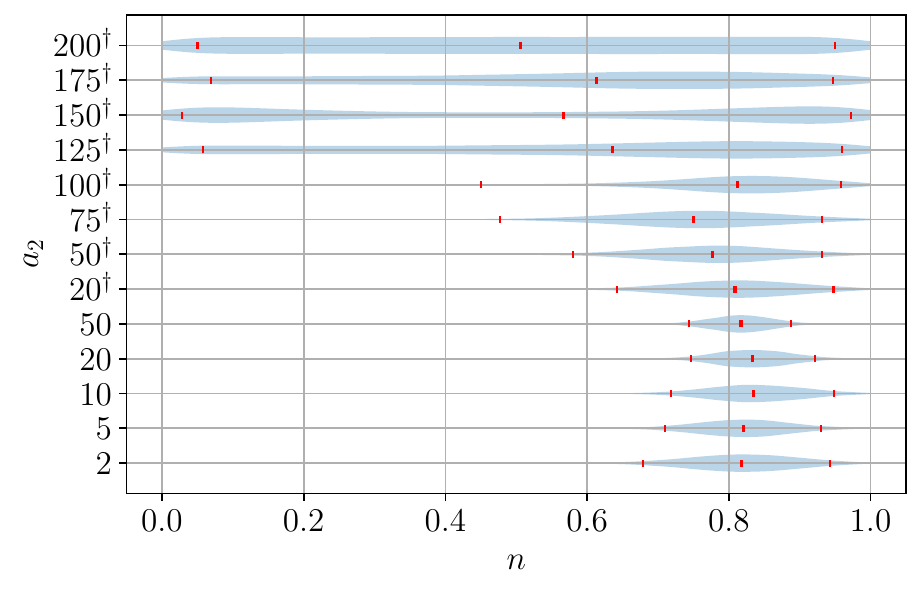}
    \caption{Recovered posteriors for the Morse factor when using a continuous prior for the case of a
    modified energy flux. The values with a dagger mean that the injections were done without waveform scaling.
    The red ticks indicate the median and 90\%
C.I.
    The four highest values are mostly uninformative, while the others all peak
    around $0.8$, with a low to no support for the physical values one obtains from lensing. Therefore,
    such deviations would not be detected when considering Bayes factors between different strong lensing
    hypotheses, but could be detected when searching for type II images with a continuous prior.}
    \label{fig:SL_continuous_modified_energy_flux}
\end{figure}

Moving to the modified energy flux, looking at the results shown in 
Figure~\ref{fig:SL_continuous_modified_energy_flux}, one sees that for the four highest amplitudes, 
the Morse factor posteriors are mostly flat. Then, when reducing the value for $a_2$, we obtain
more constrained posteriors, with values generally peaking around $n = 0.8$ and no support for
the lensing physical values. Because the main value is generally above $0.75$, these deviations in 
the Morse factor are not flagged in the discrete analyses. However, should they be found in 
the continuous case, the results would most likely trigger follow-up analyses.

\begin{figure}
    \centering
    \includegraphics[keepaspectratio, width=0.5\textwidth]{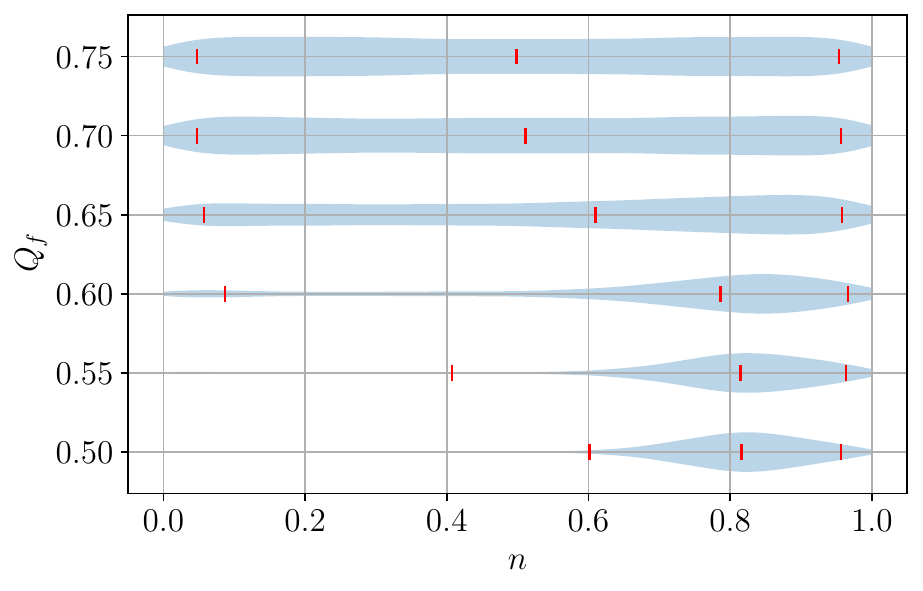}
    \caption{Morse factor posteriors for the lensing type II image searches with a continuous prior for the modified QNM spectra injections.
    The red ticks indicate the median and 90\%
C.I. 
    For the highest parameter values, the posteriors are non-informative. However, for the lower
    values, one sees the posteriors become more constrained and peak away from the GR values. Still,
    they retain (weak) compatibility with values allowed by GR ($n = 0$ and/or $1$).}
    \label{fig:SL_continuous_qnms}
\end{figure}

In the case of modified QNMs, for which the results are in Figure~\ref{fig:SL_continuous_qnms}, the picture is a bit similar to the one for the modified energy flux. The highest 
$Q_f$ values lead to a mostly uninformative posterior. The lower values lead to more constrained
posteriors with a peak around $0.8$. Here, some support is left for the lensing GR values, with
$n = 1$ still being possible for the most extreme case.

\begin{figure}
    \centering
    \includegraphics[keepaspectratio, width = 0.5\textwidth]{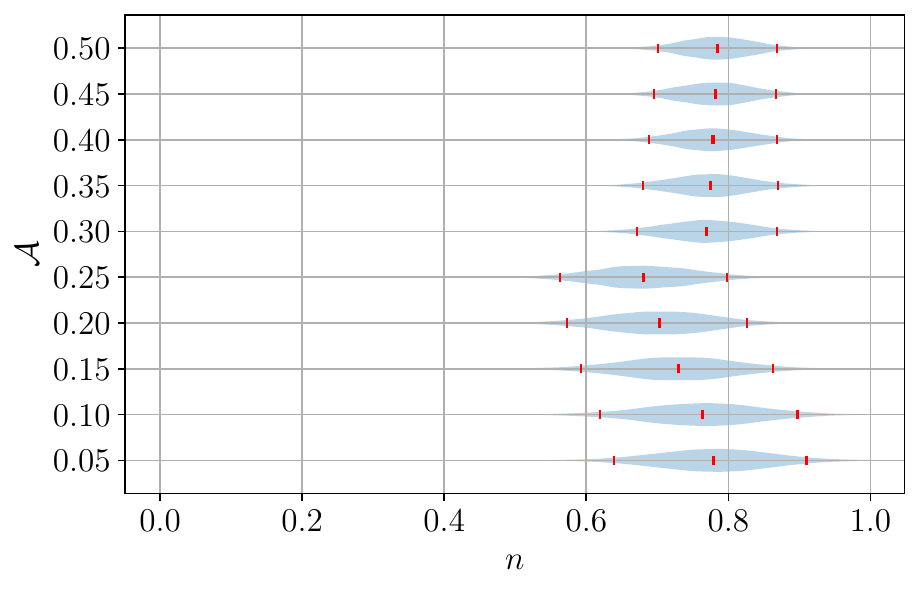}
    \caption{Morse factor posteriors for the lensing type II image searches with a continuous prior for the additional scalar polarization injections.
    The red ticks indicate the median and 90\%
C.I.
    All values of $\mathcal{A}$ lead to a deviation from the values expected for lensing in GR.
    In some cases, the posteriors have more support close to $0.5$ than to $0$ or $1$, meaning that these
    cases would be identified as possibly lensed by the discrete type II image search.}
    \label{fig:SL_continuous_scalar}
\end{figure}

For the scalar polarizations, as can be seen from Figure~\ref{fig:SL_continuous_scalar}, all the values
of the deviation parameter lead to the well-constrained posteriors peaking away from the GR values
for lensing. The exact width and median values of the posteriors change from one $\mathcal{A}$
value to the next, but, essentially, extra inquiries would be triggered in all cases when the analysis
is performed with a continuous prior. We also see that in some cases, most of the posterior weight is
well above $0.75$, which explains why the analyses done with a discrete prior are unable to find anything
particular. For those posteriors that have significant presence below $0.75$, the discrete search for
type II images had found possible lensing signatures---see Table~\ref{tab:discrete_SL_analyses}.

\begin{figure*}
    \centering
    \includegraphics[keepaspectratio, width = 0.49\textwidth]{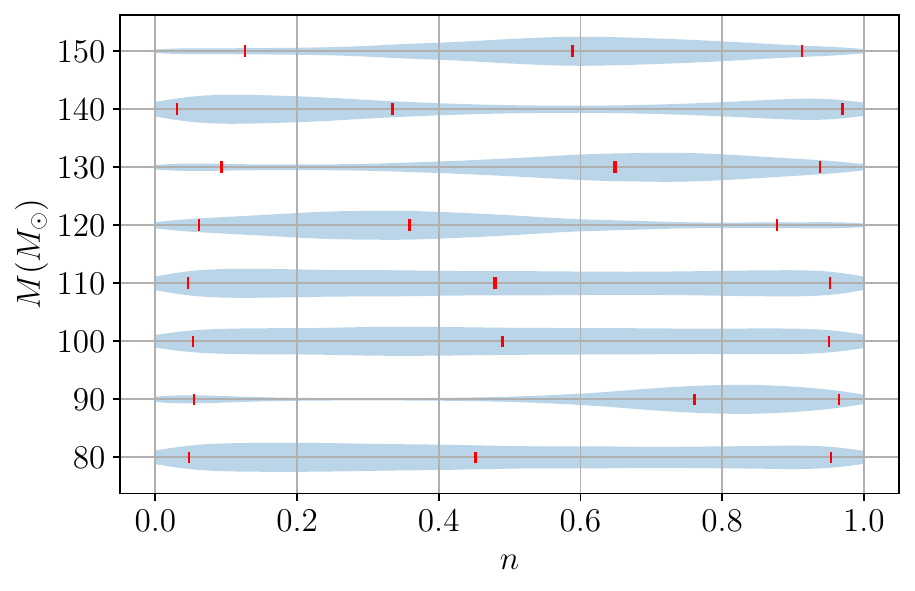}
    \includegraphics[keepaspectratio, width= 0.49\textwidth]{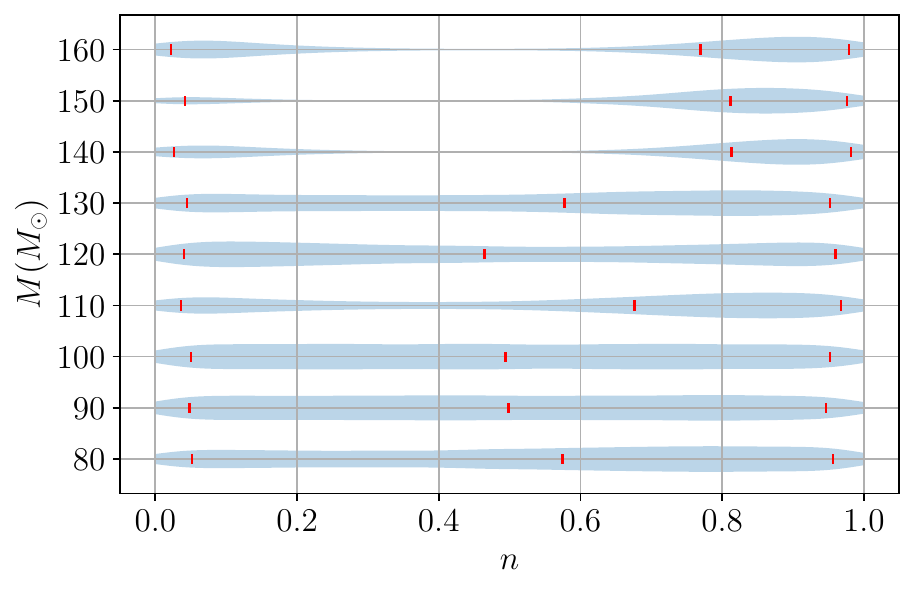}
    \caption{Recovered posteriors for the Morse factor when using a continuous prior and analyzing a binary of black
    hole mimickers. Left represents the precessing case based on the simulation from
    from~\citet{dietric_scaled_bns}. Right represents the nonspinning case based on the simulation from~\citet{ujevic_scaled_bns}.
    The red ticks indicate the median and 90\%
C.I.
    For the precessing binary, in some cases, the posteriors show a
    larger support for Morse factor values close to $0.5$, which would be consistent with a type II 
    image. For the nonspinning binary, the posteriors never show support for type II images or any values
    deviating from GR. Hence, they would not lead to any follow-up inquiries.}
    \label{fig:SL_continuous_mimickers}
\end{figure*}

For the black hole mimickers, the situation is a bit different depending on whether we consider the
precessing case following~\citet{dietric_scaled_bns} or the nonspinning case following~\citet{ujevic_scaled_bns}. For the latter, only the three
highest masses lead to informative posteriors, but they simply discard the type II hypothesis---in line
with the expectations for non-type II GR events with HOMs. The corresponding posteriors
are shown in the right panel of Figure~\ref{fig:SL_continuous_mimickers}. When considering the results
for the precessing black hole mimicker-like waveform based on~\citet{dietric_scaled_bns}, the posteriors change more from one situation 
to the other, as can be seen in the left panel of Figure~\ref{fig:SL_continuous_mimickers}. 
In many cases, they are mostly uninformative. However, for 
$M = 120, \, 130, \, \mathrm{and} \, 150 M_{\odot}$, there is a higher support for values close to $n = 0.5$.
Therefore, such examples could potentially be mistaken for type II images and would require further
investigations should they be seen.

\bibliographystyle{aasjournal}
\bibliography{tgr_lensing}

\end{document}